\documentclass[a4paper,twoside]{article_pk}
\usepackage{nohyperref}
\usepackage[authoryear]{natbib}
\bibpunct{(}{)}{;}{a}{}{,}
\usepackage{graphicx}
\usepackage{cancel}
\date{} 
\usepackage{lineno}
\newcommand{\oversim}[2]{\protect{\mbox{\lower0.5ex\vbox{%
   \baselineskip=0pt\lineskip=0.2ex
   \ialign{$\mathsurround=0pt #1\hfil##\hfil$\crcr#2\crcr\sim\crcr}}}}} 
\newcommand{\simgreat}{\mbox{$\,\mathrel{\mathpalette\oversim>}\,$}} 
\newcommand{\simless} {\mbox{$\,\mathrel{\mathpalette\oversim<}\,$}} 
\title{{\small CSIRO Publishing - PASA, submitted: 13.01.2012; accepted:
  11.03.2012; published early online: open access article available at 
http://www.publish.csiro.au/paper/AS12005.htm
  \hfill\hfill}\newline%
\large\bf\flushleft The dark matter crisis: falsification of
  the current standard model of cosmology}
\author{\parbox{\textwidth}{\flushleft
\vspace{-0.5cm}
{\it Pavel Kroupa}\\
\vspace{0.4cm}
{\small Argelander Institute for Astronomy, University of Bonn, Auf
  dem H\"ugel 71, 53121 Bonn, Germany}\\
{\small Email: pavel@astro.uni-bonn.de}}}
\begin{document}
\twocolumn[
\begin{changemargin}{.8cm}{.5cm}
\begin{minipage}{.9\textwidth}
\vspace{-1cm}
\maketitle
%
%
\small{\bf Abstract: The current standard model of cosmology (SMoC)
  requires The Dual Dwarf Galaxy Theorem to be true according to which
  two types of dwarf galaxies must exist: primordial dark-matter (DM)
  dominated (type A) dwarf galaxies, and tidal-dwarf and
  ram-pressure-dwarf (type B) galaxies void of DM. Type A dwarfs
  surround the host approximately spherically, while type B dwarfs are
  typically correlated in phase-space. Type B dwarfs must exist in any
  cosmological theory in which galaxies interact. Only one type of
  dwarf galaxy is observed to exist on the baryonic Tully-Fisher plot
  and in the radius-mass plane. The Milky Way satellite system forms a
  vast phase-space-correlated structure that includes globular
  clusters and stellar and gaseous streams. Other galaxies also have
  phase-space correlated satellite systems. Therefore, The Dual Dwarf
  Galaxy Theorem is falsified by observation and dynamically relevant
  cold or warm DM cannot exist. It is shown that the SMoC is
  incompatible with a large set of other extragalactic
  observations. Other theoretical solutions to cosmological
  observations exist. In particular, alone the empirical
  mass-discrepancy--acceleration correlation constitutes convincing
  evidence that galactic-scale dynamics must be Milgromian. Major
  problems with inflationary big bang cosmologies remain unresolved.
}


\medskip
\medskip
\end{minipage}
\end{changemargin}
]
\small
\cleardoublepage
\tableofcontents 
\cleardoublepage

\section{Introduction}
\label{sec:introd}

``{\it For if we are uncritical we shall always find what we want: we
  shall look for, and find, confirmations, and we shall look away
  from, and not see, whatever might be dangerous to our pet
  theories. In this way it is only too easy to obtain what appears to
  be overwhelming evidence in favor of a theory which, if approached
  critically, would have been refuted}'' (\citealt{Popper57}, p.~124).

That Einstein's general theory of relativity (GR,
\citealt{Einstein16}) is an excellent description of gravitational
physics has been established in the weak and strong (i.e. Solar System
and Earth, respectively) and very strong (black hole and neutron star)
field limits (e.g. \citealt{Freire10}).  Albert Einstein had developed
his field equation such that the Newtonian equations of motion be
derivable from it, and thus that it be consistent with the then
available celestial phenomena\footnote{But the new theory implied a
  perihelion shift of Mercury, as was observed but not understood with
  Newton's theory.}.  The currently popular understanding of cosmology
is based on the null hypothesis (``Hypothesis~0i'') that GR also be
valid on galactic and cosmological scales. This is a vast
extrapolation by many orders of magnitude from the well-tested scale
of planetary dynamics to the galactic and cosmological
ultra-weak-field scales, the dynamics of which were probed
\citep{RF70} only long after GR had been finalised {by Einstein
  using Newtonian, i.e. essentially Solar system constraints}. The
nature of spiral nebulae and the dimensions of the universe were
debated in 1920~by Harlow Shaply and Heber Curtis in The Great Debate,
but galactic and extragalactic distance scales were proven later
\citep{Opik22, Hubble29}.  The other ``Hypothesis~0ii'', so
fundamental that it is usually not stated, is that all present-day
matter is created as a relativistic fluid during the hot Big Bang
(BB).

The observed state of the universe at the present-epoch is such that
within the visible horizon physics appears to be identical. This
implies that every part of the visible universe had to have been in
causal contact at the BB, the geometry is extremely close to being
flat as is deduced from the position of the cosmic microwave
background (CMB) acoustic peaks, and the universe appears to be
homogeneous and isotropic on large scales.  Since there is
observational evidence suggesting that the universe began in a very
dense hot state, a disagreement with these observations emerged
because GR plus the BB would predict a highly curved inhomogeneous
universe.  Thus inflation \citep{GT80,Sato81} was postulated as an
auxiliary hypothesis (``Hypothesis~1'') in order to solve the
causality, flatness, homogeneity and isotropy problems implying a
massive expansion of volume by at least a factor of $10^{78}$ driven
by a scalar field called the ``inflaton''.

Hypothesis~0i implies that dynamics on galaxy sca\-les must be
Newtonian.  With the observation that galactic rotation curves remain
nearly flat at large radii \citep{RF70,Bosma81}, and the rapidity with
which structure emerges after the BB, new failures of the model
emerged. These were solved by introducing a second auxiliary
hypothesis (``Hypothesis~2''), namely that exotic cold (C) or warm (W)
dark matter (DM) particles be the dominant form of gravitating matter.
The mass of the DM particle defines whether it is C or W: roughly
1--10$\,$keV for WDM and above that for CDM although axions may be of
smaller mass and still be CDM \citep{BFPR84}. These postulated new
particles have to interact through gravitation and perhaps weakly in
order to decouple from the photon fluid and start to form structures
before the baryons can.

The hypothesised existence of DM particles resonated with the
contemporary extension of particle physics (see
e.g. \citealt{Aliuetal12} for an account) by string theory (see
\citealt{Smolin06} for an overview as well as criticisms) and in
particular by super-symmetry (e.g. \citealt{WZ74,WB92, WA98,Gao11} and
references therein). Super-symmetry is motivated by the ``hierarchy
problem'', because the constants of the standard model of particle
physics (SMoPP) are highly fine-tuned. For example, the weak force
being $10^{32}$ times stronger than gravity is claimed to be solved
naturally by super-symmetry. These extensions contained new excited
states that would appear in the form of additional particles beyond
the SMoPP. This work was in turn driven by the previous successful
prediction of atoms, electrons and the neutrino and anti-particles,
and by the wish to understand the SMoPP in terms of a deeper physical
description of matter in unification with GR.

The SMoPP is indeed a brilliant success in accounting for the known
particles and their excited states, but has many parameters the origin
of which remain unknown \citep{Yao06}. While no significant evidence
for a failure of the SMoPP has emerged so far, it must be incomplete
because it does not account for the oscillations of the neutrino. It
accounts for the electromagnetic, weak and strong interactions, but
fails to unify the latter with the two former and is also understood
to be incomplete because it does not account for gravitation, dark
energy, nor does it contain any viable DM particle.  Leaving aside the
issues with dark energy and DM, the tremendous success of the SMoPP
can be seen in the recent break-through achieved in
quantum-chromo-dynamical supercomputer calculations by accounting for
the Hoyle state \citep{Epelbaum11}.

Dark energy (DE) (e.g. \citealt{Bousso08, Amendola10, Bousso12,
  Afshordi12}) had to be introduced into the cosmological model as a
third auxiliary hypothesis (``Hypothesis~3'') because the
interpretation of flux and redshift data of supernova type~Ia, given
Hypothesis~0--2, suggest the universe to expand increasingly rapidly
(\citealt{Riess98, Schmidt98, Perlmutter99}, and
e.g. \citealt{Kowalski08}).  DE can be seen as the cosmological
constant $\Lambda$ inherent in Einstein's GR formulation. DE is
leading to a new era of inflation, and as BB inflation, the
corresponding particle-physics formulation is unknown.

Currently, it can be stated that inflation and DE are mathematical
ansatzes allowed by GR to solve failures of the straight-forward
combination of GR plus particle physics \citep{Brandenberger12}.  An
unsolved issue is if these ansatzes contain physics.

The resulting $\Lambda$CDM or $\Lambda$WDM model can summarizingly be
referred to as the currently popular standard model of cosmology
(SMoC). Within the SMoC, structures first form through gravitational
instabilities in the dissipationless cold or warm DM. The baryonic
matter, once sufficiently cooled, accretes into the DM potential wells
and begins to form stars. The emerging galaxies merge and interact,
forming the present-day observed cosmological structures (filaments,
galaxy clusters, galaxies and voids, \citealt{PS11} and references
therein).

The SMoC is widely held to be an excellent description of cosmological
reality. It is defined by a large number of parameters, the most
important of which define a flat space time and the energy content of
the universe to be about 4~per cent by baryonic matter, about 23~per
cent DM and about 73~per cent DE (e.g. \citealt{Kowalski08,
  FM12}). According to the SMoC the universe consists to more than
96~per cent of unknown physics.

Among the often stated great successes of the SMoC are the excellent
reproduction of the angular power spectrum of the galaxy distribution
(e.g. \citealt{Tegmark02,Hayes12}), the success in accounting for the
primordial Helium fraction through BB nucleosynthesis
(e.g. \citealt{Bludman98}) and in accounting for the CMB power
spectrum (e.g. \citealt{AD11}), whereby the latter two are not
sensitive to the validity of the SMoC as such (see
Sec.~\ref{sec:MONDcosm}).

While problems with the SMoC have been arising on galaxy scales,
typically it is held that our incomplete knowledge of baryonic physics
is responsible.  Conclusive tests of the model therefore become
difficult, since various aspects of baryonic physics can always be
invoked to argue that a given failure is not conclusive.

{\it Is it possible to test the SMoC in such a way that the test is
  independent of the details of the baryonic processes?} This
contribution details just such tests. Since the pioneering
cosmological N-body work by \cite{Aarseth79}\footnote{At the {\it
    Aarseth-Nbody meeting in Bonn} in December~2011 Sverre Aarseth
  explained over a Gl\"uhwein at the Christmas market why he did not
  continue his pioneering cosmology work: he left the field because
  the necessity of introducing dynamically relevant dark matter
  particles became too speculative in his view. Thereafter Sverre
  concentrated on developing the Aarseth N-body codes for collisional
  dynamics research.}, the cosmological N-body industry has matured to
a vast and highly active world-wide research effort. The large volume
of published output of this industry makes robust tests possible.
This numerical work combined with observations is the basis for
inferring or excluding the existence of dynamically relevant cold or
warm DM without the need to resort to direct dark-matter particle
searches. The nature of gravitation can therewith be probed in the
ultra-weak field limit.

{\it Can the SMoC be tested on the small scales of dwarf galaxies? Are
  the available simulations of high enough resolution?} Yes and yes:
The dynamics of dissipationless DM particles that orbit within the
emerging potentials is well understood within the SMoC.  The vast
world-wide effort to address the sub-structure, or missing-satellite,
problem has been producing consistent results on the distribution of
the theoretical satellite population (see Sec.~\ref{sec:fil_darkforce}
for a dialogue and Footnote~\ref{foot:paperlist} for a list of many
papers on this issue). Furthermore, the main part of the argument here
rests on the phase-space distribution of sub-structures over scales of
10s to 100s of kpc, which is a scale well resolved.

Returning to the logics of scientific theories, it ought to be
generally accepted that {\it for a logical construction to be a
  scientific theory it has to be falsifiable.  Otherwise predictions
  are not possible and the construction would not allow useful
  calculations.}  Following \cite{Popper34}, a fundamental assumption
underlying the approach taken here is that cosmological theory be
falsifiable. The classical view of philosophy that hypothesis be
proven by experiment is abandoned, because this approach would require
deduction of general rules from a number of individual cases, which is
not admissible in deductive logic. Thus, a single counter-example
suffices to disprove a hypothesis, while no number of agreements with
data can prove a hypothesis to be true.

In this contribution the SMoC is falsified using straightforward
logical arguments as detailed below and summarised in
Sec.~\ref{sec:concs}.

\section{Definition of a galaxy}
\label{sec:defgal}

The definition of a {\it galaxy} \citep{FK11,WS12} is an important
question to consider because depending on it a whole class of objects
may be excluded which may forestall further intellectual advance. For
example, we may define a tidal dwarf galaxy (TDG) to be a self-bound
system with stars and gas with baryonic mass $>10^7\,M_\odot$ formed
within a tidal arm in a galaxy--galaxy encounter. With this
definition, self-gravitating objects formed in a tidal arm but with
lower masses would not constitute TDGs and we might then not be
allowed to associate them with the dSph satellite galaxies of major
galaxies. Given the evidence presented in this contribution this would
be unphysical. Therefore, a more general definition of a galaxy needs
to be used.

Throughout this text it is implicitly assumed that a self-gravitating
object which consists of stars and perhaps gas is a {\it galaxy} if
its Newtonian median two-body relaxation time is longer than the
Hubble time, $t_{\rm rel} > \tau_{\rm H} \approx 10^4\,$Myr.  This
definition of a galaxy \citep{Kroupa98,Kroupa08,FK11} naturally
accounts for the dynamical process of energy equipartition not playing
a role in galaxies over the age of the universe.  In contrast, {\it
  star clusters} with $t_{\rm rel} < \tau_{\rm H}$, have an evolving
morphology as a result of two-body-encounter driven evolution towards
energy equipartition.  Thus, ultra-compact dwarf galaxies and
ultra-faint dwarf satellites are galaxies such that the collision-less
Boltzmann equation and the Jeans equations may be used to study their
properties. In other words, the phase-space density of stars remains
constant in such a system and, put in yet other but synonymous words,
the phase-space distribution function is incompressible in a galaxy
over the age of the universe.

Thus, any object with a mass $M$ in $M_\odot$ and a half-mass radius
$R_{0.5}$ in~pc is here taken to be a {\it galaxy} if (combining
eq.~4, 5 and 7 in \citealt{Kroupa08} for an average stellar mass of
$0.5\,M_\odot$)
\begin{equation}
  R_{0.5} > \left(10^4\;{ {\rm ln}(M) + 0.69 \over
      6.0} \right)^{2\over3} M^{-{1\over 3}}.
  \label{eq:defgal}
\end{equation}
For example, for $M=10^4\,M_\odot$, $R_{0.5}>30\,$pc. Dwarf elliptical
(dE), dwarf spheroidal (dSph) and ultra-faint dwarf (UFD) galaxies are
then indeed galaxies according to this definition.  Ultra-compact
dwarfs (UCDs) would also be galaxies (in agreement with their
classification as such by \citealt{Drinkwater98, Drinkwater04}, see
Fig.~\ref{fig:stellar_systems} below). Note that {\it all} objects
traditionally referred to by astronomers as {\it galaxies} are
galaxies according to eq.~\ref{eq:defgal}, while traditional {\it star
  clusters} are star clusters according to eq.~\ref{eq:defgal}.

\cite{Yoshida08} discovered star formation in gas clouds stripped from
a disk galaxy which is falling into a galaxy cluster (see also
\citealt{Yagi10}). These star-forming objects, which they call
``fireballs'', have physical properties next to identical to dSph
satellite galaxies ($10^6 \simless M/M_\odot \simless 10^7$,
$200\simless$ $R_{0.5}/{\rm pc}\simless 300$). He\-re these galaxies
are referred to as {\it ram-pressure dwarf galaxies} (RPDGs).

According to eq.~\ref{eq:defgal}, a galaxy is a stellar-dynamically
unevolved self-gravitating system (ignoring higher-order relaxational
processes). At older times than the current age of the universe the
boundary between star clusters and galaxies will shift to more massive
objects, which would be consistent with for example the evaporation of
stars having progressed to deplete what are present-day ``galaxies''
such that they would necessarily have to be counted as massive star
clusters then (cf. \citealt{Chilingarian11}).

An alternative definition, according to which a {\it gala\-xy} is a
self-bound stellar-dynamical object with characteristic
radius$\,\simgreat 100\,$pc \citep{Gilmore07b}, is based on the
absence of observed objects in the radius range $30-100\,$pc (the
``Gilmore gap'', e.g. Fig.~\ref{fig:stellar_systems} below). According
to this definition, TDGs would also be classified as galaxies.  UCDs
would be star-clusters (compare with
\citealt{Drinkwater00,Mieske02,Mieske12}, see also
\citealt{Hilker99}).

\section{Rigorous predictions of the SMoC}
\label{sec:pred}

The following are robust predictions of the SMoC as discussed in the
following sub-sections:

\vspace{2mm}

\noindent {\sc Robust SMoC predictions:} 
\begin{description}
\item I) Each MW-sized galaxy contains hundreds of DM-dominated
  satellite galaxies (``type~A'' dwarfs) within the virial radius of
  its DM halo.

\item II) Due to their mostly individual in-fall histories type~A dwarfs
  are distributed approximately spherically about the host,
  following the distribution of host-halo DM particles.

\item III) A significant fraction of MW-sized galaxies that had
  previous encounters with other galaxies are surrounded by TDGs
  (``type-B'' dwarf galaxies).  Rich galaxy clusters should also
  contain RPDGs (Sec.~\ref{sec:defgal}) which are also of type~B.
  Type~B dwarfs do not contain significant amounts of DM.

\item IV) Due to energy and angular momentum conservation type~B dwarfs
  are typically distributed in phase-space correlated structures about
  their host galaxies. 

\item V) The number of type~B dwarfs is comparable to the number
  of dE galaxies. 

\end{description}

\noindent Note that predictions~I and~II follow from a conjecture made by
\cite{Zwicky37}: Interpreting his observational data in terms of
Newtonian dynamics, he suggested that galaxies must be significantly
more massive as their motions in the Coma cluster of galaxies are too
rapid. This can be framed today as a Conjecture \citep{Kroupa10}:

\vspace{2mm}

\noindent {\sc Conjecture 1:}\\
{\it Galaxies contain DM, and by implication this DM must be cold or
  warm because hot DM would not condense to galaxy-sized structures.}

\vspace{2mm}

He also concluded from observation that new dwarf galaxies (i.e. TDGs)
form from the matter expelled during galaxy encounters
(\citealt{Zwicky56}, his p.~369).  Again, this may be stated today as
another Conjecture \citep{Kroupa10}:

\vspace{2mm}

\noindent {\sc Conjecture 2:}\\
{\it When galaxies interact TDGs form from the matter expelled during
  the encounter. These are largely DM free.}  

Predictions~III--V are related to this conjecture.

\vspace{2mm}

\noindent
\label{noteI}{\sc Note I:} Type~A dwarfs are speculative because they
rely on the existence of C/WDM particles. The existence of type~B
dwarfs, on the other hand, is observationally established (end of {\sc
  Note I}).

\subsection{Type-A dwarfs}
\label{sec:typeAdwarfs}

One prediction of the SMoC which is valid for {\it all} galaxies and
is independent of the details of baryonic physics is that {\it each
  and every primordial galaxy is surrounded by a DM halo with a
  significant amount of sub-structure in the form of self-bound DM
  sub-haloes} \citep{Klypin99,Moore99}.

As pointed out by \cite{Moore99}, ``The model predicts that the
virialized extent of the Milky Way's halo should contain about~500
satellites with ... bound masses about $>10^8\,M_\odot$ and tidally
limited sizes larger than about 1~kpc.''  The sub-structure is in the
form of individual dark-matter sub-haloes (self-bound sub-structures)
which follow a power-law mass function (\citealt{Mac11} and
\citealt{Kroupa10}, and references therein).  The sub-structured halo
of any galaxy is a necessary dynamical consequence of Hypothesis~2
(Sec.~\ref{sec:introd}), through gravitationally self-bound structures
developing on all scales from dissipationless cold or warm DM
particles in an expanding universe. 

The fraction of mass within self-bound sub-struc\-tur\-es increases
with radial distance, from 0.5~per cent within a distance of~35~kpc to
about 14~per cent mass fraction within the whole MW-like host halo,
while near the radius of the host halo about 30~per cent of the mass
is in self-bound sub-structures \citep{Mac11}.  \cite{Kravtsov04} find
that about 10~per~cent of the sub-haloes with present-day masses
$\simless 10^8-10^9\,M_\odot$ had substantially larger masses and
circular velocities when they formed at redshifts $z\simgreat
2$. Tidal stripping of DM satellites is therefore not a process of
significance for the whole population of self-bound sub-structures.

These sub-haloes are distributed in a spheroidal and nearly isotropic
manner within the host halo (fig. 12 in \citealt{Metz07}), as is
explicitly demonstrated by \cite{Pawlowski12}.  Within its virialised
region, the spatial distribution of sub-structure in a present-day DM
host halo closely follows that of its DM distribution
\citep{Diemand08}.  Statistically, the anisotro\-py of DM haloes
amounts to not more than about 15~per cent \citep{Aubert04}.  CDM
models predict the host DM haloes to be oblate with flattening
increasing with increasing mass and radius \citep{Combes02,
  Merrifield02}.  The ratio of minor to major axis of the DM density
distribution has the value $q_d = 0.7 \pm 0.17$ for MW sized haloes
within the virial radius. The intermediate-to-major-axis ratio is
$q_d' \simgreat 0.7$ \citep{Bullock02}.  When dissipative baryonic
physics is taken into account the haloes become more axis-symmetric
(larger $q_d'$) and more flattened, $q_d = 0.5 \pm 0.15$ within the
virial radius. The minor axis is co-linear with the angular momentum
of the baryonic disk \citep{Dubinski94}.  

Concerning the MW, the empirical evidence is that its DM halo may be
somewhat flattened (oblate) with $q_d \simgreat 0.8$ within $R
\simless 60\,$kpc \citep{OM00, OM01, Ibata01, Majewski03, Martinez04}.
For example, \cite{RPT07} find an oblate DM halo to fit the Magellanic
Clouds and Stream.  \cite{KRH10} discover, on analysing the GD-1
stellar stream, that within about 15~kpc from the MW center its halo
is essentially spherical. Beyond this distance the shape is likely to
be more oblate \citep{Bullock02,RPT07}, and invoking continuity shows
that the axis ratio $q_d$ cannot change drastically
(e.g. \citealt{VC11}). The theoretical sub-structure distribution
around MW-type hosts must therefore be quite isotropic
\citep{Ghigna98,ZB03,Diemand04}

Turning to the warm-dark matter SMoC, $\Lambda$WDM, it has been shown
that the spatial and kinematical distribution of sub-haloes cannot be
distinguished from those of the $\Lambda$CDM models
\citep{Bullock02,Knebe08}. In $\Lambda$WDM cosmologies, the sub-haloes
are slightly more spherically distributed than in $\Lambda$CDM
cosmologies \citep{Bullock02}. The number of WDM sub-structures is
reduced in comparison with CDM haloes.

Therewith we can use $\Lambda$CDM and $\Lambda$WDM synonymously when
discussing the spatial properties of satellite galaxy
distributions. {\it The {\sc Robust Prediction} of the CDM or WDM SMoC
  is thus that the sub-haloes within each host halo are distributed
  nearly isotropically following the host halo density distribution.} 

A challenge facing the SMoC is to quantify how merely a small fraction
of the sub-haloes become luminous to appear as primordial (type~A)
dwarfs, while the rest avoids forming stars. For example, the MW is
supposed to have many hundred to thousands of DM sub-haloes while
only~24 satellite galaxies have been discovered. This {\it missing
  satellite problem} \citep{Klypin99,Moore99} is deemed to have been
solved through the adjustment of various baryonic processes stopping
star-formation in most sub-haloes (see Sec.~\ref{sec:fil_darkforce}
for a dialogue and a list of many contributions dealing with this
problem in Footnote~\ref{foot:paperlist}). According to this large
body of work those dwarf galaxies that form in some of the many DM
sub-haloes are DM dominated in their optical regions.  Most of the
research effort (e.g. \citealt{deLucia12} for a review) has dealt with
the missing satellite problem by avoiding the {\it disk of satellites
  problem} (\citealt{Kroupa05}, and Failure~8 in
Sec.~\ref{sec:failures_list}).

\subsection{Type-B dwarfs}
\label{sec:typeBdwarfs}

Zwicky's observation that new dwarf galaxies form from the material
ejected as tidal tails when galaxies interact (Conjecture~2 in
Sec.~\ref{sec:pred}) has been confirmed many times since the seminal
paper by \cite{Mirabel92} who reported such an event for the first
time in detail.\\

\noindent Three implications follow from Conjecture~2:\\

\subsubsection{Implication 1}

TDGs and RPDGs cannot contain a dynamically significant amount of DM
\citep{BH92,Bournaud10} because the phase-space DM particles occupy is
far larger than the cross section for capture by type~B dwarfs.  The
phase space occupancy of gas and stars in the progenitors of the tidal
debris from which TDGs are born or of the ram-pressure-stripped gas
clouds from which the RPDGs form is that of a dynamically cold, thin
disk which is very different from the dynamically hot, quasi-spherical
halo of dark matter.  The tidal and ram-pressure disruption process is
very efficient at segregating the two components, because particles of
similar phase-space occupancy retain this occupancy.  So there should
be basically no DM left attached specifically to tidal or ram-pressure
stripped debris or any TDGs or RPDGs that form therefrom.  The DM
particles have virialised velocities too large to be trapped in the
small forming baryonic potentials of TDGs and RPDGs (for a review of
the formation of TDGs see \citealt{Bournaud10}).

Can a TDG or RPDG be accreted onto a pre-existing DM sub-halo? No. It
would have to intercept the sub-halo in coordinate space and in
velocity space. The latter is extremely unlikely because the relative
velocity of the TDG or RPDG and the sub-halo is similar to the
velocity dispersion of DM particles in the host halo such that a
sub-halo with a circular velocity below 50~km/sec will be unnoticed by
the TDG or RPDG moving faster than 100~km/sec.\\

\subsubsection{Implication 2}
\label{sec:impl2}

Type~B dwarfs form as a population in individual galaxy--galaxy
encounters. As a consequence they are correlated in phase-space
because their orbital angular momenta retain a memory of their
formation. 

\subsubsection {Implication 3}
\label{sec:impl3}

Assuming the SMoC to be correct within which larger galaxies form from
the mergers of smaller objects, it follows that type~B dwarfs may be a
prominent contribution to the satellite dwarf galaxy population:

\cite{OT00} computed the expected population of TDGs within the
SMoC. They adopted a structure formation merger tree and assumed each
gas-rich encounter only produces~1--2 long lived TDGs that evolve from
dwarf irregular (dIrr) gas-rich galaxies to dwarf elliptical (dE)
galaxies in the tidal environment around a host galaxy or within the
group or cluster. \cite{OT00} thereby discovered that so many TDGs
would be generated over a Hubble time to account for all dE
galaxies. The morphology--density relation of galaxies, according to
which rich groups or clusters have more dwarfs, emerges naturally as
well.  That dIrr galaxies do evolve to dE and dwarf spheroidal (dSph)
galaxies in a tidal environment has been demonstrated by
\cite{Mayer01}.  An impressive example of how a number of TDGs formed
around Andromeda, which bears rather clear signatures of past
interaction events, is shown in the simulations by \cite{Hammer10}.

The estimate by \cite{OT00} is, however, a lower-limit on the number
of dwarf galaxies in galaxy clusters, because they did not take into
account the formation of RPDGs (Sec.~\ref{sec:defgal}).

As demonstrated by simulations (\citealt{Wetzstein07}, see also
\citealt{Bournaud10, Bournaud11}), the number of TDGs formed scales
with the gas-fraction in the interacting galaxies.  That TDGs form
profusely at high redshift from interactions of gas-rich galaxies is
implied by the high-resolution simulations by \cite{Bournaud11}.  At
early cosmological epochs, when the forming galaxies were very gas
rich, closer together and interacting more often in small groups than
today especially in the then emerging clusters of galaxies, the
formation rate of TDGs is likely to have been significantly higher
than today, perhaps by orders of magnitude per encounter. Indeed, it
may even not be possible to discern the formation of primordial dwarf
galaxies from TDG formation as they probably occurred simultaneously
and not independently of each other. Clearly, there is much scope for
further seminal research in this area.\\

\subsubsection{Longevity of type~B dwarfs}

Once formed, can TDGs or RPDGs vanish? No: they have masses $\simless
10^9\,M_\odot$ and dynamical friction on the DM halo of the host
galaxy will therefore not significantly shrink their orbits over a
Hubble time. Unless such a type~B dwarf is on a radial orbit, it will
remain close to its original orbit apart from precession.
\cite{Kroupa97}, \cite{KK98} and \cite{Casas12} have demonstrated,
using high-resolution computations, that DM-free TDGs with an initial
stellar mass of $10^7\,M_\odot$ readily survive for times comparable
to a Hubble time on eccentric orbits in the tidal field of a host
galaxy.  Quite stable, dSph-like solutions with remnant masses of
$10^4-10^5\,M_\odot$ appear in such models \citep{MK07, Casas12}.  And
\cite{Recchi07} have shown, using chemo-dynamical calculations, that
DM-free gas-rich TDGs (or RPDGs for that matter) are not destroyed
through the star-formation process and that they self-enrich
chemically in agreement with observations of dwarf satellites. Type~B
dwarfs that retain their gas would appear as dIrr or low-mass disk
galaxies \citep{Hunter00}.

There is much observational data on the formation of new and on the
existence of older TDGs (Sec.~\ref{sec:extragal}, \ref{sec:dEorigin}
and e.g. \citealt{Pawlowski12b, Dabringhausen12}), and the formation
of RPD\-Gs has also been documented \citep{Yoshida08} and studied
(\citealt{Yagi10}, Sec.~\ref{sec:defgal}).  Such observational work
sometimes concludes that only a small fraction of satellite galaxies
may typically be of type~B (e.g. \citealt{Kaviraj12}). However, the
observational census of young type~B dwarfs is currently flux-limited
such that the detections are limited to the present-day universe and
the low-mass ($<10^7\,M_\odot$) type~B dwarfs are not detected.
Examples of low-mass TDGs that have formed in tidal arms are the about
5~Myr old star-cluster complexes in the Tadpole galaxy and the three
about 1--2~Gyr-old dSph-like TDGs in NGC$\,$5557
(Sec.~\ref{sec:extragal}).

\section{The Dual Dwarf Galaxy \\ Theorem}
\label{sec:dual_theorem}

Within the SMoC there are thus exactly two competing hypothesis for
the origin and nature of dwarf galaxies:

\vspace{2mm}

\noindent {\sc Hypothesis~A:}
\[
{\rm dSph/dE/dIrr} \Longleftrightarrow {\rm DM \; halo}
\]
({\it type~A} dwarfs)

\vspace{2mm}

\noindent and

\vspace{2mm}

\noindent {\sc Hypothesis~B:}
\[
{\rm dSph/dE/dIrr} \Longleftrightarrow  {\rm TDG/RPDG}
\]
({\it type~B} dwarfs). Remember that by Implication~1
(Sec.~\ref{sec:typeBdwarfs}) type~B dwarfs do not contain much DM.

\vspace{2mm}

Within the SMoC it is proven that larger structures form
hierarchically bottom-up from merging smaller
sub-structures. Therefore, mergers and encounters between galaxies are
logically implied events that shape all larger galaxies
(e.g. \citealt{Bournaud11,Martig12}) such that, by {\sc Implication~2}
(Sec.~\ref{sec:typeBdwarfs}), type~B dwarfs appear in significant
numbers. The following theorem has thus been proven:

\vspace{2mm}

\noindent {\sc The Dual Dwarf Galaxy Theorem:}
\[ {\rm SMOC} \; \Longrightarrow \; \exists \; {\rm type\;A} \land {\rm
  type\;B\; dwarfs}.
\]
It states that if the SMoC is true then type~A and type~B dwarf
galaxies exist at the same time. 

\vspace{2mm}

\noindent
\label{noteII} {\sc Note II:} In {\it any} realistic cosmological
theory a weak form of {\sc The Dual Dwarf Galaxy Theorem} must be true
because galaxies form directly from the cooling gas after the BB and
TDGs also form when these {\it primordial galaxies} interact (there
are thus two types of galaxy).  RPDGs can only start forming once
massive galaxy clusters have assembled.  In a (non-Einsteinian)
cosmological theory in which the laws of motion were to be such that
the dynamical properties of {\it all} galaxies are identical without
the existence of DM, no difference in the dynamical properties of
isolated\footnote{In some non-Newtonian theories a satellite galaxy
  may show dynamical differences to the same but isolated dwarf due to
  the {\it external field effect} (see \citealt{FM12} for details).}
primordial galaxies, TDGs and RPDGs would be evident. Type~A dwarfs,
as defined above, would not exist, but primordial galaxies would exist
in addition to the type~B dwarfs.  Thus, dynamically, primordial and
type~B dwarfs would be identical, but type~B dwarfs formed at a late
cosmological epoch may be evident through unusual chemical and stellar
population properties (end of {\sc Note II}).

\vspace{2mm}

In the SMoC, in which C/WDM plays a central role in structure
formation, primordial dwarf galaxies are identical to the above type~A
galaxies and significant dynamical differences to type~B dwarfs are
expected:

If type~A dwarfs exist (i.e. if the SMoC were valid), then they form
within pre-collapsed DM haloes such that the accretion and feedback as
well as environmental physics conditions are different from the
formation of type~B dwarfs without DM and within an expanding gas-rich
tidal arm. In comparison to DM-free dwarf galaxies, galaxies that
derive from DM haloes must show distinctly different morphological
properties (rotation curves, masses, radii, density distributions) as
well as different stellar-populations with distinct age and
chemical-element distributions. Since the latter two are difficult to
quantify we concentrate here on dynamical and morphological properties
\footnote{TDGs may form from pre-enriched material and during their
  formation they may capture stars from the interacting galaxies
  because stars and gas occupy a similar phase-space in disk galaxies.
  It is therefore expected that TDGs forming today will be metal
  enriched compared to primordial (type~A) dwarfs. Such cases have
  indeed been observed \citep{Duc98, MCA12}. However, TDGs may also
  form from the outer metal-poor material of gas-rich disk galaxies
  and they will then commence to self-enrich therewith following the
  usual metallicity--luminosity relation of galaxies. Since the
  majority of TDGs is expected to have formed early in the universe
  before major enrichment of the gas through star formation, such TDGs
  will today appear as normal self-enriched dwarf galaxies. Such TDGs,
  which do follow the metallicity--luminosity relation, may have been
  observed by \cite{Reverte07}. {\it Thus, the metal-rich criterion is
    sufficient to identify dwarf galaxies as being TDGs, but if a
    dwarf galaxy lies on the metallicity--luminosity relation of
    galaxies it cannot be discarded as being a TDG.}  }.

A robust prediction of the SMoC is therefore that there {\it must}
exist DM-dominated and DM-free dwarf galaxies. And, by
Sec.~\ref{sec:typeAdwarfs} type~A (DM-dominated) dwarfs are
distributed spheroidally around their host galaxy, tracing its DM
halo.  Type~B dwarfs, on the other hand, form as a population in
individual galaxy--galaxy encounters. As a consequence they show
correlations in phase space (Sec.~\ref{sec:impl2}).  {\it Therefore,
  in addition to an expected dynamical and morphological difference
  between type~A and type~B dwarfs, the SMoC predicts them to have
  different distributions in phase-space.}

It now becomes possible to test the SMoC at a fundamental level by
studying which of the above two hypothesis may be falsifiable. Note
that for a given dwarf galaxy both Hypothesis~A and~B cannot be valid
simultaneously in the SMoC.

\section{Falsification  of the SMoC}
\label{sec:disproof}

Falsification of {\sc The Dual Dwarf Galaxy Theorem} would invalidate
the SMoC to be a model of the real universe. The procedure followed
here is to test of which type (A or B) the observed dwarf galaxies
are. {\it Firstly}, rotationally supported dwarf galaxies are
considered to see if the two types of observed dwarfs (dIrr/dwarf-disk
vs rotating gas-rich TDGs) do show the necessary dynamical
differences. This is achieved by resorting to the baryonic
Tully-Fisher (BTF) relation. {\it Secondly}, the pressure-supported
dwarf galaxies are considered to see if dSph and dE galaxies differ
from the known TDGs.  Later (Secs.~\ref{sec:HypA} and~\ref{sec:LocG})
the phase-space occupancy and the physical properties of the known
observed MW satellite galaxies are considered to ensure logical
self-consistency of the deduction.

\subsection{Firstly: rotationally supported dwarf galaxies}

Assuming the SMoC to be true it follows by
{\sc The Dual Dwarf Galaxy Theorem} that TDGs cannot lie on the
BTF relation defined by DM-dominated galaxies. Thus,
\[
{\rm SMoC} \; \Longrightarrow \; {\rm BTF}_{\rm dIrr} \ne {\rm BTF}_{\rm TDG}. 
\]

\noindent If, for rotationally supported gas-rich dwarf galaxies

\vspace{2mm}

\noindent {\sc Hypothesis~B},
\[
{\rm dIrr/dwarf\;disk\;galaxies} = {\rm TDG\;or\;RPDG},
\]
\noindent were true, it would follow that the implication,
\cancel{SMoC} (not SMoC), would be true. The following first of two
falsification theorems can now be stated: \vspace{2mm}

\noindent {\sc The First SMoC Falsification Theorem:}
\[
{\rm BTF}_{\rm dIrr} = {\rm BTF}_{\rm TDG} \; \Longrightarrow \; \cancel {\rm SMoC}.
\]
\noindent It states that if TDGs lie on the same BTF relation defined
by primordial (DM-dominated) galaxies then the logically implied
conclusion is that the SMoC is ruled out to be a representation of the
real world.

\subsection{Secondly: pressure supported dwarf galaxies}

Turning now to dE/dSph satellite galaxies: Assuming the SMoC to be
true it follows by {\sc The Dual Dwarf Galaxy Theorem} that TDGs
cannot have the same dynamical and morphological properties as
primordial DM-dominated galaxies. Thus,
\[
{\rm SMoC} \; \Longrightarrow \; {\rm dE/dSph/UFD} \ne {\rm TDG}. 
\]

\noindent If, for pressure-supported dwarf galaxies

\vspace{2mm}

\noindent {\sc Hypothesis~B},
\[
{\rm dE/dSph/UFD} = {\rm TDG},
\]
\noindent were true, it would follow that the implication,
\cancel{SMoC}, would be true.  Thus, the SMoC is proven wrong if the
following second falsification theorem is true:

\vspace{2mm}

\noindent {\sc The Second SMoC Falsification Theorem:}
\[ {\rm dE/dSph/UFD} = {\rm TDG\;or\;RPDG} \; \Longrightarrow \;
\cancel {\rm SMoC}.
\]
\noindent It states that if the dE/dSph/UFD satellite galaxies of the
MW are ancient TDGs and are of the only kind then the logically
implied conclusion is that the SMoC is ruled out to be a
representation of the real world.  This would be the case because if
the MW were to have no dark matter dominated satellite galaxies then
the model is falsified\footnote{We are not considering the trivial ad
  hock solution that {\it all} DM sub-haloes with mass $M_{\rm
    DM}<10^{10}\,M_\odot$ did not form stars and thus remain dark,
  because there is no known physical process that could arrange for
  this to be the case (see \citealt{Ferrero11}).}.

\subsection{Procedure and Logical consisten\-cy}
\label{sec:logconsis}

It now remains to be shown that {\sc The First and Second SMoC
  Falsification Theorems} hold. 

In the real world there are only two logically possible outcomes of
testing these theorems: Either they are both falsified (such that the
SMoC is consistent with reality), or they are both true (such that the
SMoC is falsified as a representation of reality). It is not permitted
to have one Falsification Theorem being true and the other one
false. 

Once {\sc The Dual Dwarf Galaxy Theorem} has been falsified such that
both SMoC Falsification Theorems hold, logical consistency with this
result should imply real data to show discord with the SMoC using
other tests.  Internal logical {\it in}consistency in the present
argument would emerge if such data would indicate excellent agreement
with the SMoC predictions.  The larger part of this contribution is
devoted to studying how the observed universe, where excellent data do
exist, matches to the SMoC.

The argument presented here must be logically sou\-nd. To ensure
logical consistency we therefore cannot rely on measured high
dynamical mass-to-light ratios, $M/L$, as a diagnostic for the
presence of DM. Why is this?

It would be a circular argument: By adopting Hypothesis~0i (GR is
valid) we are forced to introduce auxiliary Hypothesis~2 (DM exists)
due to the mass-discrepancy observed in galaxies. When a
mass-discre\-pancy is observed (as it is in the BTF data of normal
galaxies and in dSph satellite galaxies of the MW) then taking this to
be evidence for DM constitutes a circular argument. It is however
permissible to compare normal/primordial galaxies and TDGs to test
{\sc The Dual Dwarf Galaxy Theorem}.

Here the foundations of gravitational theory in the ultra weak field
limit are being tested.  Rotation curves and large dynamical $M/L$
ratia in MW satellites may also be explained by non-Newtonian theories
(e.g. \citealt{BM00, Angus08, McGaugh10, Hernandez10, FM12}) and tidal
effects \citep{Kroupa97}, so large dynamical $M/L$ values do not have
a unique fundamental solution.

\section{
  Extragalactic evidence: How many types of dwarf galaxies are there
  in reality?}
\label{sec:types}

It has been shown that the SMoC predicts there to be two fundamentally
different types of dwarf galaxy. Which types are there in reality?

Observed dwarf galaxies with stellar masses $M_{\rm
  star}<10^{10}\,M_\odot$ come in two types: dIrr galaxies which are
gas dominated and rotationally supported, and gas-poor dE and dSph
galaxies which are largely pressure (i.e. random motion) supported
\citep{FB94,Mateo98,Hunter00,Dabringhausen08,Forbes08,Lisker09,
  Misgeld11,Swaters11}.  Satellite galaxies are, to a large extend, of
the gas-poor type, which is naturally understood as a result of gas
being stripped from initial gaseous dIrr-type satellite galaxies
\citep{Mayer01}.

\subsection{Rotationally supported dIrr/dwarf-disk galaxies}
\label{sec:BTF}

\begin{figure}[ht]
\begin{center}
\includegraphics[scale=0.4, angle=0]{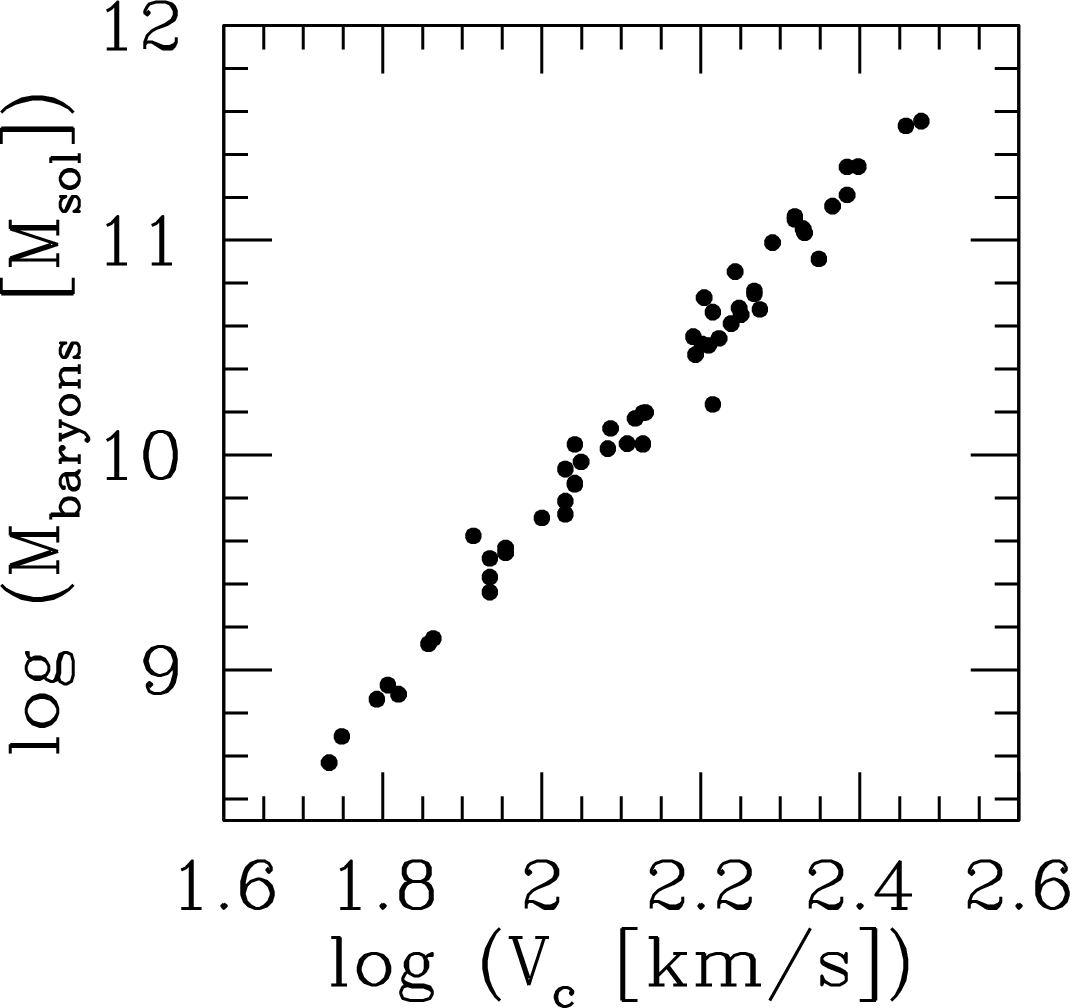}
\caption{The measured mass of all baryons, $M_{\rm baryons}$, in a
  rotationally supported galaxy is plotted in dependence of its
  measured circular rotation velocity $V_c$ (black dots). The
  measurements form a tight correlation, the baryonic Tully-Fisher
  relation of DM dominated galaxies \citep{McGaugh05, Gentile07,
    Trachternach09, McGaugh11, McGaugh12, Desmond12}. This figure was
  kindly provided by Gianfranco Gentile. }\label{fig:BTF_noTDGs}
\end{center}
\end{figure}

To differentiate DM-dominated type~A dwarf galaxies from type~B dwarfs
that contain little or no DM can be achieved by comparing their
internal kinematical state. Type~B dwarfs of similar baryonic mass,
$M_{\rm baryons}$, as type~A dwarfs must have significantly slower
motions of their stellar and gas components. A measure of the DM
content is the asymptotically flat circular velocity,
$V_c$. Fig.~\ref{fig:BTF_noTDGs} shows $M_{\rm baryons}$ vs $V_c$ data
\citep{McGaugh05} for primordial (i.e. DM-dominated) galaxies if
the SMoC were true \citep{Desmond12}.

From the figure it is evident that the data form an excellent
correlation over orders of magnitude in baryonic mass down to $V_c
\approx 15\,$km/sec $\approx 5\times 10^{6}\,M_\odot$
(Fig.~\ref{fig:BTF_MOND} below).\footnote{Galaxies having such a well
  defined BTF relation is a major challenge for the SMoC
  (e.g. \citealt{McGaugh11, PS11, McGaugh12, Desmond12}): For each
  individual DM halo of a given DM mass (i.e. for a given $V_c$), the
  assembly history is different, and a wide range of baryonic content
  would be expected and thus a significant variation of $M_{\rm
    baryons}$ \citep{Disney08}.}

{\it There is thus no evidence for the existence of multiple types of
  rotationally supported dwarf galaxies.}

\subsection{Pressure supported dE/dSph galaxies}
\label{sec:RM}

\begin{figure*}[ht]
\begin{center}
\includegraphics[scale=0.57, angle=0]{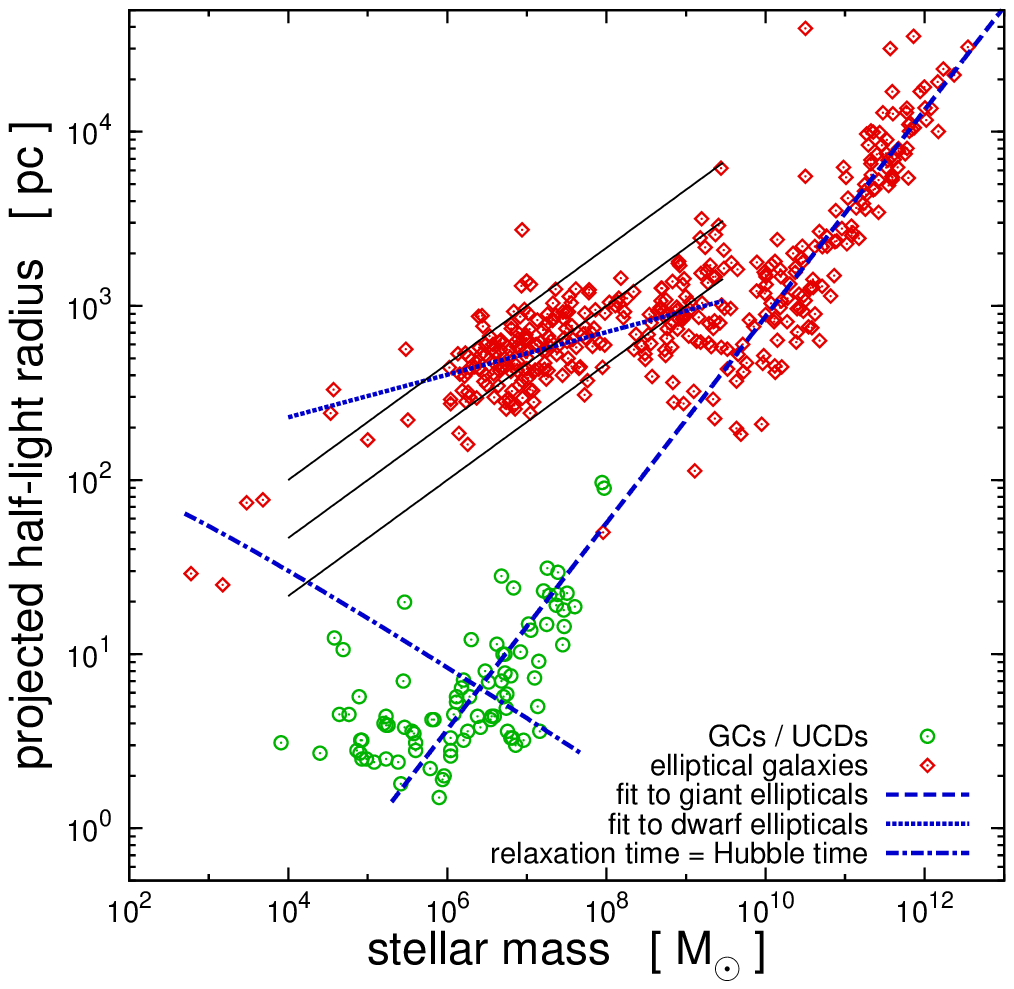}
\vspace{-5mm}
\caption{The types of pressure supported stellar systems: projected
  half-mass radius, $R_{0.5}$, vs stellar mass, $M_{\rm star}$.  Green
  circles are star clusters, $M_{\rm stars} \le 2\times
  10^6\,M_\odot$, and ultra-compact dwarfs (UCDs), $M_{\rm star} >
  2\times 10^6\,M_\odot$, red crosses are elliptical (E, $M_{\rm star}
  > 3\times 10^9\,M_\odot$), dE ($10^7< M_{\rm star}\le 3\times
  10^9\,M_\odot$), dSph ($10^4< M_{\rm star}\le 10^7\,M_\odot$) and
  UFD ($M_{\rm star} \le 10^4\,M_\odot$) galaxies The thin dotted line
  is a fit (eq.~\ref{eq:MRrel}) to dSph and dE galaxies, $b_1=0.122$
  for $10^4 \le M_{\rm star}/M_\odot \le 3 \times 10^9$.  The thin
  solid lines are constant-volume density relations for $1, 0.1$ and
  $0.01\,M_\odot/$pc$^3$ from bottom to top.  The dashed line is a fit
  (eq.~\ref{eq:MRrel}) to the E galaxies, $b_1=0.593$. It is
  extrapolated into the UCD regime and fits there as well, remarkably.
  Any object above the dash-dotted line (eq.~\ref{eq:defgal}, ignoring
  differences in projected quantities) is classified as being a {\it
    galaxy} (Sec.~\ref{sec:defgal}).  This figure was provided by
  Joerg Dabringhausen, and similar figures are available in
  \cite{Dabringhausen08, Forbes08} and \cite{Misgeld11}.  }
\label{fig:stellar_systems}
\end{center}
\end{figure*}

The next question to be addressed is if there are two types of
pressure supported dwarf galaxies, namely those derived from DM
sub-haloes and those derived from TDGs
(Sec.~\ref{sec:pred}). Fig.~\ref{fig:stellar_systems} shows an
overview of the distribution of pressure-sup\-ported stellar systems
in the radius--stellar mass plot.
Taking $R_{0.5}$ to be the projected half-light radius and $M_{\rm
  star}$ to be the stellar mass of the satellite galaxy and writing
the radius--mass relation for pressure-supported stellar systems as
\begin{equation}
{\rm log}_{10}R_{0.5} = b_0 + b_1\,{\rm log}_{10}
\left( M_{\rm star}/M_\odot\right),
\label{eq:MRrel}
\end{equation}
it follows for E galaxies ($M_{\rm star}>3\times10^9\,M_\odot$) that
$b_1=0.593\pm0.027, b_0=-2.99\pm0.30$, as already shown by
\cite{Dabringhausen08}. That this relation extends into the
ultra-compact dwarf (UCD) mass regime, which constitutes an
extrapolation by at least three orders of magnitude in mass, is
noteworthy. It may mean that the genesis of~E galaxies and of UCDs may
have followed the same physical principles, i.e. a rapid dissipational
infrared-opacity-limited collapse \citep{Murray09}, possibly first
into sub-clumps which then merge \citep{Bruens11}.  {Note that the
  rare UCDs are deemed to be related to star clusters
  \citep{Mieske12}. However, UCDs could be identified as the TDGs,
  while dE, dSph and UFD galaxies might be seen as the type~A
  dwarfs. This being a false identification will become apparent in
  Sec.~\ref{sec:HypA_RM}.}

Concerning the dwarf galaxies (dE, dSph, UFD), it is evident that
there is one single branch. Applying
eq.~\ref{eq:MRrel} over the mass range $10^4 \le M_{\rm star}/M_\odot
\le 3 \times 10^9$, the data yield $b_1=0.122\pm0.011$ and
$b_0=1.87\pm0.10$ for dE and dSph galaxies.  As is evident from
Fig.~\ref{fig:stellar_systems}, dSph satellites ($M_{\rm star}
\simless 10^7\,M_\odot$) are an extension of the dE galaxy population
($M_{\rm star} \simgreat$ $10^7\,M_\odot$). This is particularly
emphasised by \cite{Forbes11}. Ultra-faint dwarf (UFD) satellite
galaxies with $M_{\rm star}\simless\,$ $10^4\,$ $M_\odot$ form an
extension of the dSph sequence to lower masses
\citep{Strigari08}.\footnote{On the issue of what constitutes a galaxy
  in view of the faintest satellites recently discovered, as opposed
  to being classified as star clusters, see Sec.~\ref{sec:defgal}.}
\cite{FB94} discuss dE and dSph galaxies as part of the same family.

{\it Thus, two fundamentally different types of satellite galaxies, as
  ought to exist if {\sc The Dual Dwarf Galaxy Theorem} were true
  (Sec.~\ref{sec:dual_theorem}), do not appear to be present.}

\section{Are the known dwarf galaxies of type~A or of type~B?}

It has emerged that for $M_{\rm baryons}<10^{10}\,M_\odot$ there is no
evidence for the existence of galaxy populations with two different
dynamical or morphological properties. This constitutes a
falsification of {\sc The Dual Dwarf Galaxy Theorem}. But perhaps the
observational data only contain dwarfs of the one type~A?  To achieve
a rigorous falsification, TDGs and RPDGs (type~B dwarfs) need to be
directly compared with the putative type~A dwarfs.

The question to be answered now is which type of dwarf galaxies do we
have?  Are dIrr/dwarf-disk, dE and dSph satellite galaxies DM
dominated (i.e. of type~A) or are they TDGs (i.e. of type~B)?  From
past work (see Sec.~\ref{sec:typeBdwarfs}) it is already established
that type~B dwarfs (TDGs), once formed, mostly do not dissolve but
remain on orbits about their host for at least a Hubble time. Since
they are observed to form and because galaxies are known to interact
in the real universe they must be around. On the other hand, the
existence of type~A dwarfs depends solely on the truth of Hypothesis~2
which has until now not been verified. {\it Type~A dwarfs are
  therefore speculative objects, while type~B dwarfs are known to form
  and to survive.}

The following Sections~\ref{sec:HypA_BTF} and~\ref{sec:HypA_RM}
compare the hitherto known ``normal'' dwarf galaxies, which have
popularly but speculatively been assumed to be of type~A, with
observed TDGs (type~B dwarfs).

\section{Hypothesis~A: dIrr and dwarf disk galaxies are DM dominated}
\label{sec:HypA_BTF}

\begin{figure}[ht]
\begin{center}
\includegraphics[scale=0.4, angle=0]{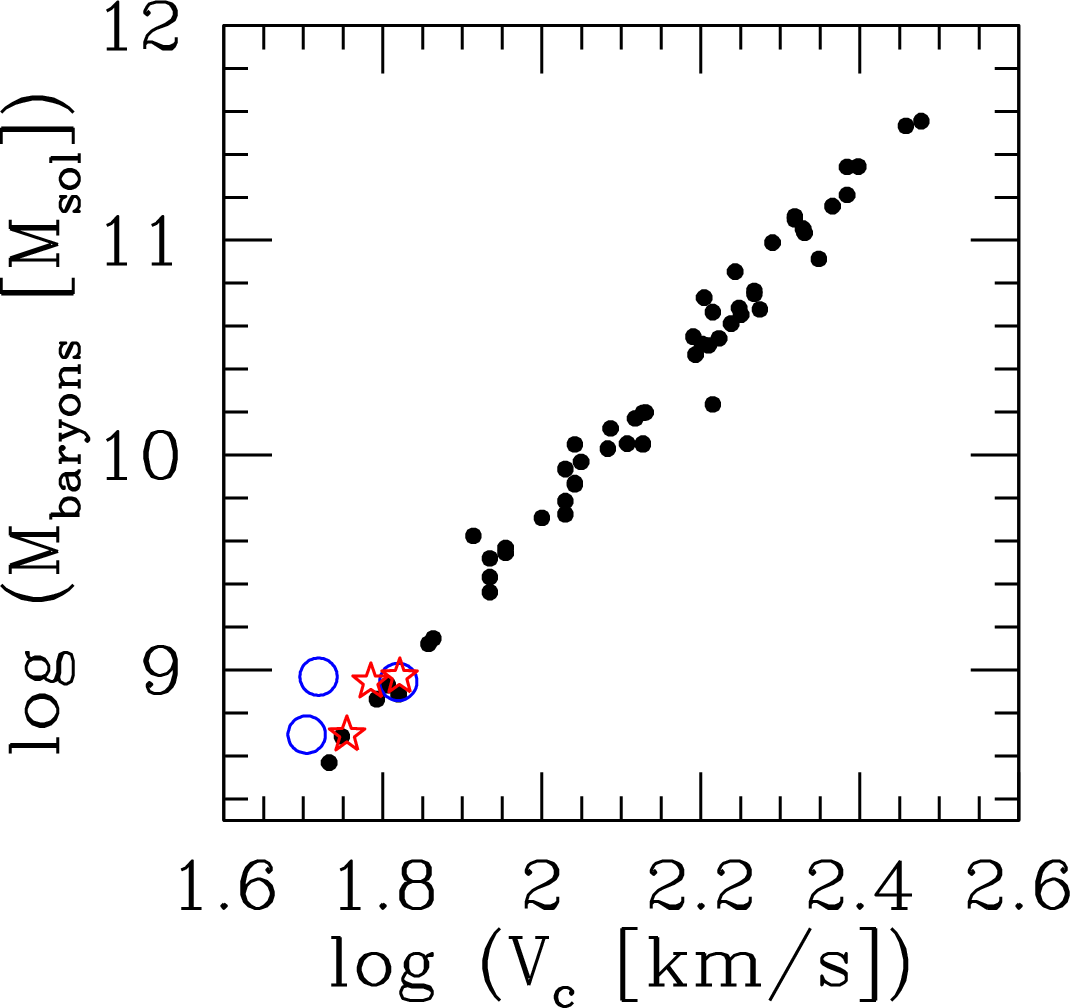}
\caption{As Fig.~\ref{fig:BTF_noTDGs} but with $V_c$ from rotation
  curves of three young TDGs observed by \cite{Bournaud07}
  over-plotted as blue open circles and red open stars.  These TDGs
  belong to the post merger host galaxy NGC$\,$5291, the tidal arms of
  which are seen at an inclination of~45$^{\rm o}$
  (Sec.~\ref{sec:extragal}).  The blue open circles assume the same
  inclination of $i=45^{\rm o}$ for the disks of the TDGs while the
  red stars are for a free inclination (\citealt{Gentile07}).
}\label{fig:BTF_withTDGs}
\end{center}
\end{figure}

If normal dwarf galaxies are DM dominated then the young TDGs of the
same baryonic mass and dimension should have smaller rotational
velocities. That is, in the BTF diagramme (Fig.~\ref{fig:BTF_noTDGs})
the latter should lie significantly to the left of the former. But for
three TDGs rotation curves have been measured, and all three coincide
with the DM-dominated BTF relation (Fig.~\ref{fig:BTF_withTDGs}).

The trivial solution that unobservable gas and/or not virialised young
structures makes up the DM effect evident in the TDG galaxies is
untenable because it would require a strong fine-tuning and chance
configuration between the gas content, the gas flows and $V_c$, to
conspire in each of the three cases to move the TDG onto the BTF
relation of DM dominated galaxies.

It is thus evident that type~B dwarfs (the TDGs) lie on the same
relation as the type~A dwarfs. That is, only one dynamical type of
rotating dwarf galaxy appears to exist. In other words, type~B dwarfs
are identical to type~A dwarfs, ${\rm BTF}_{\rm dIrr} = \; {\rm
  BTF}_{\rm TDG}$, and observed rotationally supported dwarf galaxies
falsify {\sc The Dual Dwarf Galaxy Theorem} in logical consistency
with Sec.~\ref{sec:BTF} and~\ref{sec:RM}. By the {\sc First SMoC
  Falsification Theorem} this implies the SMoC to be falsified.

\section{Hypothesis~A: dE, dSph and UFD galaxies are DM
  dominated}
\label{sec:HypA_RM}

\begin{figure*}[ht]
\begin{center}
\includegraphics[scale=0.57, angle=0]{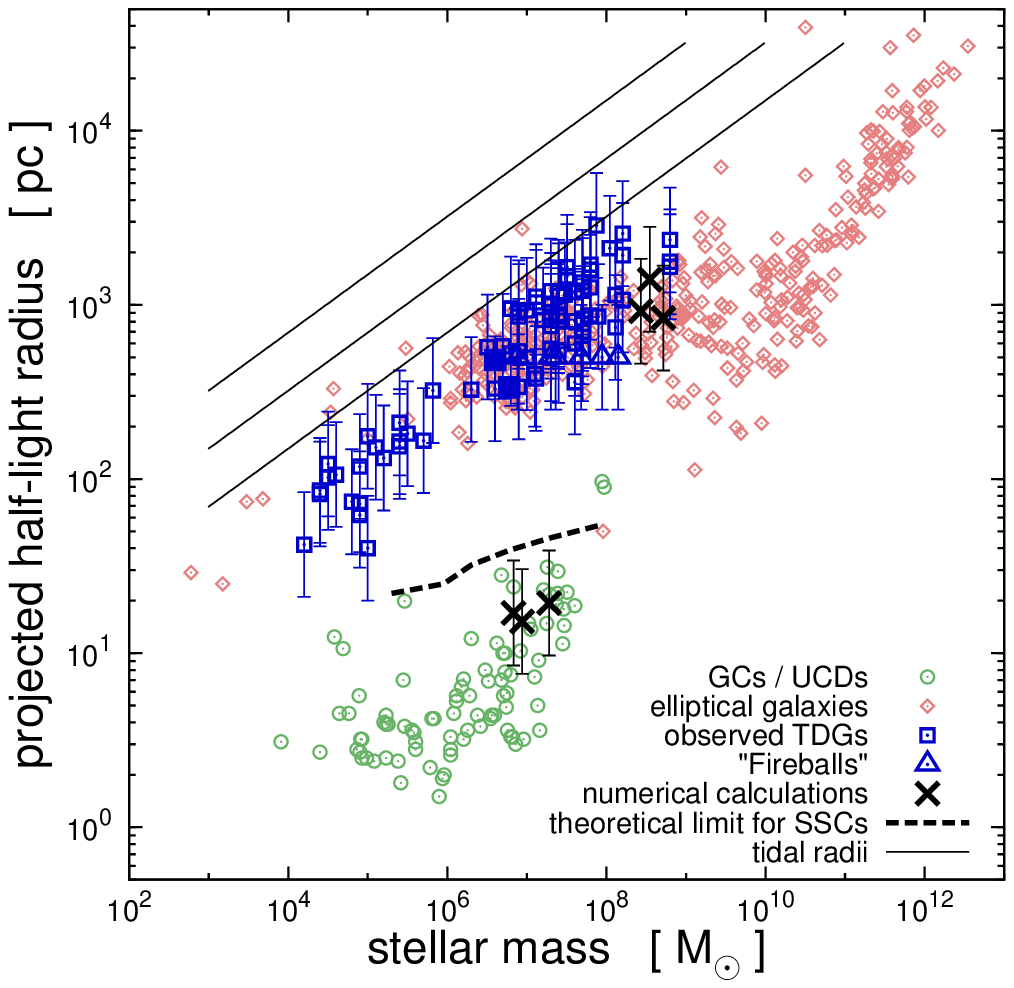}
\vspace{-5mm}
\caption{As Fig.~\ref{fig:stellar_systems} but with type~B dwarfs
  (TDGs and RPDGs) over-plotted (blue squares and triangles).  The two
  older TDGs from \cite{Galianni10} are shown as the solid blue
  squares. They lie along the dog-leg stream in a
  phase-space-correlated structure and are typical dSph satellite
  galaxies (Sec.~\ref{sec:extragal}).  Younger TDGs, which contain
  gas, are shown as the blue squares, whereby the blue triangles are
  RPDGs. The lower end of each blue ``error-bar'' is the present
  location of these objects, the position of the blue square or
  triangle is the radius of the type~B dwarf after adiabatic removal
  of 50~per cent of its mass (gas blow out or stripping), and the
  upper end of each ``error-bar'' shows the location after adiabatic
  removal of 75~per cent of the gas mass assuming the stellar mass
  does not change.  The black symbols denote the star clusters and
  TDGs formed in the high-resolution simulation of \cite{Bournaud08}.
  This figure demonstrates that dE and dSph galaxies, which have been
  until now thought to be hosted by DM haloes, in actuality appear to
  coincide with type~B dwarfs. That is, dEs and dSphs appear to be
  DM-free TDGs, thus proving {\sc The Second SMoC Falsification
    Theorem} (Sec.~\ref{sec:disproof}).  {The three diagonal solid
    lines are, from top to bottom, Newtonian tidal radii
    (eq.~\ref{eq:rtid} below) for host-galaxy masses $M_{\rm
      host}=10^{10}, 10^{11}, 10^{12}\,M_\odot$.}  The dashed curve is
  the observed upper limit for extended star clusters and UCDs
  reproduced by simulations (SSCs, \citealt{Bruens11}).  For details
  see \cite{Dabringhausen12}.  This figure is provided by Joerg
  Dabringhausen.}
  \label{fig:TDGs}
\end{center}
\end{figure*}

Because the physics of the formation of type~A (DM-dominated) galaxies
differs significantly from the formation of type~B dwarfs, they
should show different radii at a given baryonic mass. That is, known
TDGs should not follow the same radius--mass relation as dE, dSph and
UFD galaxies. This is tested in Fig.~\ref{fig:TDGs}.
It is evident that type~B dwarfs (the TDGs) lie on the same relation
as the type~A dwarfs. That is, only one dynamical type of
pressure-supported dwarf galaxy appears to exist.  In other words,
type~B dwarfs are identical to type~A dwarfs, ${\rm dSph} = \; {\rm
  TDG}$, and observed pressure supported dwarf galaxies falsify {\sc
  The Dual Dwarf Galaxy Theorem} in logical consistency with
Sec.~\ref{sec:BTF}, \ref{sec:RM}~and~\ref{sec:HypA_BTF}.  By the {\sc
  Second SMoC Falsification Theorem} this implies the SMoC to be
falsified. This deduction is logically consistent with the observed
dynamical evidence that dE galaxies do not contain much if any DM.

For a further discussion on the origin of dE galaxies 
see Sec.~\ref{sec:dEorigin}.

\section{Hypothesis~A: MW dSph satellites are DM substructures}
\label{sec:HypA}

According to Sections~\ref{sec:HypA_BTF} and~\ref{sec:HypA_RM} the
SMoC is falsified through both the {\sc First} and {\sc Second SMoC
  Falsification Theorems}. If this is true, then other observational
data concerning the properties and distribution of dwarf galaxies must
be consistent with this deduction.

In the following Sections~\ref{sec:LG} to~\ref{sec:log_inc} further
(auxilliary) tests of Hypothesis~A (dSph are embedded in DM
sub-haloes, Sec.~\ref{sec:dual_theorem}) are provided. Before
continuing with these tests, it is useful to first establish the
observational facts on the phase-space distribution of satellite
galaxies around the MW, because the MW is our primary auxilliary test
case: if the SMoC is falsified (Sections~\ref{sec:HypA_BTF}
and~\ref{sec:HypA_RM}) then the excellent MW data should be conform to
this.

\subsection{The distribution of satellite galaxies, globular clusters
  and streams in a vast polar structure (VPOS) around the MW}
\label{sec:VPOS}

An important test of the nature and origin of the MW satellite
galaxies is provided by their distribution in phase-space. Detailed
predictions have been made on this in the framework of the SMoC
(Sec.~\ref{sec:pred}).  If they were to be dwarfs of type~A then they
would have independent formation and evolution histories since each
would have formed within its own DM halo independently of the other DM
haloes. The rare cases that two DM haloes interact can be neglected
here given the vastness of the available phase-space (6-dim. volume of
roughly $250^3\,$kpc$^3 \times 400^3\,$km$^3$/sec$^3$) in comparison
with their small sizes. If, on the other hand, the satellites are of
type~B and stem from one encounter that involved the young MW then
they would be highly correlated in phase-space (Sec.~\ref{sec:pred}).

Here the following question is addresed: How are the satellite galaxies
and the globular clusters and strea\-ms in the outer halo of the MW
distributed in phase space?

\subsubsection{Phase-space distribution}
\label{sec:hypAphasespace}

The highly anisotropic distribution of the known dSph satellites, of
the two Magellanic Cloud satellite galaxies and of some globular
clusters, as well as the association with the Magellanic Stream about
the MW in a vast band on the Galactic sky, had been noted more than
thirty years ago \citep{LyndenBell76, Kunkel79}. But the contradiction
with the expectation from the later-adopted SMoC was emphasised for
the first time by \cite{Kroupa05}.  The anisotropic distribution is a
disk-like structure (the disk of satellites, DoS) with a
root-mean-square height of 10-30 kpc which lies nearly perpendicularly
to the plane of the MW. This distribution of the 9~''classical''
(i.e. brightest dSph satellites) has a likelihood of occurrence of
0.5~per cent if the parent distribution were a spherical DM host halo.

A number of subsequent research papers continuously enhanced the
discrepancy, and \cite{Metz07} showed that even oblate and prolate DM
host haloes of the MW do not match the observed satellite anisotropy.
It was found that Andromeda also has a non-isotropic satellite
distribution seen edge-on (\citealt{Karachentsev96, KG06, Metz07}, see
also fig.~1 in \citealt{Tollerud11}). The orbital planes of those
satellites that have observational constraints on their orbital angular
momenta suggest the MW satellite system to be a rotational disk-like
structure \citep{Metz08}. 

While already highly significant, the discrepancy with the
expectations from the SMoC became more significant with the addition
of the~13 new ultra-faint dSph satellite galaxies which independently
define the same phase-space correlation \citep{Metz09, Kroupa10}, a
result which is not affected by an incomplete survey sky coverage
(Sec.~\ref{sec:UFDs}).

Is the DoS a physical structure of the MW? If it is an unlikely chance
occurrence among the~9 classical dSph satellites, or if the 13~UFD
satellite galaxies are not physically related to the classical dwarfs,
then the UFDs cannot be distributed in the same DoS. Furthermore, if
the DoS is not physical, then the orbital angular momenta of the
satellite galaxies would not align with the normal vector of the
DoS. And, if the DoS is not physical then no other objects or
structures (e.g. globular clusters, stellar and gaseous streams) ought
to show a similar alignment. 

Instead, a VPOS surrounding the MW emerges which contains a highly
significant overabundance of all mentioned components
\citep{Pawlowski12b}. The individual components of this VPOS are
discussed next, and Figs~\ref{fig:VPOSedge}--\ref{fig:flyby_dir} below
visualise how these fit together and how a single model can account
for this structure. The normal vector to the VPOS is referred to as
NPOS.  A detailed analysis and discussion of the VPOS data is provided
by \cite{Pawlowski12b}.

\subsubsection{The UFD satellites}
\label{sec:UFDs}

The ultra-faint dwarfs -- UFDs -- have different discovery histories
than the classical satellites. The latter were discovered mostly on
photographic plates prior to about the year~2000 and their census is
complete over most of the sky apart in the regions obscured by the MW
disk.  The UFDs on the other hand were discovered using the robotic
Sloan Digitised Sky Survey (SDSS) after~2000. The sky coverage is not
complete, but the coverage extends over most part of the northern
hemisphere therewith being a cone rather than a slab with small
regions having also been surveyed in the southern Galactic hemisphere
(see fig.~1 in \citealt{Metz07}). {\it If there had been any
  observational bias that might have led to the discovery of those
  classical dSph satellites that, by an as yet unknown reason, lie in
  a DoS, then the UFDs clearly cannot be subject to the same bias.}

Fitting a plane to the classical satellites yields the well-known
DoS. The DoS is oriented such that when we look towards the Galactic
centre from the Sun, the DoS is seen nearly face on.  In
Galactocentric angular coordinates the normal of the classical DoS
points towards $l=157^{\rm o}.6 \pm 1^{\rm o}.1, b=-12^{\rm o}.0\pm
0^{\rm o}.5$ with a RMS height of $\Delta_{\rm dSph}=18.3\,$kpc.
Fitting a plane only to the~13 UFDs yields a DoS which is next to
identical to the classical DoS ($l=151^{\rm o}.4 \pm 2^{\rm o}.0,
b=9^{\rm o}.1\pm1^{\rm o}.0$ with $\Delta_{\rm UFD}= 28.6\,$kpc,
\citealt{Kroupa10}).

{\it Thus, the parent phase-space distributions of the classical
  dwarfs and of the UFDs can be taken to be equal.  Therefore they
  have a common origin}, because if this were not to be the case an
unnatural coincidence would need to be postulated without a known
physical mechanism.  The DoS normal vector of the combined population
points towards $l=156^{\rm o}.4 \pm 1^{\rm o}.8, b=-2^{\rm o}.2\pm
0^{\rm o}.6$ with $\Delta= 28.9\,$kpc.

\subsubsection{The globular clusters}

Considering the GCs of the MW these can be differentiated into the
bulge and disk (BD) GCs, into the old halo (OH) GCs and into the young
halo (YH) GCs (\citealt{Mackey05} for the classification;
\citealt{Harris96}, 2003~update, for positional data). The same
disk-fitting algorithm used to quantify the DoS of the classical dSph
satellites and of the UFDs can be applied to obtain the best-fitting
planar description of the three GC populations \citep{Pawlowski12b}.

The BDGCs lead to a planar fit with a normal oriented towards $(l,
b)=(175^{\rm o},-85.7^{\rm o})$, i.e. towards the Southern Galactic
Pole.  This is the exactly expected orientation for a component which
is associated with the MW disk and bulge.  For the OHGCs, on the other
hand, no good plane solution is found. Again, this is exactly as
expected because the OHGCs form a spheroidal distribution.

The YHGCs lead to a well defined disk of GCs (DoYHGCs) which is
indistinguishable to the satellite DoS. Its normal points
merely~13$^{\rm o}$ away from the DoS normal.  The probability of a
randomly oriented vector being as close to the DoS normal direction as
is the DoYHGC normal vector is about~2.5~per cent.

Sub-dividing the YHGC population of 30~YHGCs into those~20 within
20~kpc and those~10 beyond~25~kpc leads to two independently obtained
planar fits, each of which is well aligned with the DoS of the dSph
and UFD satellites (Fig.~\ref{fig:flyby_dir} below).

{\it Thus, the inner and outer YHGCs, the classical dSph satellites
  and the UFD satellites independently of each other define the same
  vast polar disk-like structure about the MW.} This is remarkable and
cannot be due to observational bias.

\subsubsection{Stellar and gaseous streams}

Furthermore, the known stellar and gaseous streams within and around the
MW can be analysed in terms of their orientations. Using a method
introduced in \cite{Pawlowski12b} to calculate the normal to the plane
defined by two points on the stream and the MW centre, it is possible
to study the directions of the normals to the 14~known stellar and gas
streams. It turns out that half of the stream normals cluster around
the above two DoS and the two DoYHGCs, an alignment which has a
likelihood of 0.34~per cent {\it if} the streams were randomly
oriented. The actual chance of finding the degree of orientation
evident in the MW streams is smaller, because it would be expected
that the streams predominantly map the continuous addition of material
into the MW disk. That is, the stream normals ought to be
preferentially oriented towards the poles of the MW.

\subsubsection{Combined likelihood}
\label{sec:comb}

The chance that the normals of the disks fitted to the classical dSph
satellites, the UFD satellites, the inner and outer YHGCs as well as
to the stellar and gaseous streams all cluster around the same region
on the Galactic sky is smaller than 2.5~per cent $\times$~0.34~per
cent $=8.5\times10^{-5}$ {\it if} they were physically unrelated.  It
is significantly smaller still because this number only considers the
YHGCs and streams.

\subsubsection{Orbital angular momenta of satellite galaxies}

A consistency check on the physical reality of the VPOS is provided by
the motions of its constituents. These need to be confined within the
VPOS for it to be a physical structure.  At present only the motions
of the nearest satellite galaxies are known.

Proper-motion measurements of the innermost~6 classical dSph and of
the LMC and SMC satellites have shown that the majority have orbital
angular momenta about the MW that point into a direction towards the
normal to the VPOS, i.e. towards the NPOS
(e.g. \citealt{Pawlowski12b}, Fig.~\ref{fig:flyby_dir} below). The
Sculptor dSph has an orbital angular momentum direction which places
it within the DoS but on a retrograde orbit relative to the average
direction of the other orbits.

It follows that of the eight satellite galaxies with proper motion
measurements, seven appear to orbit within the DoS with one being
within the DoS but on a retrograde orbit compared to the six
others. One satellite, namely Sagittarius, orbits approximately
perpendicularly both to the DoS {\it and} to the MW disk. Sagittarius
may have been deflected onto its present highly-bound orbit. Such a
scenario has been studied for the first time by \cite{Zhao98} and will
need to be re-investigated in view of the most recent data on the
Sagittarius stream (e.g. \citealt{Carlin12}) and the orbits of the
other satellite galaxies, and in view of the question whether
Sagittarius may have originally been orbiting within the DoS.

\subsubsection{Conclusions: the VPOS is a physical structure}
\label{sec:VPOS_concs}

\begin{figure}[ht]
\begin{center}
\includegraphics[scale=0.5, angle=0]{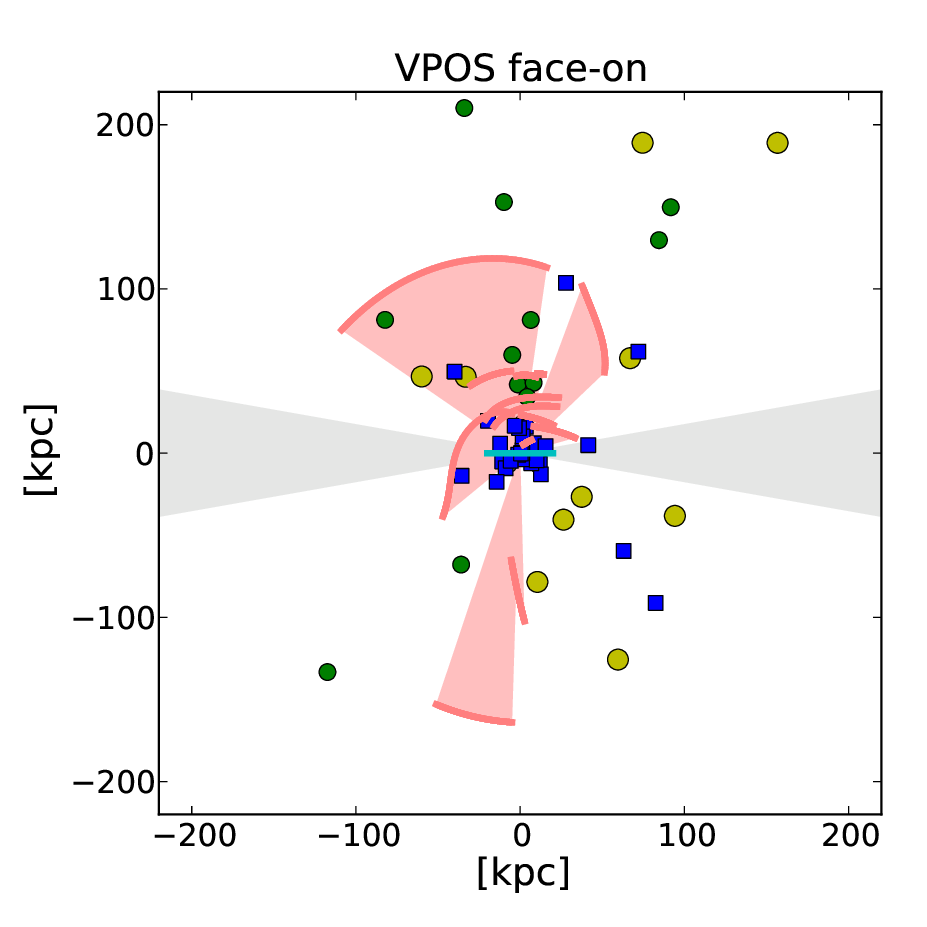}
\caption{The VPOS viewed face on. The large yellow circles are the~2
  Magellanic satellites and the 9~classical dSph satellite galaxies of
  the MW, the~13 new ultra-faint satellites discovered with the SDSS
  are shown as small green circles.  Blue squares are the YHGCs. The
  red curves are stellar and gaseous stream segments, magnified by a
  factor of three in distance to ease the visualisation. The region of
  obscuration by the MW disk is depicted as the grey equatorial
  zone. This figure is reprinted from \cite{Pawlowski12b}.
}\label{fig:VPOSface}
\end{center}
\end{figure}

\begin{figure}[ht]
\begin{center}
\includegraphics[scale=0.5, angle=0]{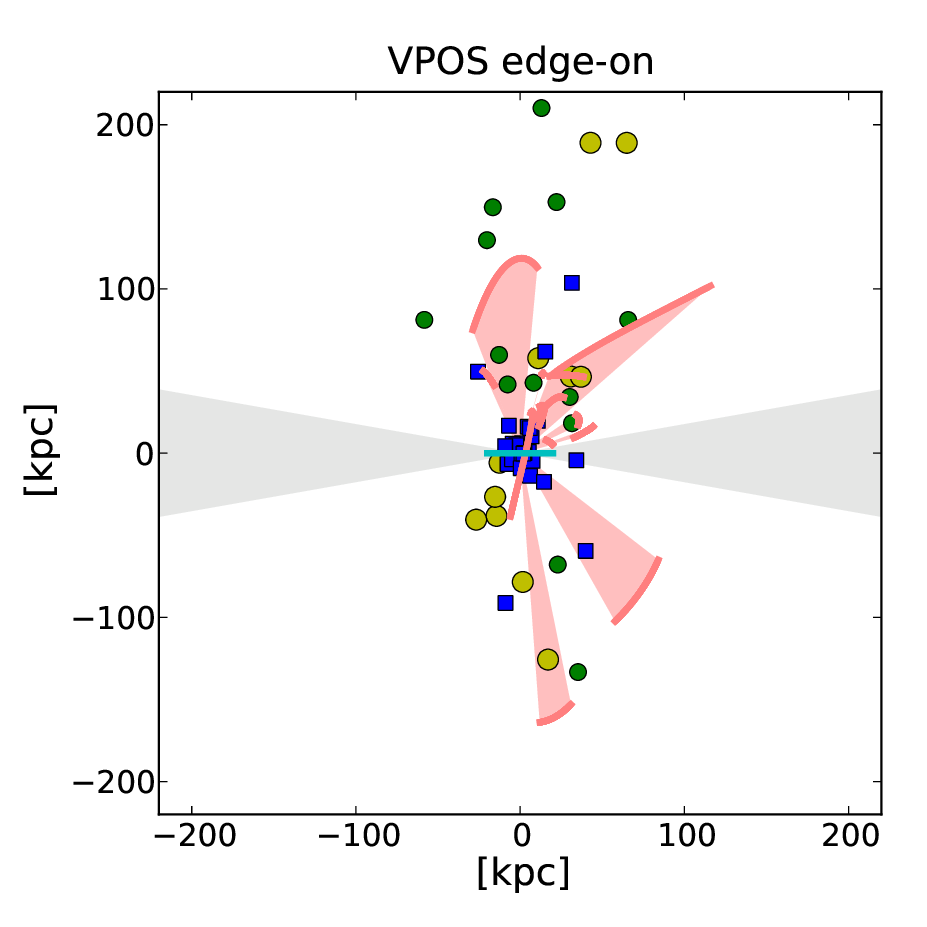}
\caption{As Fig.~\ref{fig:VPOSface} but viewing the VPOS edge on. Note
  that the streams (red curves) appear overemphasised because they are
  shown three times enlarged. 
}\label{fig:VPOSedge}
\end{center}
\end{figure}

\cite{Pawlowski12b} have thus discovered a vast polar structure
surrounding the MW. It is identified by a region on the Galactic sky
towards which the normals of the DoS, DoYHGC and half of all known
stellar and gaseous streams point. Fig.~\ref{fig:VPOSface}
and~\ref{fig:VPOSedge} show the VPOS face-on and edge-on,
respectively.\footnote{A movie visualising ``The Vast polar Structure
  around the Milky Way'' by Marcel Pawlowski is available on YouTube.} 

\begin{figure*}[ht]
\begin{center}
\includegraphics[scale=0.7, angle=0]{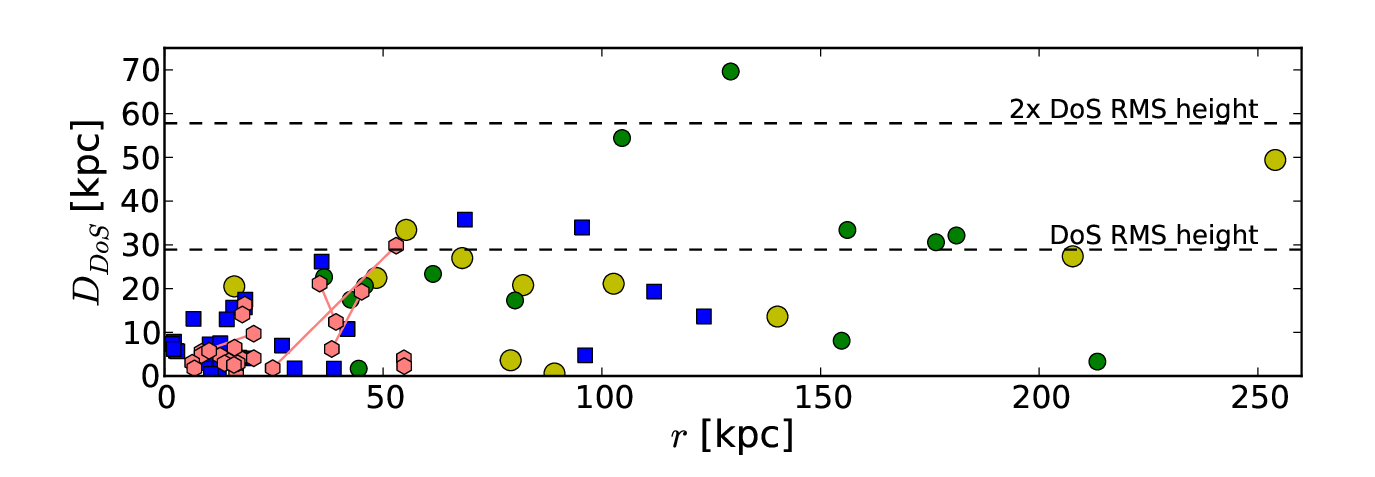}
\vspace{-5mm}
\caption{The constituents and structure of the VPOS. The distance,
  $D_{DoS}$, of the classical (bright) MW satellites (large yellow
  circles), the UFD galaxies (small green circles), the YHGCs (blue
  squares) and the two anchor (i.e. opposite end) points of the
  stellar and gaseous streams (red hexagons connected by thin red
  lines) are plotted as a function of their Galactocentric distance,
  $r$.  $D_{DoS}$ is the distance of the object from the DoS plane,
  i.e. the perpendicular distance to the edge-on DoS
  (Fig.~\ref{fig:VPOSedge}).  The DoS used here is as published in
  \cite{Kroupa10}. The DoS-parameters are (using all 24~satellite
  galaxies): the normal vector, NPOS, points to $l = 156.4^{\rm o}$,
  $b=-2^{\rm o}.2$, and the DoS is offset from the Galactic center
  by~8.2~kpc.  The DoS RMS height is $\Delta=28.9\,$kpc and is
  illustrated with the dashed lines. This figure was prepared by
  Marcel Pawlowski.  }\label{fig:DoSdistance}
\end{center}
\end{figure*}

It is useful to study how the various components are arranged in the
VPOS. As is suggested by Fig.~\ref{fig:DoSdistance}, the UFD
satellites which are fainter and thus have a smaller baryonic mass
have a somewhat larger dispersion in $D_{DoS}$ values than the
classical dSph satellites, which have larger baryonic masses.  {This
  is also evident in $\Delta_{\rm UFD} > \Delta_{\rm dSph}$
  (Sec.~\ref{sec:UFDs}).}  Is this mass segregation towards the
mid-plane of the VPOS?

Counting the number of objects out of all that have a distance,
$D_{DoS}$, within one and within two times the RMS height of the DoS,
$\Delta$: In total there are, at Galactocentric distances larger than
$r=10\,$kpc, 2~Magellanic satellites and 9~classical dSph satellites,
13~UFD satellites, 22~young halo GCs and 28~stream anchor points (74
objects, \citealt{Pawlowski12b}).  For $D_{DoS}<\Delta$ are found
17~of 24~satellites, 20~of 22 YHGCs, 25 of 28 anchor points.  For
$D_{DoS}<2\,\Delta$ we have 23 of 24 satellites, 22 of 22 YHGCs, 26 of
28 anchor points. That is, within $\Delta$ can be found 84~per cent,
and within $2\,\Delta$ are 96~per cent of all objects.

The VPOS therefore contains a variety of components and extends from
about 10~kpc out to at least~250~kpc. It has a height-to-radius ratio
of about 1:10 and therefore it constitutes a thin disk-like polar
structure.  The existence of this VPOS, or disk-like polar arrangement
of baryonic matter on a vast scale about the MW, stands beyond any
reasonable amount of doubt. It is incompatible with being derived from
accreted dark-matter sub-structures, taking the likely-hoods from
Sec.~\ref{sec:LG} below into account (0.056~per cent) as well as the
likelihood that the streams are also associated with the YHGC and
satellite galaxy distribution (less than $8.5\times10^{-3}$~per cent,
Sec.~\ref{sec:comb}).\footnote{An accretion origin of the satellite
  galaxies from DM filaments is negated by \cite{Angus11} and
  explicitly by \cite{Pawlowski12}.}

\section{Testing Hypothesis A on the Local Group}
\label{sec:LocG}

\subsection{The phase-space distribution and the properties of the
  Local Group}
\label{sec:LG}

From Sec.~\ref{sec:VPOS} it has thus become apparent that the MW is
surrounded by a vast phase-space-correlated structure, the VPOS, which
is made up of dSph and UFD satellite galaxies, inner and outer YHGCs
as well as stellar and gaseous streams.

The observed phase-space distribution of the MW satellites can be
compared to the allowed phase-space region assuming they are of
type~A.  To obtain significant ani\-so\-tro\-pies in the luminous
sub-halo distribution the following problem needs to be overcome: a
physical process needs to be found which allows star formation only in
sub-haloes that are highly correlated in phase-space, while all the
others remain dark. However, no such physical mechanism is available
within the SMoC despite many attempts
\citep{Metz09b,Pawlowski12,Pawlowski12b}.

High-resolution computations of the formation of MW-mass DM
host haloes within the SMoC have been performed by \cite{Libeskind09}
using semi-analytic modelling trim\-med to account for the observed
galaxy population in order to study the formation of MW-type galaxies
and their satellite systems. These calculations provide the following
data which have been published by the seminal work of Libeskind et
al.:

These supercomputer simulations with $10^9$ particles yield 31000~DM
haloes of mass comparable to the DM halo of the MW ($2\times 10^{11} <
M_{\rm DM}/M_\odot < 2 \times 10^{12}$). These host 3201~main galaxies
of similar luminosity as the MW (a galaxy more luminous in the V-band
than $M_V = -20$).  The remaining types of galaxies which are in
similar DM host haloes are not specified by the authors.  The relevant
sample of 3201~''MW-type galaxies'' host 436~galaxies with at least
11~luminous satellites. About 35~per cent of these have a satellite
system in which at least~3 satellites have orbital angular momenta
pointing within~30 degrees of the normal to the plane defined by
the~11 brightest satellites.

A DM halo of MW mass thus has a likelihood less than $3201/31000
\approx 10$~per cent of hosting a MW-type galaxy (defined to be a
galaxy with an absolute V-band magnitude $M_V<-20$, without
considering whether it is a major disk galaxy or a spheroid). The
majority of other galaxies also hosted by similar DM haloes are not
described further by the authors, but are galaxies fainter than
$M_V=-20$.  This appears to be in disagreement with the real
population of galaxies, since \cite{Disney08} have shown that the
galaxy population is remarkably invariant at any luminosity. The
observed uniformity is a significant failure of the cosmological
model, because of the large variation expected within the SMoC: Each
DM host halo has a different merger history (this is the {\it
  invariant baryonic galaxy problem}, \citealt{Kroupa10}). Further, of
the~31000 host haloes $436/31000 = 1.4$~per cent have a host galaxy of
MW luminosity {\it and} at least~11 luminous satellites. Of the
original sample of~31000 host haloes, about~0.4~per cent have these
properties {\it and} at least~3 satellites orbit within~30 degrees of
the normal to the plane defined by the~11 brightest satellites.
According to these numbers, and if the SMoC were valid, then the MW
and its phase-space correlated bright satellites would be a highly
significant exception of likelihood~0.4~per cent. This likelihood is
lower still because neither the thinness of the model DoS nor the
orientation of the DoS, being polar relative to the disk of the host
galaxy, are taken into account.

The Local Group however contains the MW and Andromeda. Andromeda is a
galaxy similar to the MW (but somewhat more complex,
\citealt{Hammer10}) and it also hosts $>11$ luminous satellites, in
full conformity with the {\it invariant baryonic galaxy property} of
the real universe. Thus, given a DM halo of MW/Andromeda mass, the
chance of obtaining an Andromeda-type galaxy within it is $1.4$~per
cent, by the above reasoning. The combined likelihood of having,
within the Local Group, two independent MW-type DM haloes hosting a MW
{\it and} Andromeda galaxy with their satellite systems is thus $<1.4
\times 0.4=\,$ $0.056$~per cent.

{\it The SMoC can thus be discarded with better than 99.9~per cent
confidence, by this one test alone.} 

But are we not merely making ever more precise demands to the point
that yes, the MW is a unique case just as each and every galaxy is
(e.g. \citealt{Hammer07})?  The above argument rests on generic
properties of the Local Group in how likely it is for a group of two
major galaxies to contain, in the SMoC, two similar MW-type galaxies
which have similar satellite systems whereby at least one of them has
an anisotropic satellite distribution. Nevertheless, this one test
alone would not suffice to discard the SMoC, because it can always be
argued that the Local Group happens to be an exception given the
unique properties we are interested in.

Ignoring the falsification of the SMoC through {\sc The Dual Dwarf
  Galaxy Theorem} (Sec.~\ref{sec:HypA_BTF} and~\ref{sec:HypA_RM}), a
relevant question that may be answered by considering the catalogues
of numerical SMoC simulations is how often groups of galaxies occur in
the model which have properties similar to those of the Local Group by
consisting of two major and similar disk galaxies. Such
Local-Group-type systems are common in the real universe with the
majority of galaxies being disk galaxies in such groups
\citep{Karachentsev96, Marino10}. The {\it invariant baryonic galaxy
  problem} discussed above would indicate that in the SMoC such groups
would be rare. Indeed, \cite{FR11} also arrive at small likelihoods of
finding a Local-Group equivalent if the SMoC were true (their Sec.~5.3
and~6).

In the present context, the following statement by \cite{Libeskind11}
is interesting: ``While the planarity of MW satellites is no longer
deemed a threat to the standard modell, its origin has evaded a
definitive understanding.''  \cite{Lovell11}, who also address the
anisotropy problem using the numerical data from the Aquarius SMoC
simulation, write: ``All six Aquarius haloes contain statistically
significant populations of sub-halo orbits that are aligned with the
main halo spin. All haloes possess a population of sub-haloes that
rotates in the same direction as the main halo and three of them
possess, in addition, a population that rotates in the opposite
direction. These configurations arise from the filamentary accretion
of sub-haloes. Quasi-planar distributions of coherently rotating
satellites, such as those inferred in the Milky Way and other
galaxies, arise naturally in simulations of a ΛCDM universe.''
This statement is clearly in contradiction with the above
\cite{Libeskind09} numbers, which is odd given that both papers are
published by the same SMoC research team.  Indeed, \cite{Pawlowski12}
demonstrate that the Lovell et al. contribution needs to be viewed
critically, as in essence the authors mark a sub-set of DM sub-haloes
which have orbital angular momenta aligned with the spin of the host
halo.  Considering this sub-set the authors conclude that the
disk-like distribution of MW satellites arises naturally in the
SMoC. They do not state however, why the vast number of sub-haloes on
other orbits should not play a role in establishing the satellite
population. That is, which physics would be active to select no other
than those sub-haloes to make stars which happen to be in the
disk-like sub-sample is not specified. Given that the host halo spin
tends to roughly align with the spin of the host disk galaxy, the
\cite{Lovell11} claim would suggest the MW satellite system in the
SMoC to be more in equatorial orientation in contradiction to the
observed VPOS (Sec.~\ref{sec:VPOS}).\footnote{The contributions by
  \cite{Lovell11} and \cite{Libeskind11} constitute examples of an
  overly optimistic interpretation of numerical SMoC data in view of
  the necessity to solve a major problem. See also
  Footnote~\ref{foot:fanelli}.}

Can the sub-grid parametrisation of baryonic physi\-cs be responsible
for the disagreement between model and observation? This cannot be the
case because the phase-space occupied by dark matter sub-haloes and
the star-formation processes within them are uncorrelated.  Indeed,
the large volume of published galaxy formation models up until~2011
(not counting the contribution by \citealt{Lovell11}) have all been in
mutual agreement with each other in reproducing the luminous
properties and spheroidal distribution of model satellite galaxies in
DM sub-haloes. The vast number of galaxy-formation simulations are
thus quite consistent with each other, which is an important
consistency check on the physics used in the simulations: {\it the
  reported research (see Footnote~\ref{foot:paperlist}) shows an
  internal consistency within the framework of the SMoC}.

{\it In summary:} it has thus emerged that the satellite phase-space
distribution of the MW in a VPOS extending from about~10~kpc to at
least about~250~kpc is not compatible with Hypothesis~A. This
conclusion is based on one auxiliary test, as discussed here.  If one
type of test falsifies Hypothesis~A, and if it is a robust test, then
other independent tests ought to yield the same conclusion.

In the following five additional and independent tests of Hypotheses~A
are performed for MW satellites. These can be viewed as stand-alone
tests, or as further consistency/auxiliary tests.

\subsection{The mass--luminosity data}
\label{sec:MLtest}
Assume that Hypothesis~A (Sec.~\ref{sec:dual_theorem}) is true. Then
by energy conservation the dSph satellite galaxies must show a
correlation between their luminosity, $L$, and hypothesised
dark-matter halo mass, $M_{\rm DM}$, which is deduced from
observations of the density and velocity dispersion profiles of the
dSph satellites by solving the Newtonian Jeans equation
(e.g. \citealt{Klimentowski07, Lokas11}). {\it Note that the
  statistical correlation between $L$ and $M_{\rm DM}$ does not rely
  on the details of baryonic physical processes, since the binding
  energy of the structure dictates what can form within it by whatever
  process, as long as the processes are generically the same in all
  satellites} (i.e. gas physics and stellar feedback, ionisation from
outside; tides do not play a major role for the population of
satellites as shown in Sec.~\ref{sec:pred}). That such a correlation
exists among galaxies (interpreting their matter content within the
SMoC) is very well established (e.g. \citealt{Leauthaud12}).

However, it has already been shown that the dSph satellite galaxies of
the MW violate the expected correlation \citep{Mateo93, Strigari08,
  Wolf10}. By solving the Newtonian Jeans equation the DM mass,
$M_{0.3}$, within the central 300~pc radius of each satellite can be
calculated; $M_{0.3}$ is a measure of $M_{\rm DM}$ by virtue of the
properties of the SMoC.  \cite{Kroupa10} test all available SMoC
models of satellite galaxies for the existence of a positive
correlation between model luminosity and model DM mass\footnote{The
  tests were conducted using models that had been computed before the
  tests were conducted. At the present time this is not possible any
  longer, because new satellite-galaxy models within the SMoC may be
  influenced by the posterior need to solve the $\kappa=0$
  problem.}. This correlation is quantified by the slope, $\kappa$, of
the log(DM mass)--log(luminosity) relation. The eight tested SMoC
models include the physics of feedback, inhomogeneous re-ionisation,
in-fall, CDM and WDM and semi-analytical models as well as stellar
population synthesis models, and each one has $\kappa>0.12$. The
observational data, on the other hand, have $\kappa<0.11$ at the
99.7~per cent (3-sigma) confidence level. In other words, as
\cite{Wolf10} state: ``...  all of the Milky Way dwarf spheroidal
galaxies (MW dSph) are consistent with having formed within a halo of
a mass of approximately $3 \times 10^9\,M_\odot$, assuming a $\Lambda$
cold dark matter cosmology. The faintest MW dSph seem to have formed
in dark matter haloes that are at least as massive as those of the
brightest MW dSph, despite the almost five orders of magnitude spread
in luminosity between them.''

Both, the observational data and the models thus agree at a confidence
level of 0.5~per cent.

\cite{Tollerud11} perform a resolved-star spectroscopic survey of~15
dSph satellites of Andromeda and find that the luminosities of these
satellites are independent of their DM mass, as is the case for the MW
satellites. Thus, for Andromeda, $\kappa\approx0$ as well.

{\it In summary:} The hypothesis that the SMoC models of dSph
satellite galaxies represent the real dSph satellite galaxies can thus
be discarded with a confidence of 99.5~per cent. {\it In other words,
  the hypothesis that the satellites reside in DM haloes made of C/WDM
  particles appears to be unphysical}, in full consistency with the
conclusion of Sec.~\ref{sec:LG} above.

\subsection{The mass function of dSph DM haloes}
\label{sec:MF}

There are various aspects of this test: 

Firstly, the missing satellite problem is well known: 24~dSph
satellites (counting both, the 11 bright, ``classical'', satellites
discovered mostly on photographic pla\-tes, and the 13~UFD satellites
discovered with the SDSS) have been found while hundreds are expected.
It is popularly (there exists a vast number of research papers on this
problem) claimed to be solved within the SMoC (\citealt{Kroupa10} and
references therein, see also e.g. \citealt{Font11}) as the
parametrisation of sub-grid baryonic physics is tuned to reproduce the
small number of observed dSph satellite galaxies. If the SMoC were
true then even within the solar neighbourhood there ought to be
hundreds of concentrated dark matter clumps
\citep{Diemand08}. According to these state-of-the-art SMoC
computations there ought to be about 150 additional faint satellite
galaxies within the MW DM halo which must be discovered
(e.g. \citealt{Bovill11}).  According to \cite{BoylanKolchin11} up to
a ``factor of~5 to~20 times as many faint galaxies could remain
undetected at present owing to incomplete sky coverage, luminosity
bias, and surface brightness limits.''

Secondly, assuming the SMoC to be true and each dSph satellite galaxy
to be embedded in a DM halo, the form of the mass function (MF) of
these observed luminous DM haloes is not in agreement with the
theoretically expected MF of luminous sub-haloes. Including the Large
and Small Magellanic Clouds (LMC, SMC, respectively) leads to the
following result: the hypothesis that the MW satellites stem from the
SMoC can be discarded with more than 96~per cent confidence, but a
more realistic assessment implies exclusion at the 99~per cent level
\citep{Kroupa10}. In particular, the observed sample of satellites has
a significant overabundance of $M_{0.3}\approx 1.5\times
10^7\,M_\odot$ DM haloes.

Thirdly, as documented in fig.~2 in \cite{Kroupa10}, all DM sub-haloes
containing dSph satellites have a mass $M_{300}<2.5\times
10^7\,M_\odot$, while 15~per cent of the sub-haloes in the MW DM halo
ought to be more massive within~300~pc according to the $\Lambda$CDM
model. The MW halo is thus missing a substantial fraction (15~per
cent) of its massive sub-haloes.  \cite{Bovill11} emphasise this
failure of the SMoC for the first time. \cite{BoylanKolchin11}
continue to point out that the MW should host at least six sub-haloes
that had maximum circular velocities in excess of 30~km/sec but are
incompatible with any known MW satellite (including the Magellanic
Clouds) having a V-band luminosity $L_V > 10^5\,L_\odot$.

To solve this problem, \cite{BoylanKolchin11} suggest that galaxy
formation in DM haloes with a mass $\simless 10^{10}\,M_\odot$ may
become stochastic, or that the MW is an exception.  However, this is
not conform to known physical laws. The suggestion that the MW (and by
implication) Andromeda (which has the same problem) are exceptions is
ruled out by \cite{SW11} who study a large ensemble of MW-type
galaxies from the SDSS confirming the significant deficit of bright
satellites around the hosts if the SMoC were true.

Fourthly, in modelling galaxy formation within the SMoC it has to be
assumed that the galaxy formation efficiency decreases sharply with
decreasing DM halo mass {because the DM halo mass function rises
  steeply with decreasing mass}.  Effectively below a threshold mass
$\approx 10^{10}\,M_\odot$ galaxies do not form
\citep{Ferrero11}. These authors demonstrate that almost one half of
dwarf galaxies with stellar mass in the range $10^6 < M_{\rm
  stars}/M_\odot < 10^7$ are in haloes with masses substantially below
that threshold. They emphasise that this is not easily accommodated
within the SMoC. In their abstract they state ``Extending galaxy
formation to haloes well below $10^{10}\,M_\odot$ would lead to severe
disagreement with the low mass end of the galaxy stellar mass
function; at the same time, the extremely low stellar mass of the
systems involved make it unlikely that baryonic effects may be
responsible for reducing their dark matter content.''

{\it In summary:} The number and DM halo mass distribution of MW
satellite galaxies is in highly significant disagreement with the
expectations from the SMoC and there is no physically known process
that may be able to solve the disagreements. {\it In brief, the
  concept that the satellites reside in DM haloes made of C/WDM
  particles breaks down}, in full consistency with the conclusions of
Sec.~\ref{sec:LG}--\ref{sec:MLtest} above.

\subsection{The morphological appearance of dSph satellite galaxies}
\label{sec:morph}

Globular star clusters (GCs) have a stellar velocity dispersion
$\sigma\approx 10\,$km/sec. For a typical diameter of $2\,r\approx
8\,$pc, this implies that the crossing time scale is about 1~Myr. Any
internal sub-structure is thus phase-mixed away on a time scale of a
few~Myr, which is why GCs appear as perfectly smooth, symmetric and
spheroidal stellar systems despite being immersed in the tidal field
of the MW.

The dSph satellite galaxies also have $\sigma\approx 10\,$km/sec but
radii of about 300~pc. If this velocity dispersion is related to their
DM halo mass, which it must be if the SMoC were correct, then any
internal sub-structure would phase-mix away within 100~Myr. Given
their ages of about~10~Gyr, it follows that the satellites ought to
appear as smooth and symmetric as GCs.

Tidal forces from the MW cannot be effective in distorting the
satellites, as most of them are at Galactocentric distances
$D>50\,$kpc, if each is surrounded by an extensive (radii $>{\rm
  few}\,$kpc) DM halo weighing about $10^9\,M_\odot$. Computational
work has shown about~10~per cent of the DM sub-haloes to be affected
by tidal forces (Sec.~\ref{sec:pred}). A smaller fraction will be
affected so severely that the innermost~1~kpc regions that contain the
stars would be distorted by tides.

In this context, \cite{Hayashi03} write ``We apply these results to
substructure in the Milky Way and conclude that the dark matter haloes
surrounding its dwarf spheroidal (dSph) satellites have circular
velocity curves that peak well beyond the luminous radius at
velocities significantly higher than expected from the stellar
line-of-sight velocity dispersion. Our modeling suggests that the true
tidal radii of dSph lie well beyond the putative tidal cutoff observed
in the surface brightness profile, suggesting that the latter are not
really tidal in origin but rather features in the light profile of
limited dynamical relevance.''  Following on from this,
\cite{Walcher03} emphasise a statement by \cite{Stoehr02}: ``Although
there is no problem accommodating a single disrupting object like
Sagittarius, it would become uncomfortable if tidal stripping were
detected unambiguously in other systems.''  Note that the inner region
of a satellite is affected by tides after significant tidal
destruction of its outer parts \citep{Kazantzidis04}. 

Thus, a small fraction, far less than ten~per cent, of the satellites
may show morphological evidence for tidal affects such as being
flattened or somewhat distorted. Indeed, the Sagittarius satellite
galaxy, at $D\approx 16\,$kpc, is the best known example of a
satellite galaxy being strongly affected by the MW tidal field.

From the sample of 24~dSph satellite galaxies, too many show
non-spherical and in many cases also asymmetric morphologies. Ursa
Minor ($D\approx 65\,$kpc) is a well known case with internal
sub-structure and a highly flattened and asymmetric appearance
\citep{Kleyna98}. It counts as one of the most DM dominated galaxy
with a mass-to-light ratio, $M/L\approx 70$, but its internal
structure is incompatible with the existence of a CDM halo
\citep{Kleyna03}. The Fornax dSph satellite ($D\approx 140\,$kpc) also
shows significant distortions by appearing flattened, asymmetric and
with twisted isophotes \citep{Demers94, Walcher03}. The Carina dSph
satellite ($D\approx 93\,$kpc) similarly shows a flattening with one
side appearing to be more compact than the other, and it has isophotes
that are not ellipsoidal \citep{Walcher03}. Among the faintest
satellites, Hercules is highly elliptical and somewhat amorphous
\citep{Coleman07}. \cite{Sand11} find that a large fraction of the
faint satellites, which are supposedly the most DM dominated dwarf
galaxies, show tidal signatures.  Noteworthy is that this problem
appears to persist for Andromeda satellites too: Andromeda~X is found
to be highly elongated at a distance of~170~kpc from its host galaxy
\citep{Brasseur11}.

\cite{McGaugh10} study the internal dynamics and the elongation of
most of the MW and Andromeda satellites finding very strong evidence
for the majority being affected by tidal forces. From their fig.~5 it
is seen that of the 24~satellites plotted, 75~per cent have an
ellipticity larger than~0.3, whereby the ellipticity correlates
strongly with the distance from the host galaxy. {That even the
  classical (bright) dSph satellites of the MW have substantial
  ellipticities is evident from table~1 in \cite{Lokas11}.}

{\it In summary:} While a homogeneous statistical study of the
morphological appearance of each dSph satellite is wanted, the above
examples and reasults already demonstrate that {\it the notion that
  the satellites are immersed in DM haloes appears to be unphysical}
because there are too many satellites with distorted morphologies.
This is in full consistency with the conclusions of
Sec.~\ref{sec:LG}--\ref{sec:MF} above.

\subsection{Orbital decay of MW satellites}
\label{sec:orbdecay}

The conventional Newtonian interpretation of the MW dSph satellite
galaxies is that they are hosted within DM-sub-haloes. With this
assumption, Jeans modelling implies them to have similar DM halo
masses of $10^9\,M_\odot$ \citep{Mateo93,Strigari08,Wolf10}. This
appears to be the case for the Andromeda satellites as well
\citep{Tollerud11}.

To account for the existence of the DoS, and if the satellites were of
type~A (i.e. hosted in DM sub-haloes), then they would have been
accreted onto the growing DM halo of the MW from a DM filament that
would need to connect to the MW DM halo.  They would thus have fallen
in from large distances, and dynamical friction would have decayed and
circularised their initial orbits to the present-day orbits about the
MW. \cite{Angus11} demonstrate that for this to be viable, the
satellites with measured proper motions must have been significantly
more massive than $10^9\,M_\odot$. This is in contradiction to the
Jeans modelling.

{\it In summary:} There is therefore no consistent combined solution
of the existence of the DoS, the orbital angular momenta and masses of
the dSph satellites within the framework of the SMoC. {\it dSph
  satellites cannot be DM sub-haloes} in full consistency with the
conclusions of Sec.~\ref{sec:LG}--\ref{sec:morph} above.

\subsection{Further logical inconsistencies}
\label{sec:log_inc}

All independent tests concerning dSph satellite galaxies
(Sec.~\ref{sec:LG}--\ref{sec:orbdecay}) yield the same result:
{\it Hypothesis~A is incompatible with the data}.

While Hypothesis~A can by now be taken to have been disproven, it is
nevertheless of use to point out the following mutually excluding
results based on excellent high resolution simulations of the
formation of MW-type galaxies and their satellites within the SMoC
framework:

{In a detailed discussion of the problem at hand,} \cite{Deason11}
write ``The satellite galaxies have been accreted relatively
recently'' (at a redshift of $z<1$) in order to account for their
disk-like distribution in the VPOS.

\cite{Nichols11} model the evolution of initially gas rich satellite
galaxies. Since these are today essentially free of gas, even out to
about 250~kpc, the authors conclude ``This model of evolution is able
to explain the observed radial distribution of gas-deficient and
gas-rich dwarfs around the Galaxy and M31 if the dwarfs fell in at
high redshift ($z \approx 3-10$).''

The observed MW satellite system is arranged in a large polar
disk-like structure {\it and} the satellites are void of gas.
Both, recent and long-past accretion into the MW halo of the same
satellites is not physically possible. 

Apart from this inconsistency arrived at in the SMoC, infall of a
group of dwarf galaxies as the origin of the phase-space correlation
is ruled out by the following reasons \citep{Metz09b}:

Firstly, the putative group would have had to have been compact with a
diameter of less than about 30~kpc to account for the thinness of the
VPOS (Fig.~\ref{fig:VPOSedge}). But all known groups of dwarf galaxies
have diameters of a few hundred~kpc. It would thus be necessary to
postulate that the MW accreted a group of a type which does not exist
any longer. This however would be an inferior hypothesis because in
order to ``solve'' the DoS problem the existence of an unobserved
ad-hock type of dwarf group would need to be postulated.

Secondly, there would then be no endemic luminous DM sub-haloes of the
MW. The missing satellite problem would then become a catastrophic
failure, since {\it all} simulations of major galaxy formation require
them to have hundreds of luminous satellites that have individual
in-fall histories.

\section{Conclusions on testing Hypothesis A}

While there is a strong notion and
peer-pressure\footnote{\label{foot:fanelli}Perhaps of relevance in
  this context is the research on the sociology of science by
  \cite{Fanelli10} who's abstract reads ``The growing competition and
  "publish or perish" culture in academia might conflict with the
  objectivity and integrity of research, because it forces scientists
  to produce ``publishable'' results at all costs. Papers are less
  likely to be published and to be cited if they report ``negative''
  results (results that fail to support the tested
  hypothesis). Therefore, if publication pressures increase scientific
  bias, the frequency of "positive" results in the literature should
  be higher in the more competitive and "productive" academic
  environments.'' The study of \cite{Fanelli10} finds that ``...these
  results support the hypothesis that competitive academic
  environments increase not only scientists' productivity but also
  their bias.'' This may be the reason why the academic system of {\it
    la Grande Nation} has allowed France to take leadership on issues
  pertaining to TDG formation and non-Newtonian/non-Einsteinian
  gravitational research. In France intellectual freedom is highly
  cherished and fostered in an academic system without major
  hierarchies. Germany, being a heavily hierarchical academic
  environment in which the majority of resources are controlled
  long-term by merely a few, and the USA, being an extremely
  competitive and research-grant-driven environment, do not leave much
  room for such research at the present. In these environments, often
  the mere {\it opinion} of a few is crucial for the success in
  obtaining research money. A good example of the Fanelli effect can
  be seen in fig.~44 of \cite{FM12} where the constraints on the
  measured cosmological baryon density are plotted in dependence of
  time: Prior to the CMB data (around the year~2000) the independent
  measurements yielded a low density.  After the CMB acoustic peak
  data became available and after it emerged that the previously
  measured cosmological baryon density was consistently too low to be
  consistent with the SMoC and the acoustic peak data, the new
  independent constraints typically and magically began yielding
  density measurements consistent with the measured acoustic peaks for
  the SMoC to be correct.} that Hypothesis~0i must be valid on
galactic and cosmological scales, all tests (Sec.~\ref{sec:HypA_BTF},
\ref{sec:HypA_RM}, Sec.~\ref{sec:VPOS}
and~\ref{sec:LG}--\ref{sec:log_inc}) fail Hypothesis~A consistently
and without exception.

It could well have been different: We could have had the situation
that one of the tests fails, but that others show consistency of the
models with the data. For example, the dSph satellites could have had
$\kappa > 0$, about the correct mass distribution and morphological
appearance.  In this case the challenge would have to have been to
understand the remaining failure given the otherwise
agreement. However, the consistent failure, always in the same sense
that the observational data are in conflict with Hypothesis~A,
i.e. the existence of dwarf-galaxy-hosting DM haloes, is so grave that
a remedy cannot be found within the SMoC.  

\vspace{2mm}

\noindent {\it In summary:}
\begin{enumerate}
\item Young gas-rich rotationally supported TDGs lie on the BTF
  relation defined by DM-dominated dwarf and normal galaxies. This
  cannot be the case if DM defines the rotation velocities of the dIrr
  galaxies. Thus ${\rm BTF}_{\rm dIRR} = {\rm BTF}_{\rm TDG}$ which
  implies \cancel{SMoC}.

\item Intermediate-age TDGs lie on the radius--mass relation of dE and
  dSph galaxies. This cannot be the case if dE and dSph galaxies
  formed in a mass-dominating DM halo.  Thus ${\rm dE, dSph} = {\rm
    TDG}$ (or RPDG) which implies \cancel{SMoC}.

\item The dSph satellites of the MW (and to a certain degree also of
  Andromeda) have a highly significant phase-space correlation which,
  for the MW, is a vast polar structure. This VPOS is inconsistent
  with the satellites being accreted individually or with them being
  endemic DM sub-haloes.

\item The dSph satellites of the MW and Andromeda have constant DM
  sub-halo masses over many orders of magnitude of luminosity in
  violation of the necessary correlation between the two quantities if
  the satellites were in their own DM sub-haloes.

\item The DM-mass function of the observed MW satellite galaxies does
  not agree with the theoretical one of luminous sub-haloes derived
  from the SMoC. The suggestion that star formation in DM haloes less
  massive than $10^{10}\,M_\odot$ becomes stochastic has no physical
  basis.

\item Too many dSph satellites show morphological distortions which
  ought not to be the case if they were embedded in their own
  mass-dominating DM sub-haloes.

\item Within the same framework of the SMoC the gas-free dSph
  satellites loose their gas if they were accreted at a redshift $z>3$
  while they may appear in a disk-like VPOS distribution if they
  accreted at $z<1$. The MW satellites are both, gas poor and in a
  disk-like VPOS.

\item Finally, the emergence in the SMoC of a group of galaxies with
  the generic properties of the Local Group (two similar spirals, each
  with at least 11~luminous satellites) is negligibly small.

\end{enumerate}

Therefore, {\it Hypothesis~A needs to be discarded}. That is, dwarf
galaxies cannot be of type~A and therefore they cannot be embedded in
DM sub-haloes. By Sec.~\ref{sec:dual_theorem}, the allowed alternative
is then for dwarf satellite galaxies to be TDGs, as is indeed already
suggested by the points~1. and~2. above.

\section{Hypothesis B: the dE galaxies and dSph satellites are ancient
  TDGs and RPDs (i.e. of type~B)}

In the above Sec.~\ref{sec:VPOS} it has been shown that the MW is
surrounded by a VPOS which is a physical structural part of the MW
made up of all known satellite galaxies, the young halo globular
clusters and half of all known stellar and gaseous streams. The
existence of this structure is perfectly consistent with the
falsification of {\sc The Dual Dwarf Galaxy Theorem} in
Sec.~\ref{sec:HypA_BTF}--\ref{sec:LocG}. As suggested in
Sec.~\ref{sec:dual_theorem}, the only available alternative would be
for the VPOS to be a remnant tidal structure that formed when the
young MW had encountered another galaxy about 10--11~Gyr ago.

How does the theoretical and observational evidence on dE galaxies in
general, and dSph satellites in particular stand up to this
interpretation? Are there other extragalactic satellite galaxy systems
which also show phase-space correlations? If the MW were unique and
thus an exception confidence in the interpretation of dE galaxies and
of dSph and UFD satellite galaxies as TDGs would be compromised.

\subsection{The formation of TDGs and associated GCs}
\label{sec:VPOS_origin}

It is well known that star clusters and TDGs form in one and the same
galaxy--galaxy encounter.  This is observed in many cases, but TDGs
with masses lower than about $10^7\,M_\odot$ can usually not be
detected.\footnote{This can lead to the erroneous conclusion that TDGs
  contribute only a small fraction to the dwarf galaxy
  population. While this is true for the TDGs being formed in the
  present-day universe, the number of low-mass TDGs formed during the
  early cosmological era would have been sufficient to account for the
  observed population of dSph and dE galaxies (Sec.~\ref{sec:impl3},
  Sec.~\ref{sec:extragal}).}  An example of a young TDG with a mass
near $10^6\,M_\odot$ is the complex of star clusters in the tidal arm
of the Tadpole galaxy which has a half-light radius near 160~pc
\citep{Tran03}, while other cluster complexes in the arm are of lower
mass still. Such objects may evolve through the merging of the
clusters to spheroidal dwarf galaxies \citep{Kroupa98} such that the
progenitors of the faint MW satellites may have looked similar to
these Tadpole objects.

The highest-ever-resolution particle-mesh computation by
\cite{Bournaud08} of a wet galaxy--galaxy encounter has a gas fraction
of 17~per cent and a resolution (or cell-length) which is $cl=32\,$pc
within Galactocentric distances of $D=25\,$kpc, $cl=64\,$pc within
$D=50\,$kpc and $cl=128\,$pc for $D>50\,$kpc. This pioneering work
demonstrates that star-cluster sized bound objects with masses in the
range $10^5-10^6\,M_\odot$ form readily. Also, a few TDGs with masses
$10^8-10^9\,M_\odot$ are formed. These are rotating dIrr gas-rich
galaxies with diameters of a few~kpc (Fig.~\ref{fig:TDGs}).

It is important to note that less-massive TDGs cannot form in these
simulations because the resolution is still too poor at $D>50\,$kpc,
while the tidal stresses at smaller $D$ only allow compact star
clusters to emerge in the simulations.  These have resolution-given
radii of tens of~pc and form due to the pressured colliding inter
stellar media of the two galaxies. At $D>50\,$kpc gas accretes from
the expanding tidal arms into gravitationally unstable regions forming
rotation-supported gas-rich dwarf-galaxies.

The phase-space density of matter in the tidal arm is comparable to
that of the pre-collision galactic disk from where it stems because
the flow in phase-space is incompressible for two-body
relaxation-free, i.e. colli\-sion-\-less, stellar-dynamical
systems. Therefore, regions whi\-ch become self-gravitating within the
tidal arm due to density variations along it, should have,
approximately, the matter density of galactic disks. Within these
gravitationally decoupled dIrr galaxies star-formation proceeds as in
any other dwarf galaxy in a distribution of star-formation events
which can, for all practical purposes, be described as embedded star
clusters \citep{LL03} with a maximum mass which correlates with the
star-formation rate of the TDG \citep{WKL04, Kroupa11}.

The average density of young TDGs must therefore be comparable to
roughly $0.01-1\,M_\odot/$pc$^3$, which is the baryonic matter density
in disk galaxies. It is noteworthy that the dE and dSph satellite
galaxies have such densities: Returning to the observable properties
of dwarf galaxies, a constant density relation, $\rho \propto M_{\rm
  star}/$ $R_{0.5}^3$, implies $b_1=1/3$. As is evident from
Fig.~\ref{fig:stellar_systems}, dE galaxies with $M_{\rm star} \approx
10^9\,M_\odot$ have densities around $1\,M_\odot$/pc$^3$, while
less-massive dE and the dSph satellites have densities of
$0.01\,M_\odot$/pc$^3$.  This is similar to the typical density of
baryonic matter within a disk galaxy. For example, the present-day
density of baryonic matter nearby to the Sun is about
$1\,M_\odot/$pc$^3$, with an exponential decrease to larger
Galactocentric distances. Star-formation activity is observed to be
taking place at Galactocentric distances out to about 25~kpc. For a
radial exponential disk scale-length of 3.5~kpc
(e.g. \citealt{Sale10}) the average density at 25~kpc becomes
$0.01\,M_\odot/$ pc$^3$.

A lower density cutoff for TDGs is given by the necessity for
self-gravitation to be sufficiently strong to overcome tidal shear.
That is, 
\begin{equation}
M_{\rm TDG}/R_{\rm max, TDG}^3 \approx M_{\rm host}/D^3,
\end{equation}
is required, where $M_{\rm TDG}$ and $R_{\rm max, TDG}$ are the mass
and maximal radius of the TDG, while $M_{\rm host}$ is the mass of the
host galaxy and $D$ the distance of the TDG to the host. For typical
birth distances of $D=10^5\,$pc and $M_{\rm host}\approx
10^{10}\,M_\odot$ it follows that
\begin{equation}
  R_{\rm max, TDG}/{\rm pc} \approx 10^{5/3}\,(M_{\rm TDG}/M_\odot)^{1/3}.
\end{equation}
Such objects are not likely to form since regions of the extend
implied are not likely to be filled with matter, the tidal tails
usually being more confined.  But if they would form, then they would
not readily be found by observation because they have low projected
densities. For example, a TDG with a mass of $10^9\,M_\odot$ would
have $R_{\rm max, TDG}\approx 46\,$kpc. Such an object would not
likely survive its first perigalactic passage, but sub-regions of
sufficient self-binding energy may. A low-density stellar population
of such a dimension has recently been discovered nearby the host
galaxy NGC$\,$7531 with a high-sensitivity survey using small
telescopes (panel~E in fig.~1 of \citealt{Martinez10}). The stellar
structure nearby the dwarf galaxy NGC$\,$4449 may also be of this
category (fig.~1 in \citealt{Martinez11}).

{Taking the Newtonian tidal radius estimate from \cite{BT87},
\begin{equation}
R_{\rm tid} = R_{\rm max, TD} / 3^{1/3},
\label{eq:rtid}
\end{equation}
it can be seen from Fig.~\ref{fig:TDGs} that the upper envelope of dE
data corresponds to $M_{\rm host}\approx 10^{12}\,M_\odot$ which is
consistent with the majority of dE galaxies having been formed in
galaxy clusters, taking the red-shift evolution of their mass into
account.  }

\subsection{The physical nature of the  MW VPOS}

The VPOS is composed of satellite galaxies, globular clusters and
stellar and gaseous streams (Sec.~\ref{sec:VPOS}). Can the VPOS be an
ancient remnant of a major tidal arm created about 10--11~Gyr ago
around the MW?

The pioneering work of \cite{Bournaud08} demonstrates the simultaneous
formation of star clusters and TDGs during galaxy-galaxy encounters
while it cannot yet reach the resolution relevant to the MW dSph
satellite galaxies. The number of satellite galaxies formed also
scales with the gas fraction, and it is expected to have been higher
about 10~Gyr ago when the gas fractions where higher
\citep{Wetzstein07}.

Streams then arise from these DM-free TDGs and star clusters 
as they dissolve over time through energy-equipartition driven
evaporation of stars in the collisional systems (i.e. the star
clusters) that have median two-body relaxation times shorter than a
Hubble time and/or through time-variable tidal fields
\citep{Kroupa97,Kuepper10}.

The gaseous streams in the VPOS may be either ancient remnants from
the original tidal material from the encounter or gas that has been
ram-pressure strip\-ped or otherwise from the satellite
galaxies. Indeed, the Magellanic Stream, being well aligned with the
VPOS, is such a young structure.

That the UFD satellites have a larger spread away from the DoS than
the classical satellites (Sec.~\ref{sec:VPOS_concs}) would be
consistent with UFDs having formed in lower-density tidal material
which was spatially more extended than the denser
material. Theoretical results on this suggestion do not exist yet.

Can a structure such as the VPOS be obtained naturally in
galaxy--galaxy encounters?  A large number of galaxy--galaxy
encounters have been calculated by \cite{Pawlowski11} to study the
phase-space distribution of tidal material expelled during the
encounter or merger. This work has demonstrated that the VPOS can be
naturally understood as the remnant of tidal material expelled during
a flyby-encounter of the young scaled-down MW and another similar
young galaxy about 10~Gyr ago. An encounter with a smaller young
galaxy is also possible.  Interesting in this context is that about
10~Gyr ago the scaled-down young MW would have appeared similar to M33
today. A bulge forms in such a major encounter and a disk can regrow
(e.g. \citealt{Hammer05, Bournaud05, Hammer07, Hammer09}, 
\citealt{Wei10}, \citealt{Bournaud11,Martig12}).

Thus, as argued below, consistency with the observed properties of the
MW is achieved, since the Galactic thin disk is younger than about
10~Gyr while a significant fraction of the bulge is old, as are the
constituents of the VPOS (e.g. \citealt{Tsujimoto12} for constraints
on the formation of the MW bulge).

In some of the galaxy--galaxy encounters by \cite{Pawlowski11} the
tidal material populates phase-space next-to identically to the
currently available information on the MW VPOS. Documented for the
first time by \cite{Pawlowski11} is the natural emergence of
counter-rotating tidal material in excellent agreement with the
counter-orbiting Sculptor dSph sa\-te\-llite. These models show how
the ratio of pro- and counter-rotating tidal debris about the MW
significantly constrains the allowed encounter.

\begin{figure*}[ht]
\begin{center}
\includegraphics[scale=0.8, angle=0]{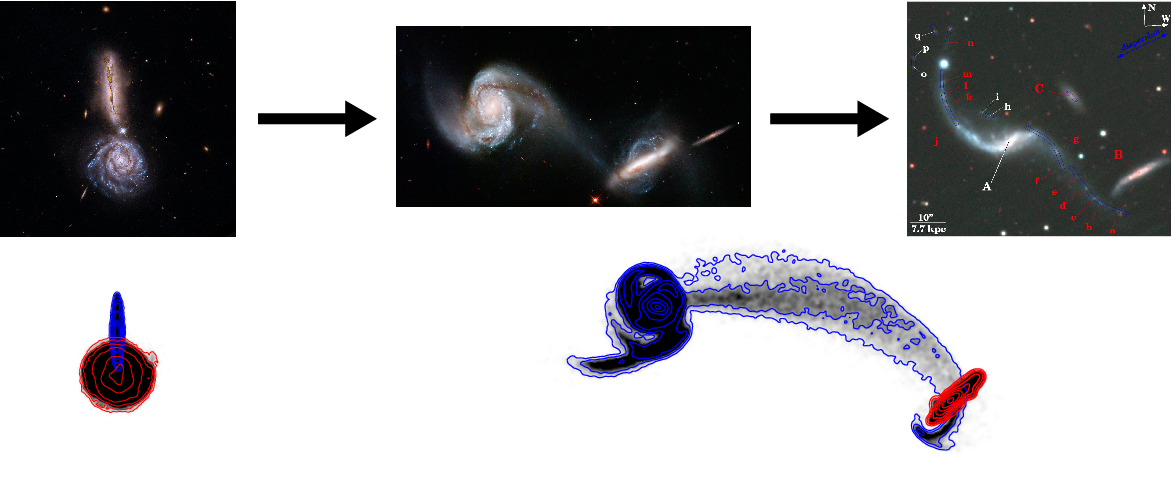}
\vspace{-6mm}
\caption{A sequence of real and model flybys.  The upper left image is
  Arp~302, the upper central one is Arp~87 and the upper right image
  shows the Dentist Chair galaxy with TDG candidates
  \citep{Weilbacher02}.  \cite{Whitmore90} describe Arp~87 as possibly
  being related to polar-ring galaxies.  The lower images are particle
  densities from the computed flyby model~5deg200vel of
  \cite{Pawlowski11}.  The model shows a similar encounter morphology
  as in the real galaxies in the upper row. It is a flyby encounter
  between two equal galaxies, each being a down-scaled MW galaxy about
  10~Gyr ago therewith being similar to M33 today.  The encounter is
  polar and occurs with a relative velocity of 200~km/sec. It forms a
  VPOS around the MW (red) galaxy (Fig.~\ref{fig:flyby_dir}) and has
  not been designed to give a particularly good fit to the VPOS.  This
  figure is reprinted from \cite{Pawlowski12b}.  Image credits:
  Arp~302: NASA/STScI/NRAO/ Evans et al.; Arp~87: NASA, ESA, and the
  Hubble Heritage Team (STScI/AURA); the ``Dentist Chair'' Galaxy AM
  1353-272 \citep{Weilbacher02}.  }\label{fig:flyby}
\end{center}
\end{figure*}

\begin{figure*}[ht]
\begin{center}
\includegraphics[scale=0.5, angle=0]{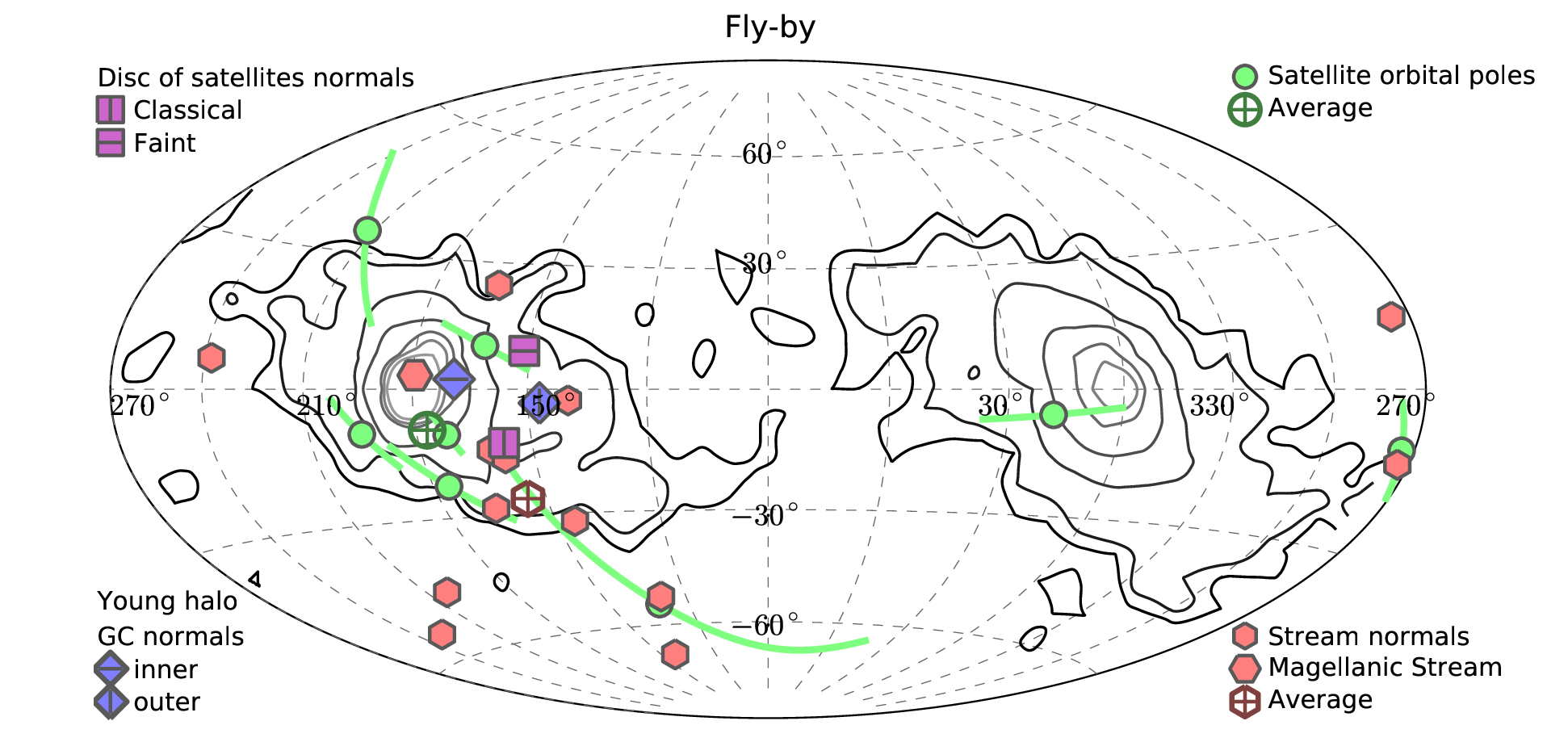}
\caption{The VPOS as a tidally created structure. Directions on the
  Galactic sky of normals to the DoSs of the classical (11 bright) and
  new (13 UFD) satellite galaxies, and to the disks of 20~inner and
  10~outer young halo globular clusters. The directions of orbital
  angular momenta (orbital poles) of satellite galaxies with measured
  proper motions are also shown, as are the normals to stellar and
  gaseous streams. Fly-by model~5deg200vel (Fig.~\ref{fig:flyby})
  produces tidal debris which orbits the MW. The density of orbital
  angular momenta directions of particles in the model at distances
  larger than 20~kpc at the final snapshot at~10~Gyr is over-plotted
  as contours which include 95, 90, 80, 70...~per cent of all
  particles.  Note the secondary peak of pole directions near
  $l=0^{\rm o}$ where the orbital pole of the Sculptor dSph satellite
  lies in this coordinate system. It is on an orbit in
  counter-rotation to the bulk of the other satellites. Sagittarius,
  which is on an orbit perpendicular to the MW disk and to the VPOS is
  seen near $l=270^{\rm o}$.  The coordinate system used here is such
  that $(l,b)=(180^{\rm o},0^{\rm o})$ points towards the initial
  orbital pole of the infalling galaxy and $b=90^{\rm o}$ is the spin
  direction of the target galaxy, which is the young MW in this model.
  For details see \cite{Pawlowski12b}.  This figure is reprinted from
  \cite{Pawlowski12b}.  }\label{fig:flyby_dir}
\end{center}
\end{figure*}

The phase-space constraints provided by the VPOS allow a
re-construction of the events that played a role in forming the young
MW. The encounter had to have been near-polar relative to the young MW
and the incoming galaxy must have been close to
edge-on. Fig.~\ref{fig:flyby} shows a sequence of images of such
interacting pairs in the present universe.  In Fig.~\ref{fig:flyby} is
also shown a time-sequence of a model fly-by encounter involving the
young MW and a young galaxy of similar mass as the young MW. The
material pulled out from the incoming galaxy spreads about the MW in a
model VPOS similar to the real VPOS (Fig.~\ref{fig:VPOSedge}). The
distribution of orbital poles of tidal debris on the Galactic sky that
this model produces is shown in Fig.~\ref{fig:flyby_dir}.

Striking is that the directions of orbital angular momenta of the
tidal debris in the model populate the same diagonal region (from the
upper left to below centre in Fig.~\ref{fig:flyby_dir}) on the
Galactic sky as the actual real VPOS does. The existence of
counter-rotating tidal debris is also evident, coinciding well with
the orbital-angular-momentum direction of Sculptor. The model shown
has not been created to match the data particularly well, and other
models computed by \cite{Pawlowski11, Pawlowski12b} also match,
demonstrating that the generic properties of the VPOS can be readily
accounted for by a wide class of encounter models.  Its detailed
properties can however allow a reconstruction of the events that
shaped the MW about 10~Gyr ago.

\subsection{The VPOS, mergers, bulges and disk regrowth}

If the young MW would have had a major encounter that lead to the
present-day VPOS, would such an encounter not have destroyed the MW
disk? Would it leave visible morphological evidence? This is an
important question, and it turns out that the MW bulge may hold
clues. 

\begin{figure}[ht]
\begin{center}
\includegraphics[scale=0.37, angle=0]{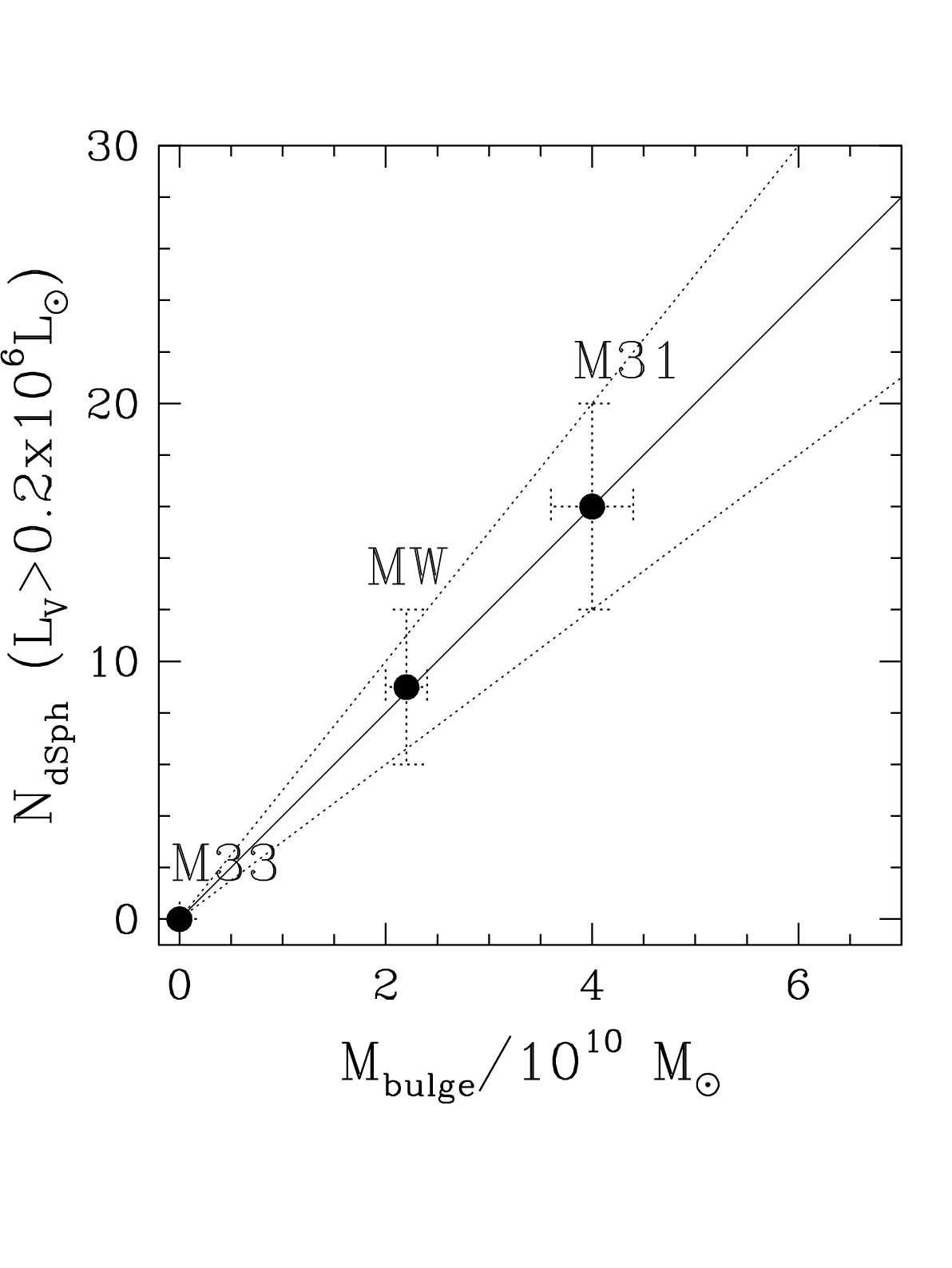}
\vspace{-15mm}
\caption{The number of dSph and dE satellite galaxies more luminous
  than $0.2\times10^6\,L_\odot$ is plotted as a function of the bulge
  mass of the host galaxy. Only satellites within a distance of
  270~kpc of the MW and M31 are used. The solid line is the deduced
  correlation between the number of satellites and the bulge mass.
  The upper and the lower dotted lines illustrate the relative
  uncertainty assumed in the Monte Carlo experiment. This figure is
  adapted from \cite{Kroupa10}.  }\label{fig:Mbulge_Nsat}
\end{center}
\end{figure}

Within the Local Group there exists a near-to-perfectly linear
correlation between bulge mass and the number of satellite galaxies
(Fig.~\ref{fig:Mbulge_Nsat}): M33 has no bulge and no known satellite
galaxies, while Andromeda has a more massive bulge and more satellites
than the MW and is in general more complex with a probably more recent
merger event than the MW \citep{Hammer10}.

Evidently, the validity of such a correlation needs to be tested with
galaxies beyond the Local Group.  On the basis of extragalactic
observational data, \cite{Karachentsevetal05} note, but do not
quantify, the existence of a correlation between the bulge luminosity
and the number of associated satellite galaxies such that galaxies
without a bulge have no known dSph companions, such as M101.
\cite{Karachentsevetal05} also point out that the number of known dSph
satellites increases with the tidal environment.  In effect,
Fig.~\ref{fig:Mbulge_Nsat} shows a correlation between the colour or
bulge-to-disk ratio of the host galaxy and the number of its
satellites. That redder host galaxies do have more satellite galaxies
than bluer hosts has indeed been discovered using the SDSS of isolated
bright host galaxies \citep{WW12}.

A correlation as evident in Fig.~\ref{fig:Mbulge_Nsat} ought to arise
naturally {\it if} the majority of satellite galaxies are ancient
TDGs, because bulges form in major galaxy--galaxy encounters
\citep{Hammer05}: Due to time-variable changes in the potentials of
each galaxy its gaseous component is channeled onto radial orbits
whe\-re it dissipates and forms stars rapidly forming a central
spheroidal component on a dynamical time-scale (a few
$10^8\,$yr). When the encounter or merger is over, the bulges may
regrow disks from accreting gas \citep{Hammer05, Hammer07, Hammer09,
  Bournaud11}. The combination of chemical and age constraints
available on the growth of the MW bulge (e.g. \citealt{Tsujimoto12})
with corresponding constraints available for the thick disk and the
VPOS constituents should allow a detailed re-construction of the early
encounter event which would have occurred after the MW old halo
spheroid formed \citep{MK10}.

It is known that present-day disk galaxies are sustaining their SFRs
through on-going gas accretion. \\ That disk galaxies regrow their
disks after significant encounters which may produce thickened older
disk components has been demonstrated in models
(e.g. \citealt{Bournaud05, Bournaud11}, see also
\citealt{Reshetnikov97}).
 
{\it The existence of the VPOS with counter-rotating constituents, and the
MW having a bulge and a thickened disk are thus understandable as
structures created in an early encounter between the young MW and
another galaxy.}

\subsection{Recreating the event}

It will be interesting to investigate if this entire MW structure {of
  VPOS, bulge and thick disk} can be created using one self-consistent
simulation. This work is likely to re-construct quite precisely the
events that occurred about 10~Gyr ago (compare with
\citealt{Hammer07}), and it may also identify which the other galaxy
involved may have been. Indeed, currently there are two galaxies that
may have been involved: the LMC is on about the right orbit already
\citep{Pawlowski11}. And alternatively, Andromeda is also close to the
DoS in projection (fig.~1 in \citealt{Metz07}) and is currently
approaching the MW again. If Andromeda was the early fly-by candidate,
then the LMC may be a massive TDG (compare with a similar suggestion
by \citealt{YH10}).

Galaxy interactions, bulge formation and the associated emergence of
TDGs would have been most common in the early universe when the young
galaxies were spaced closer to each other and when they were
presumably more gas rich than today. This is consistent with the
bulges and satellite galaxies being typically old.

It is thus proven that VPOSs emerge naturally from galaxy--galaxy
encounters, and that they allow a reconstruction of the encounter. It
needs to be studied how unique such a reconstruction is. That is,
which range of initial conditions (galaxy mass ratios, relative
inclinations and orbital angular momenta) are allowed given the
properties and constituents of the real VPOS.

\subsection{Was the young MW a polar ring \\ galaxy?}

According to the evidence uncovered in the course of this work
\citep{Pawlowski12b}, the MW has a VPOS which is naturally explained
as the remnant of an ancient encounter that must have occurred about
10--11~Gyr ago between the young MW and another perhaps similar
gas-rich galaxy.  The correlation between bulges and the number of
satellites evident in the Local Group is consistent with bulges being
produced during galaxy encounters. That galaxy encounters readily
produce {\it polar rings} has been demonstrated by
\cite{Bournaud03}. While the likelihood of creating polar ring
galaxies from mergers is small, according to \cite{Bournaud03} fly-by
encounters readily produce polar rings through the accretion of gas
from the passing donor to the host galaxy (Fig.~\ref{fig:flyby}). The
fly-by event would have pulled a tidal arm out of the passing galaxy
which fell onto the young MW thereby forming the polar ring. At the
same time, star-formation throughout the tidal arm would have produced
the young halo globular clusters (YHGCs), and the TDG precursors of
the present-day ancient dSph and UDF galaxies out to distances of 100s
of kpc. As the tidal arm swept across, MW satellites and possibly
YHGCs on counter-rotating orbits would have been generated
(Sec.~\ref{sec:VPOS_origin}).  The tidal perturbation would have lead
to bar formation in the young MW which would have formed a
bulge-component as well as a thickened disk. Previous observational
evidence for the MW possibly being a polar ring galaxy were presented
by \cite{Haud88}.

\subsection{Other extragalactic satellite-\\stream  alignments}
\label{sec:extragal}

If the MW VPOS is the remnant of an ancient tidal structure that was
created in a galaxy--galaxy encounter about 10--11~Gyr ago, is there
evidence for other galaxies also having alignments in their satellite
systems? If the MW were unique, confidence in this scenario would be
eroded.

In the local universe there are a number of known host galaxies with
stream--TDG satellite alignments: (1)~The NGC$\,$1097 ``dog leg'' stellar
stream with two typical dSph satellite galaxies
\citep{Galianni10}. (2)~The about 300~kpc long tidal stream with at
least three dSph satellite galaxies in the 1--2~Gyr old
post-interaction NGC$\,$5557 \citep{Duc11}.

Furthermore, NGC$\,$5291 \citep{Bournaud07} is surrounded by a vast
gaseous tidal ring with many embedded young and gas-rich TDGs,
altogether amounting to a vast phase-space correlated structure
containing gaseous streams and young dIrr galaxies.

Notwithstanding these examples, the Tadpole gala\-xy is an ongoing
merger with a tidal tail with many star clusters within it. Many of
the clusters are clustered in young (4--5$\,$Myr old) cluster
complexes (CCs).  The most prominent one of mass
$1.3\times10^6\,M_\odot$ and half-light radius of $160\,$pc
\citep{Tran03} may evolve into a low-mass UCD \citep{Bruens11}, or
even a typical dSph satellite if the gas and clusters surrounding the
CC were to be taken into account as a larger structure.

Finally, the Dentist Chair is an example of an interacting galaxy with
tidal tails which contain many TDG candidates in a highly phase-space
correlated overall structure \citep{Weilbacher02}.

The above examples are extragalactic systems in the local universe
which are surrounded by prominent young to intermediate-age correlated
phase-space structures which include gaseous streams, star clusters
and TDGs. Such vast structures evolve over many orbital times but
remain evident for longer than a Hubble time in phase space due to the
conservation of orbital angular momentum and energy.

{\it Thus, phase-space correlated assemblages of stellar and gaseous
streams, young TDGs as well as old dSph satellite galaxies exist in
the Local Universe.} These are probably not rare, given that such
systems are hard to find because of their low surface densities
and the faintness of the satellite galaxies. Due to the observational
flux limits, an observational bias towards discovering fairly massive
($>10^8\,M_\odot$) satellite galaxies exists implying that fainter
structures are likely to be even more common. An important task will
be to survey as many nearby galaxies as possible for the faintest
streams and possibly associated faint satellite galaxies to quantify
the frequency of occurrence of such correlated systems.

That the MW has a VPOS, while Andromeda also shows an anisotropic
satellite distribution with strong evidence for a chain-like
arrangement of a substantial number of its satellites
(\citealt{Karachentsev96, KG06, Metz07}, notably fig.~1 in
\citealt{Tollerud11}) already suggest that phase-space correlated
satellite populations may be quite common around disk galaxies with
bulges. This is consistent with dSph satellite galaxies being mostly
old TDGs \citep{Casas12}.

\subsection{On the origin of dE galaxies}
\label{sec:dEorigin}

In Sec.~\ref{sec:HypA_RM} the coincidence of dE galaxies with observed
and model TDGs in the radius--baryonic-mass diagram\-me has already
emerged, suggesting that the majority if not all dwarf satellite
galaxies may be TDGs, whereby RPDGs may also play a role in the dwarf
population of galaxy clusters. This is consistent with some of the
observed extragalactic dSph satellites being in phase-space correlated
structures (Sec.~\ref{sec:extragal}).

\cite{OT00} have demonstrated with conservative assumptions that the
number of long-lived TDGs produced over cosmological times due to
galaxy--galaxy encounters within the SMoC amounts to the number of
observed dE galaxies. They have also demonstrated that the
morphology--density relation is reproduced: poor groups of galaxies
end up having fewer dwarf galaxy members than rich groups and clusters
of galaxies. 

Also, the stellar mass-to-light ratios of dE galaxies are fully
consistent with them not having DM (e.g.  \citealt{Lisker09} and
references therein), which is expected for this class of object since
TDGs cannot capture significant amounts of DM even if it were to exist
(Sec.~\ref{sec:typeBdwarfs}). Concerning the putative DM content of dE
galaxies, the clash with the SMoC is so significant that some authors
speculate baryonic processes to be responsible for pushing out the DM
to radii where it is dynamically unimportant (\citealt{Forbes11}, see
also Footnote~\ref{foot:fanelli}). However, none of the realistic
galaxy evolution or formation simulations has ever resulted in the DM
being pushed out to the degree required.

The majority of TDGs would have been produced in the young universe
and are thus metal poor. The presently born TDGs are a minority since
galaxy--galaxy encounters are today rarer and the galaxies are not as
gas rich as in the cosmological past.  Therefore the metallicity
criterion for distinguishing TDGs from normal dwarf galaxies cannot be
applied as a robust test for TDG status. TDGs, once they decouple,
begin their own chemo-dynamical evolution and thus follow the
mass--metallicity relation \citep{Recchi07}. That even low-mass TDGs
survive for a Hubble time despite being on eccentric orbits about
their host galaxy has been shown by \cite{Kroupa97} and
\cite{Casas12}.

From the above it would thus appear that the existing data and
theoretical work are consistent with dE galaxies being old TDGs.

\section{A dialogue: the discovery of ultra thin dark matter
  filaments and of the dark force}
\label{sec:fil_darkforce}

The SMoC is highly regarded and for many as established as the SMoPP
(Sec.~\ref{sec:introd}). Given that is seems unlikely for the vast
majority of contemporary astronomers to be interpreting the data so
wrongly, it may therefore be that we are here missing some essential
aspect of the SMoC. Perhaps the SMoC is valid after all, and there are
unknown baryonic processes which would account for, among the other
issues, (i) TDGs lying on the DM-defined BTF relation, (ii) TDGs
coinciding with dE galaxies in the radius--baryonic-mass diagramme,
(iii) dE galaxies not having evidence for DM, (iv) the existence of
the VPOS around the MW, (v) correlated phase-space structures composed
of satellite galaxies about other host galaxies, and (vi) the
existence of a host-galaxy-bulge-mass---number-of-satellites
correlation. Here is a dialogue which is based on true conversations
that occurred in November~2011 in Bonn and January~2012 in Vienna:

\begin{description}

\item {\it SMoC enthusiast 1}: ``OK, I can agree that the MW satellite
  dwarf galaxies are TDGs.''

\item {\it Dynamicist}: But then there are no DM-dominated satellite
  galaxies in DM haloes, and this would invalidate the SMoC since it
  requires each larger galaxy to be accompanied by hundreds of DM
  satellites.

\item {\it SMoC enthusiast 1}: No, this does not invalidate the SMoC
  because I can just turn up re-ionisation and other effects such that
  all satellite dark matter sub-haloes with mass about $\simless
  10^{10}\,M_\odot$ remain dark.

\item {\it Dynamicist}: But would you not be going into extremes of
  fine tuning?  And, no existing galaxy-formation and evolution
  simulation has ever sho\-wn that re-ionisation can be made so
  destructive that {\it all} DM sub-haloes with about
  $<10^{10}\,M_\odot$ remain dark. Also, what about the more isolated
  dwarf-galaxy groups \citep{Tully06}?

\item {\it SMoC enthusiast 2}: Well, we have the freedom to recompute
  the models and the lack of such satellites constrains the physical
  processes that played an important role during early galaxy
  assembly. We can in actuality constrain the relevant sub-grids
  physics in order to match the observations. 

\item {\it Dynamicist}: But here is the final nail into the coffin:
  You agree that the satellite galaxies are TDGs. But TDGs cannot have
  DM, as has been emphasised many times. Now, the satellites of the MW
  have large dynamical M/L ratia ($>10$ up to a few hundred in some
  cases). So this clearly disproves the SMoC because in it TDGs cannot
  contain DM.  The solution is to move away from DM on galaxy scales
  and to accept that gravity in non-New\-toni\-an. Then we can
  understand the satellites as being TDGs. And they appear to be
  dominated by DM if we interpret the motions of their stars in
  Newtonian dynamics, but in actuality what we are seeing is
  non-Newtonian dynamics. That Milgromian dynamics plus tidal effects
  is a good solution to the satellites has already been shown by
  \cite{BM00, Angus08, Kroupa10} and \cite{McGaugh10}, so this appears
  to be the right research path to be taken.

\item {\it SMoC enthusiast 1}: No, I still disagree. The high
  dynamical M/L ratia suggest that TDGs, once formed, connect to thin
  dark-matter filaments from which they accrete kinematically cold DM.

\item {\it Dynamicist}: Now I have you cornered: There are at least
  two problems with your postulated thin DM filaments: (1)~100~Myr old
  TDGs show convincing evidence for having DM
  \citep{Bournaud07,Gentile07}. At this age, there is not enough time
  for them to connect to the postulated thin DM filaments which must
  originate from outside the virial radius of the host DM halo (if it
  were to exist). 

\item {\it SMoC enthusiast 2}: Yes but these young TDGs you are
  talking about are different, they are coming from only one galaxy
  progenitor, and this galaxy happens to probably have a lot of cold
  molecular gas which makes up the missing mass in these young
  TDGs. The dynamical M/L ratia are much smaller than those of some of
  the old satellites of the Milky Way you are referring to. So
  actually two different dominant types of DM (exotic and gas) would
  be involved in these two age groups. Nothing is wrong with that.

\item {\it Dynamicist}: But what about problem~(2): the fact that
  these young TDGs fall onto the baryonic Tully-Fisher relation
  \citep{Gentile07}? This clearly cannot be the case if they are a
  fundamentally different type of galaxies than the normal
  DM-dominated ones.

\item {\it SMoC enthusiast 3}: Look, it is generally known that the
  SMoC cannot be tested on the small scales of dwarf galaxies because
  the baryonic physics is uncertain and the simulated models lack the
  resolution.

\item {\it Dynamicist}: I would not agree with this. After all, the DM
  structures that form at a mass scale of $\simgreat 10^8\,M_\odot$
  due to the dissipationless gravitating DM particles are rather well
  understood.  Se e.g. the work of \cite{Diemand08}.  Are you then
  implying that the large body of research output from the SMoC
  community over the past 5~to~10~years claiming to solve the missing
  satellite problem is essentially unbelievable? It seems that this
  statement would suggest that these {\it many} published
  results\footnote{\label{foot:paperlist} E.g. listing mostly more
    recent papers: \cite{DekelSilk86, DekelWoo03, Tollerud08,
      DOnghia08, Li08, MKM09, Koposovetal09, OF09, Kirby09, Shaya2009,
      Busha09, Primack09, MF09, Maccioetal10, Cooper10, Deason11,
      Nichols11, Font11}.}  cannot be resorted to in order to test the
  SMoC?  If this were the case, then the whole model becomes
  untestable.

\item{\it SMoC enthusiast 3}: You are resting your tests on the MW,
  but this is not permissible. One cannot test the whole SMoC just on
  one single galaxy. We are {\it within} the MW and the data are
  poor. The surveys suffer from serious obscuration and there are
  whole regions of the sky where we cannot even look at properly, like
  in the disk of the MW. You need to use other galaxies.

\item {\it Dynamicist}: We are testing also Andromeda, but it is true
  that as we are acquiring ever better data on the MW we are finding
  that the discrepancies with the SMoC expectations are
  increasing. The surveys are sufficiently complete to allow robust
  tests. For example, the SDSS is a cone such that the
  disk-like distribution of ultra-faint satellites cannot be due to
  selective surveying. The surveys for the brighter satellites and the
  young halo globular clusters have a quite complete sky coverage. The
  tests do not rely on a single property of the MW but on generic
  features that together signal a failure.  Obviously the MW {\it
    must} be part of the SMoC, as it is within this universe.

\item {\it SMoC enthusiast 4}: Listening to this conversation I think
  you are wrong: the MW satellites are not TDGs. Instead, the SMoC is
  mostly valid but needs to be slightly modified. Similarly to the
  argument by \cite{PN10}, who also emphasise the SMoC to be an
  overall successful description of cosmology, I would argue that the
  problems of the SMoC on galactic scales shows that DM particles must
  couple to the baryons or to each other differently than only through
  gravitation.

  There may well be a {\it dark force} (e.g. \citealt{Nusser05}) which
  allows DM particles and baryons to couple in such a way as to
  enhance structure formation, as demanded by \cite{PN10}, and at the
  same time to reduce the missing satellite problem and enforce the
  satellites forming a DoS. We could postulate that this dark force
  becomes active when the thermal temperature of the baryons has
  decayed sufficiently, so that structure formation in the early
  universe is not affected. The dark force may have a number of
  components that couple differently to the different constituents,
  and all of these may be time variable (e.g.
  \citealt{Aarssen12}). So I see absolutely no reason to discard the
  SMoC in favour of a purely speculative and ad-hock modification of
  Newtonian dynamics in the ultra-weak field limit, such as what
  Milgromian dynamics, i.e. MOND, is.

\end{description}

\section{Conclusion on testing Hypothesis A and B}

Hypothesis~A (the MW satellite galaxies are DM sub-structures) has
been ruled out, while Hypothesis~B (the satellites are ancient TDGs)
stands up to all available constraints and tests.  According to {\sc
  The First} and {\sc Second SMoC Falsification Theorems}
(Sec.~\ref{sec:disproof}) the SMoC must be discarded and C/W\-DM
cannot exist.

\subsection{Why is the existence of cold or warm DM
  ruled out?}
\label{sec:noDM}

By having shown that only one type of dwarf galaxy exists, and that
this type has all the required properties associated with the known
type~B dwarfs, {\sc The Dual Dwarf Galaxy Theorem} has been falsified,
and C/\-WDM particles cannot exist leading to the SMoC being
falsified. But why does the falsification of the ``Dual Theorem''
imply that C/WDM particles cannot exist?

The reason is as follows: With the falsification of the Dual Theorem
the existence of C/WDM is immediately ruled out because by the
satellite galaxies being TDGs (i.e. type~B dwarfs) none of the C/WDM
sub-haloes would have been able to have formed stars in stark
contradiction to the calculations. Amending the physics of the C/WDM
particles to allow this to have occurred would already be a non-SMoC
model. However, such an approach would also be violated because the
dSph satellite galaxies are observed to show large dynamical M/L
ratia, i.e. they appear to be dominated by DM. Thus, if, by virtue of
their phase-space correlation they are TDGs, then the only logically
allowed solution is to discard the SMoC entirely and to consider
modified gravity models. This conclusion is fully consistent with the
known young rotating TDGs lying on the BTF relation
(Sec.~\ref{sec:noDMbutMG} below), and by the existence of the MDA
correlation (Fig.~\ref{fig:acc_discr} below), both of which cannot be
understood as a result of the physics of DM particles.

Finally, the extensive effort world-wide to detect DM particles in
terrestrial experiments has so far not been successful
(e.g. \citealt{Baudis12}). For example, the CRESST-II DM search has
reported a possible detection of a CDM particle signal
\citep{Angloher11}, but their fig.~13 also shows this putative signal
to be in the parameter region excluded by the CDMS-II \citep{Ahmed10}
and XENON100 \citep{Aprile11} DM-particle experiments.  The search for
a DM-particle-annihilation or DM-particle-decay signature from regions
where high DM densities are measured assuming Newtonian dynamics to be
valid has also been unsuccessful (e.g. the MW satellite galaxy Segue~1
has the highest DM density known but no DM signal has been detected,
\citealt{Aliuetal12}).

Increasing loss of confidence is suffered by the experiments having to
postulate ever decreasing interaction cross sections for the putative
DM particles, significantly below and away from the originally
favoured ones.  This is at the same time a fallacy of the adopted
procedure: The existence of DM particles can never be disproven by
direct experiment because ever lighter particles and/or ever smaller
interaction cross sections just below the current detection threshold
may be postulated for every non-detection. There exists no falsifiable
prediction concerning the DM particles.

\subsection{Can a modified SMoC be constructed?}
\label{sec:modSMoC}

That the SMoC needs to be discarded as a model of the real universe is
true even if cold DM filaments or dark forces
(Sec.~\ref{sec:fil_darkforce}) were to exist because structure
formation simulations would have to be repeated with these
ingredients. That is, the currently available cosmological models
would need to be revised substantially. But the revisions would be
many, since many new degrees of freedom appear with the notion of a
multi-component dark force
(Sec.~\ref{sec:fil_darkforce}). Predictability of this model would not
be given any longer, since any new discordant observation would be
accounted for, at least in principle, by new parameters in the dark
sector.

A simpler and more elegant option may be obtained by considering
non-Newton\-ian alternatives and therewith the foundations of the
SMoC.

\section{The SMoC is falsified,  \\ long live  . . . ?}
\label{sec:alternative}

As detailed in Sec.~\ref{sec:introd} the foundation of the SMoC is
Hypothesis~0i. Since Albert Einstein constrained his ansatz on
gravitation by solar system {(i.e. Newtonian)} dynamics, it is useful
to reconsider this assumption.

\subsection{The MDA correlation and
  solar system constraints}
\label{sec:MDA}

The need to introduce dynamically relevant DM on galactic scales arose
because the assumption that Einstein's field equation
\citep{Einstein16} be valid on galactic and cosmological scales led to
a failure of it as soon as kinematical measurements in galaxies and of
galaxies in galaxy clusters became available long after~1916 if it is
assumed that only baryonic matter exists \citep{Zwicky37,RF70,FG79,
  BFPR84}.  But the speculation that exotic DM particles exist that
are to be dynamically relevant in galaxies cannot be understood within
the SMoPP, have not been discovered by direct experiment despite a
highly significant effort world-wide over the past decades to detect
them, and lead to the contradictions with astronomical observations
that constitute the falsification of the SMoC above.  With the failure
of the SMoC it has now become evident that Einstein's ansatz may need
additions in the dark-physics sector (Sec.~\ref{sec:fil_darkforce}).

\begin{figure*}[ht]
\begin{center}
\hspace{0mm}\includegraphics[scale=0.8, angle=0]{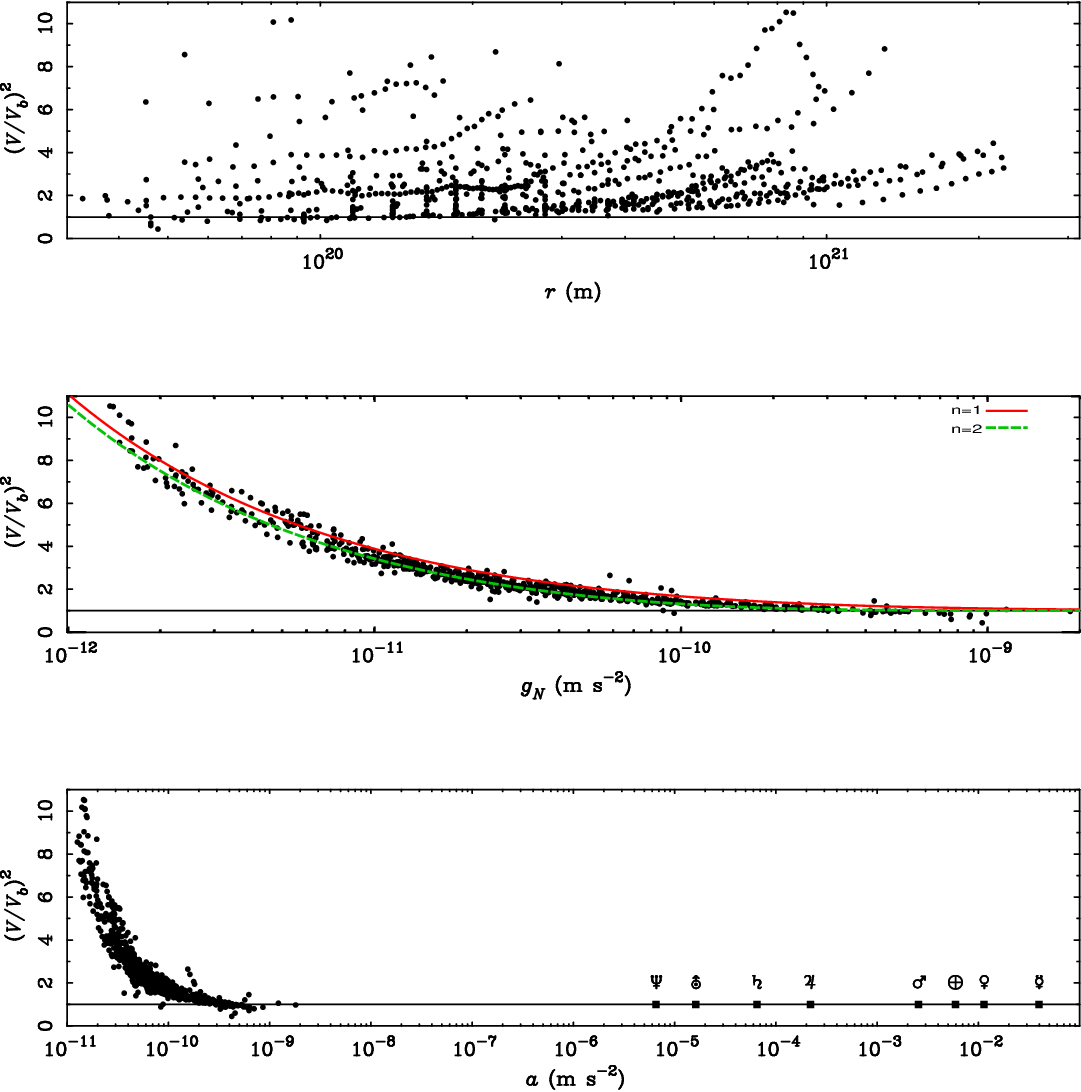}
\vspace{-2mm}
\caption{Mass-discrepancy--acceleration (MDA) data for hundreds of
  measurements in different disk galaxies (black dots).  The
  mass-discrepancy (MD) data are defined as $(V/V_b)^2$. Here $V$
  ($=V_c$ in the text) is the observed circular velocity in a disk
  galaxy while $V_b$ is the circular velocity the galaxy ought to have
  at the same radius given the observed mass of baryonic matter within
  that radius $r$. The upper panel shows the MD data as a function of
  $r$, while the middle and lower panels show the same data in
  dependence of $a$ and $g_N$. Here, $a$ is the centripetal
  acceleration, $a=V^2/r$, while $g_N = V_b^2/r$ is the acceleration
  predicted by Newtonian dynamics given the observed mass of baryonic
  matter. Evidence for DM appears exclusively only when a MD exists,
  i.e. when $V>V_b$. It is evident that there is no systematic
  behaviour of the MD with $r$, but that a well defined correlation
  exists between the MD and $a$ and $g_N$. This shows that if cold or
  warm DM were to exist, then it would need to have the property for
  accounting for this MDA correlation.  That the MD appears only at
  significantly smaller accelerations than occur in the Solar System
  is well evident in the lowest panel. High-precision tests of gravity
  that have been confirming the Einsteinian/Newtonian theory have only
  been possible in the Solar System or near neutron stars, while
  gravity in the ultra-weak field limit is probed on galaxy scales
  which were not available to Einstein in~1916.  While the SMoC has
  not allowed reproduction of the MDA data, the MDA correlation is
  accounted for excellently by Milgromian dynamics (the thin solid red
  and dashed green curves are eq.~\ref{eq:MDA}).  Adapted with kind
  permission from \cite{FM12}. This figure was prepared by Stacy
  McGaugh and Fabian L\"ughausen.}\label{fig:acc_discr}
\end{center}
\end{figure*}

The discrepancy between Newtonian dynamics and the dynamics observed
in galaxies is concisely documented as the mass-discrepancy introduced
by \cite{McGaugh04}. The mass-discrepancy (MD) in a disk galaxy is the
ratio between the observed circular velocity squared, $V^2(=V_c^2)$,
and the circular velocity squared, $V_b^2$, expected from the observed
amount of baryonic matter assuming Newtonian dynamics is
applicable. The observational data are plotted in
Fig.~\ref{fig:acc_discr}.

The observed mass-discrepancy data follow a well defined correlation
with acceleration. This is the {\it McGaugh
  mass-discrepancy--acceleration (MDA) correlation}. The MDA data show
that the discrepancy and thus evidence for DM only appears when the
observed (true) acceleration, $a$, is smaller than a critical
acceleration $a_0$,
\begin{equation}
a < a_0 = 1.12\times 10^{-10}\,{\rm m/sec}^2 = 3.6\,{\rm pc/Myr}^2, 
\end{equation}
{\it which cannot be accounted for by the SMoC
  because the physics of DM particles does not depend on
  $a$}. The critical acceleration $a_0$ constitutes a constant of
nature. It is constrained by e.g. only one single rotation curve of
one galaxy. \cite{McGaugh98} had already pointed out that understanding
the MDA correlation within the SMoC ``leads to troublesome fine-tuning
problems''.

Fig.~\ref{fig:acc_discr} demonstrates the excellent agreement between
the prediction\footnote{This is a true {\it prediction} because the
  data did not exist when Milgrom formulated a non-Newtonian approach
  in 1983. Alas, modern cosmological jargon often uses the word
  ``prediction'' to mean a value calculated within the SMoC in order
  to account for existing data.} of Milgromian dynamics and the
data. According to Milgrom's suggestion, the gravitational force
acting on a mass $m$ which experiences the acceleration $a$ is
\begin{equation}
F = \nu(|g_N/a0|)\, m\, g_N = m\,a,
\label{eq:mond}
\end{equation}
where $g_N$ is the Newtonian acceleration and the function which
describes the transition from classical Einsteinian/Newtonian dynamics
to Milgromian dynamics can be written (\citealt{FM12})
\begin{equation}
\nu(y) = \left( {1 \over 2} 
\left(1 + \left(1+ {4 \over y^{n}}\right)^{1\over2} \right)
\right)^{1\over n}, 
\label{eq:nu}
\end{equation}
where $y=g_N/a_0$ and $\nu(y) \rightarrow 1$ for $y\gg 1$ and $\nu(y)
\rightarrow y^{-{1\over2}}$ for $y \ll 1$. When $n=1$ we have the
``simple $\nu$ function''.

For a circular Newtonian orbit the velocity squared is $V_b^2(r) = g_N
\, r$. With $V_c^2=a\,r$ being the actual observed circular velocity
and $a=\nu(y)\,g_N$, the MD becomes
\begin{equation} {V_c^2(r) \over V_b^2(r)} = \nu(|g_N/a_0|).
\label{eq:MDA}
\end{equation}

\begin{figure}[ht]
\begin{center}
\hspace{0mm}\includegraphics[scale=0.55, angle=-90]{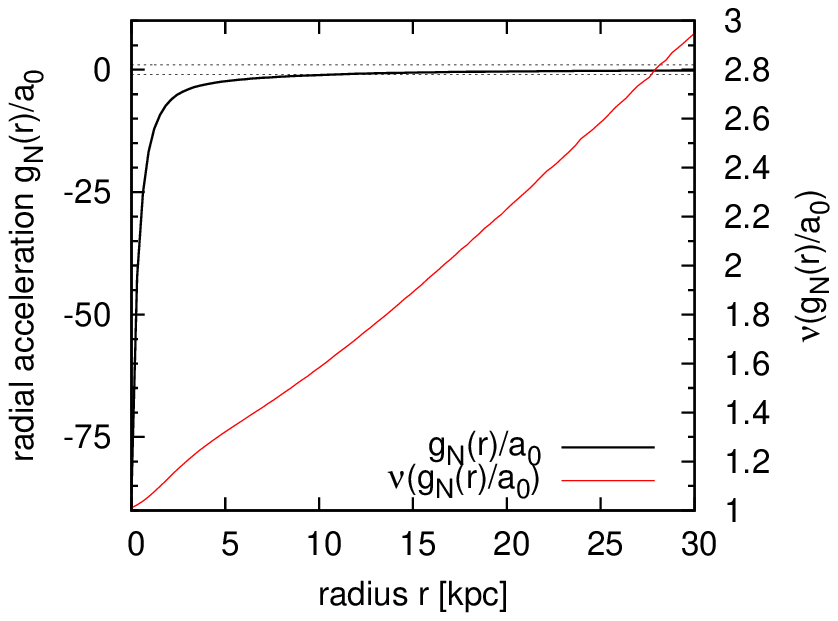}
\vspace{0mm}
\caption{The radial acceleration in units of $a_0$ is plotted as a
  function of Galactocentric distance, $r$, for a model of the MW
  (steeply rising black curve).  The horizontal dotted lines are $\pm
  a_0$, while the slowly rising (red) solid line is $\nu(y)$
  (eq.~\ref{eq:nu}).  This figure was prepared by Fabian
  L\"ughausen.}\label{fig:MWacc}
\end{center}
\end{figure}

Concerning the MW, it is useful to graph the Milgromian radial
acceleration as a function of $r$ (Fig.~\ref{fig:MWacc}). The MW model
has a baryonic Plummer bulge of mass $M_{\rm bulge}=3.4\times
10^{10}\,M_\odot$ and Plummer radius $r_{\rm pl}=0.7\,$kpc, and a
Miyamoto-Nagai disc with baryonic mass $M_{\rm
  disk}=10^{11}\,M_\odot$, radius $r_0=6.5\,$kpc and scale height of
$0.26\,$kpc.
Fig.~\ref{fig:MWacc} shows that Milgromian dynamics is expected to
become evident at $r\simgreat 8\,$kpc where $\nu\simgreat 1.4$.

What is Milgromian dynamics (i.e. MOND$\,=\,$Mo\-di\-fied Newtonian
Dynamics)? The existence of transition functions is well know in
physics, notable examples being the transitions from quantum mechanics
to classical mechanics and from relativistic to classical speeds . The
Milgromian $\nu$ function can be interpreted to be such a transition
function.  \cite{Milgrom99} showed that $\nu$ may be derived by
considering quantum mechanical effects in space-time for very small
$a$. This allows Milgromian dynamics to be seen as a modification of
inertial mass $m$, as is also evident from eq.~\ref{eq:mond} through
the terms $\{\nu(y)\,m\}\,g_N= m\,a$ (see also Appendix~A in
\citealt{Kroupa10} for a description of this ansatz). Alternatively,
Einstein's assumption may have been invalid, and gravitation does not
follow the field equation in the ultra-weak field limit. In both
cases, effective gravity would be non-Newtonian and is described by
Mordehai Milgrom's formulation \citep{Milgrom83a}.

{\it Thus, the MDA correlation constitutes a consistency with the
  general failure of the SMoC discovered above (see also
  Sec.~\ref{sec:concs}).} Indeed, the postulate by \cite{Milgrom83a}
that below the acceleration scale $a_0\approx 3.6\,$pc/Myr$^2$
dynamics becomes non-Newtonian is fully verified by the latest
high-quality kinematical data in galaxies, as demonstrated in
Fig.~\ref{fig:acc_discr} and in much depth in \cite{FM12}.

\subsection{Non-Einsteinian effective gravity}

That effective gravitation may be non-Einsteinian/non-Newtonian thus
appears to be the simpler if not the sole option
(Sec.~\ref{sec:modSMoC}): by relaxing Einstein's assumption that
gravity should be conform to Newtonian dynamics in the ultra-weak field
limit and thus allowing new effective field equations to be suggested,
it becomes possible to keep particle physics entirely within the SMoPP
(Sec.~\ref{sec:introd}). This is a highly attractive option because
the SMoPP is the most successful physical theory at hand, and because
gravitation remains poorly understood as we still do not know how
matter couples to space-time and which of the two is an emergent
property.

An example of a new interpretation of gravity is given by the recent
suggestion that ``Newton's law of gravitation naturally arises in a
theory in which space emerges through a holographic scenario''
\citep{Verlinde11}. And, a scalar tensor vector gravity theory
(leading to modified gravity, or MOG) is suggested by \cite{Moffat06}
according to which, effectively, far from a source gravity is stronger
than the Newtonian prediction, while at shorter distances it is
compensated by a vector-field-generated repulsive fifth force. This
can also be viewed as a Yukawa-type modification of the gravitational
force due to a point source.  And, it is well known that a successful
theory of quantum gravity has not been discovered yet \citep{Abdo09}.

Whatever the true solution to gravitation may be, Milgrom's suggestion
of how to modify the effective gravitational force law at ultra-low
accelerations $a\simless$ $3.6$ pc/Myr$^2$ (Sec.~\ref{sec:MDA}) has
stood the test of time.  That Milgrom's dynamics can be embedded in a
generalised relativistic tensor vector scalar (TeVeS) gravity theory
(not to be confused with the above ``scalar tensor vector gravity
theory'') has been proven by \cite{Bekenstein04}.  TeVeS is derived
from the action principle and therewith respects conservation
laws. The impact of this break-through is evident in the increase in
citations to the original research paper \citep{Milgrom83a}, which has
by now accumulated about~1000 citations. Furthermore,
\cite{ModestoRandano10} study the approach of \cite{Verlinde11} and
suggest that Milgromian-like dynamics ensues once well-motivated
corrections are applied.

Given that Milgromian dynamics is the correct description of galactic
dynamics (just as Newtonian dynamics correctly accounts for Solar
system dynamics), an increased effort to embed Milgromian dynamics
within a Lorentz-covariant framework has ensued
(e.g. \citealt{Bekenstein04, Sanders05, Zlosnik07, Bruneton07, Zhao08,
  Blanchet09, Skordis09, Milgrom09}). A quasi-linear formulation of
Milgromian dynamics has been discovered only recently
\citep{Milgrom10, ZF10} which appears to allow easier access to N-body
calculations.  An interesting suggestion has been followed by
\cite{Bruneton09} who study theories in which DM is the source of the
Milgromian phenomenology by introducing an interaction term between
baryonic matter, DM and gravity.  Additional approaches to an
environmentally-dependent dark sector, where the merits of CDM on
large scales are unified with the merits of Milgromian dynamics on
galactic scales, have also been suggested and studied \citep{Zhao07,
  LiZhao09, ZhaoLi10} (but see the ``dark force'' issues in
Sec.~\ref{sec:modSMoC}).

It is expected that the coming years will be providing many new
exciting insights into gravitation and the dynamics and evolution of
galaxies as well as of larger cosmological structures.  An excellent
comprehensive treatment of this entire ansatz and an overview of the
most recent progress and research activity is available in the major
review by \cite{FM12} and in the book by \cite{Sanders10a}.

\subsection{Galactic-scale problems \\ vanish}

It is rather noteworthy that virtually all problems on galactic scales
disappear naturally within Milgrom's framework (\citealt{FM12} for a
thorough review). For example, the MW satellite galaxies and the VPOS
would naturally be TDGs since the early encounter of the MW with
another galaxy would have occurred also in this framework. The
existence of phase-space correlated structures such as the MW VPOS
would then be a natural consequence of satellite galaxies forming as
TDGs (Failure~8 in Sec.~\ref{sec:concs}).  Because there would be no
DM halo around the MW and the other galaxy, dynamical friction would
not decay their relative orbits such that the fly-by scenario between
the young MW and the young LMC or even the young Andromeda would be
feasible. Multiple encounters between two gas rich galaxies become
readily possible, each time spawning new TDG and GC populations in
correlated phase-space structures.  Also, the {\it invariant baryonic
  galaxy problem} (Failure~10 in Sec.~\ref{sec:concs}) in the SMoC
\citep{Kroupa10} disappears entirely. And, the {\it structure growth
  problem} in the SMoC \citep{PN10} is alleviated in a Milgromian
cosmology whereby the {\it Bullet Cluster problem} (Failure~17 in
Sec.~\ref{sec:concs}) of obtaining the high relative-cluster velocity
is also avoided \citep{AD11}.  With baryonic matter as the sole source
of gravitation on galactic scales, the {\it TDG mass-deficit problem}
(Failure~9 in Sec.~\ref{sec:concs}) disappears in Milgromian dynamics
\citep{Gentile07,Milgrom07}, while it is unsolvable in the SMoC: In
Milgromian dynamics TDGs and {\it all} rotating stellar systems lie on
the Tully-Fisher relation, and the Faber-Jackson relation for
pressure-supported stellar systems also emerges naturally
\citep{Sanders10b}.

\subsection{The Bullet and Train-Wreck\\ Clusters}

The Bullet Cluster is often perceived to be a disproof of Milgromian
dynamics because even in Milgromian dynamics DM is required to explain
the observed separation of the weak lensing signal and the baryonic
matter. In actuality, the Bullet Cluster is, if anything, a major
problem for the SMoC because the large relative cluster--cluster
velocity at the mass scale of the two observed clusters required to
provide the observed gas shock front cannot be attained in the SMoC
\citep{LK10,TN11}.  But such velocities arise naturally and abundantly
in a Milgromian cosmology.

Assuming the Milgromian framework to be the correct description of
effective gravitational dynamics, it has been shown that the Bullet
Cluster lensing signal can be accounted for in it \citep{AFZ06}.  If a
Milgromian cosmology is allowed to have a hot DM component then the
Bullet Cluster is well explainable \citep{AM08,Angus11}.  We know that
neutrinos oscillate, therefore they must have a mass. That mass is
small. This makes them a form of hot DM that we most definitely know
to exist. In order to explain the oscillations, particle physics
suggests the possible existence of more massive, sterile neutrinos,
which interact by gravity. If they exist they might be massive enough
to account for the missing mass in galaxy clusters in Milgromian
dynamics (and they can fit the first three acoustic peaks in the
CMB). Taking this ansatz, \cite{AFD10} demonstrate that consistency in
solving the mass-deficit in galaxy clusters and accounting for the CMB
radiation power spectrum is achieved if sterile neutrinos (SN) have a
mass near~11~eV.  They write ``we conclude that it is intriguing that
the minimum mass of SN particle that can match the CMB is the same as
the minimum mass found here to be consistent with equilibrium
configurations of Milgromian clusters of galaxies.''\footnote{As
  \cite{Angus09} emphasises, a mass of 11~eV for sterile neutrinos is
  excluded by cosmological data {\it only if} it is assumed that
  Newton's laws are correct.}

The Train-Wreck Cluster (Abell 520) has been sho\-wn to be
incompatible with the SMoC because the putative C/WDM particles have
separated from the galaxies such that a core of DM is left
\citep{Mahdavi07, Jee12}. While these authors speculate on a possible
solution such as DM possibly having a self-interaction property (but
see Sec.~\ref{sec:modSMoC}), in a Milgromian cosmological model with
HDM it is conceivable for the self-bound galaxy-cluster-sized HDM core
to dissociate itself from the baryonic matter in galaxies, which
individually would remain on the BTF relation.  {However, different
  groups analysing the same lensing data obtain different mass maps
  (see \citealt{OU08}). The monopole degeneracy, which can lead to
  false peaks in the mass map \citep{Liesenborgs08}, also affects the
  weak lensing mass reconstruction. Thus the issue remains
  inconclusive.}

Within the MOG framework, \cite{MT09a} argue to be able to account for
both the Bullet and the Train Wreck Cluster.

\subsection{Milgromian cosmology}
\label{sec:MONDcosm}

\begin{figure*}[ht]
\begin{center}
\includegraphics[scale=0.8, angle=0]{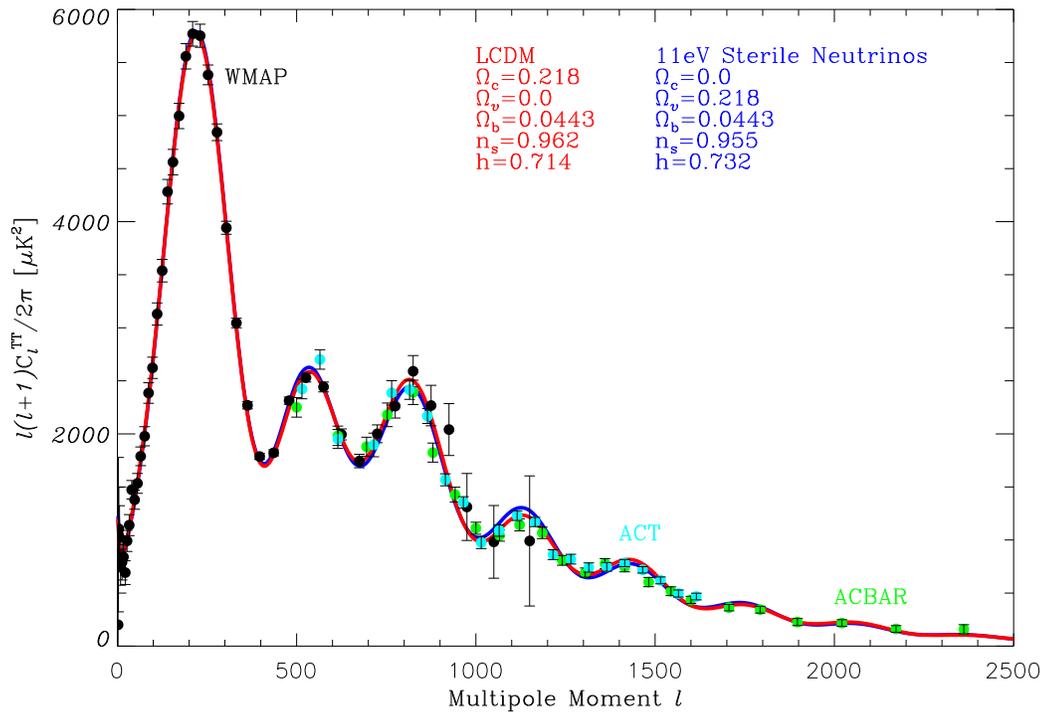}
\caption{Milgromian-based cosmological theories account for the CMB
  power spectrum just as well as the SMoC.  The CMB power spectrum as
  measured by the WMAP satellite year seven data release (filled
  circles), ACT (turquoise data) and the ACBAR 2008 data release
  (green circles).  The SMoC/$\Lambda$CDM and Milgromian dynamics
  (assuming hot DM is in the form of 11~eV sterile neutrinos) models
  are an identical representation of the CMB data, while the
  Milgromian model completely outperforms the SMoC on galactic scales.
  See \cite{AD11} for more details. (fig.~1 from \citealt{AD11} with
  kind permission from Garry Angus.)  }\label{fig:cmb_MOND}
\end{center}
\end{figure*}

A cosmological model based on Milgromian dynamics is presented by the
pioneering work of \cite{Llinares08,Angus09} and \cite{AD11}. It has
the same expansion history as the SMoC and therefore shares the same
BB physics, but it differs from the SMoC at the galactic scale where
it outperforms the SMoC comprehensively. Structures form more rapidly
\citep{Nusser02, Llinares08, AD11} as demanded by \cite{PN10} on
studying data in the Local Volume of galaxies. The structure-formation
computations are more demanding due to the gravitational theory being
non-linear which limits the currently attainable numerical
resolution. Merely ``dust'' simulations have been achieved so far in
which the baryonic matter is approximated by particles that interact
only via gravitation.  A realistic structure formation simulation
would however have to account for galaxies being purely baryonic
objects such that dissipationless physics, as dominates structure
formation in the SMoC, is not applicable. Currently such computations
within the Milgromian framework are {out of reach}.
Fig.~\ref{fig:cmb_MOND} demonstrates that this Milgromian-based
``Angus-cosmological model'' accounts for the CMB power spectrum as
well as the SMoC does.

The above thus {\it disproves} any claims to the effect that the SMoC
be the only cosmological model accounting for the CMB and data on
structure formation.  The pioneering work done by Garry Angus and
others has demonstrated that the SMoC is not unique in explaining the
CMB. {\it Therewith the final obstacle against discarding the SMoC has
  been surmounted. There is no logically consistent argument for
  adopting the SMoC over other models.}

As a challenge for the future, any alternative to the SMoC needs to be
shown to agree with the measured galaxy correlation function, as well
as with all the other observational data that have been accumulating
over time.

\section{Conclusions}
\label{sec:concs}

\subsection{Summary of SMoC falsification}

The falsification of the SMoC can be summarised with the following steps:

\begin{enumerate}

\item It is an observed as well as theoretical {\it fact} that new
  (type~B) dwarf galaxies can from in galaxy--galaxy encounters as
  TDGs and as RPDGs in galaxy clusters. If the SMoC were true then
  type~B dwarfs still could not contain much DM and many TDGs would
  have been born during the early cosmological epoch. It is emphasised
  that the dwarf galaxies discussed here have baryonic masses
  $\simgreat 10^4\,M_\odot$ and include dwarfs significantly lower in
  surface luminosity than can be observed currently near interacting
  galaxies.

\item If the SMoC were valid then this implies the existence of two
  types of dwarf galaxies: those with DM (type~A dwarfs) and type~B
  dwarfs without DM ({\sc The Dual Dwarf Galaxy Theorem},
  Sec.~\ref{sec:dual_theorem}). {\it Note that the existence of type~A
    dwarfs rests on the speculation that there is C or W DM. Type~B
    dwarfs on the other hand are observed to exist and are known to
    survive.}

\item But the observed type~B dwarfs lie on the BTF relation, which is
  supposedly defined by DM dominated galaxies (both dwarf and major).
  This proves {\sc The First SMoC Falsification Theorem}. And type~B
  dwarfs are observed to coincide morphologically with dE/dSph
  galaxies.  dE galaxies are observed to not contain DM.  This proves
  {\sc The Second SMoC Falsification Theorem}.

\item Type~A (DM) dwarfs are expected to be distributed approximately
  isotropically about their host galaxy.  Type~B dwarfs ought to
  typically form phase-space correlated populations surroun\-ding
  their host galaxy.  The MW satellites are found to be in a
  phase-space-correlated structure.  That this anisotropic
  distribution of MW satellites may be due to an anisotropic search is
  ruled out by the dSph satellites, the UFDs, the inner YHGCs, the
  outer YHGCs and stellar and gaseous streams all independently
  showing the same distribution despite completely unrelated and
  different discovery and search strategies and methods.  Seven of
  eight measured satellite motions confine these to be within the
  VPOS.  The satellites of other galaxies are found to be in
  phase-space-correlated structures as well.

  The observed properties of dE galaxies and the distribution of dwarf
  galaxies about the MW within the VPOS and in phase-space correlated
  structures about other galaxies requires these to be ancient type~B
  dwarfs.

\item Observations thus only ever show evidence for the existence of
  one type~B of dwarf galaxy, therewith falsifying {\sc The Dual
    Dwarf Galaxy Theorem}.

\item Therewith there are no type~A dwarf galaxies and there are no
  type~A satellites near the MW. C or W DM therefore cannot exist. 

\item Consistency checks show that the SMoC is in disagreement with
  other observational properties of galaxies.

\item It has been established that the SMoC is not unique in
  accounting for the CMB and BB nucleosynthesis.

\item The SMoC appears to suffer under generic failures
  (Sec.~\ref{sec:generic} below).
\end{enumerate}

It is important to seek consistency of this deduction with other
arguments: If {\sc The SMoC Falsification Theorems} are true, then the
SMoC must fail on other tests as well:

\subsection{The VPOS}

Concerning only the MW, the vast polar structure -- VPOS -- is a
physical property of the MW galaxy. 

By itself, the VPOS is not fundamentally incompatible with the SMoC
since such structures are expected to arise in any cosmological theory
which allows galaxy--galaxy interactions to occur.  The logical reason
for why the existence of the VPOS alone already does exclude the SMoC
is because the VPOS does not allow any luminous DM sub-structures to
exist around the MW. The vast output from the world-wide cosmological
simulation community has, however, shown beyond any doubt that the MW
must contain hundreds of shining DM sub-structures.

If the existence of these shining DM sub-structures around the MW is
excluded by observation, then there is no logical nor physical basis
for arguing that dark DM sub-structures exist. This is because there
is no known physical process that can avoid star formation in all DM
haloes of mass $\simless 10^{10}\,M_\odot$.

{\it If the MW has no DM sub-structures, then no other galaxy can have
  them}. This is the case because the MW is expected to have many
thousands of DM sub-structures. The chance occurrence of a major
galaxy such as the MW having no DM sub-structures is ruled out at an
extraordinarily high confidence level, as even simple Poisson
statistics demonstrate.

\subsection{Logical consistency of the SMoC falsification -- }
\label{sec:failures}

As stated above, if the SMoC is truly a false representation of
reality, then there must be many failures of it when confronted with
observational data. This is indeed the case, as summarised in the
following two subsections.

\subsubsection{Generic failures of inflationary BB models ?}
\label{sec:generic}

\cite{Starkman12} have shown the CMB fluctuations to be incompatible
with the SMoC causing major tension with standard inflationary
cosmologies. \cite{LiuLi12} find that the WMAP data completely miss
the quadrupole CMB signal posing a serious challenge to the SMoC, but
this also constrains any cosmological model.

BB nucleosynthesis, which is usually reported to successfully account
for the emergence of elements, has not been able to explain the
observed low Li abundance \citep{Coc12, FM12}.

An important problem related to Hypothesis~0ii which plagues all known
BB-based cosmological models are the {\it two missing baryon problems}
(e.g. \citealt{AB10,McGaugh2010b}): (i) From BB-nucleosynthesis and
CMB observations the cosmic average baryon-to-DM mass-density fraction
is $f_{\rm b,DM}=0.171\pm0.006$ but less than half of the baryon
density has been found. (ii) Galaxies are observed to have a
significantly smaller baryon fraction relative to the cosmic
average. Both problems remain unresolved, as it is not known in which
form the missing baryons reside nor why galaxies are so depleted in
baryons. The above are generic failures of inflationary BB
cosmologies.

That the distribution and properties of galaxies in the whole Local
Volume is incompatible with the expectations from the SMoC has been
emphasised by \cite{PN10}.

\subsubsection{A long list of Failures}
\label{sec:failures_list}

\begin{figure*}[ht]
\begin{center}
\includegraphics[scale=0.6, angle=0]{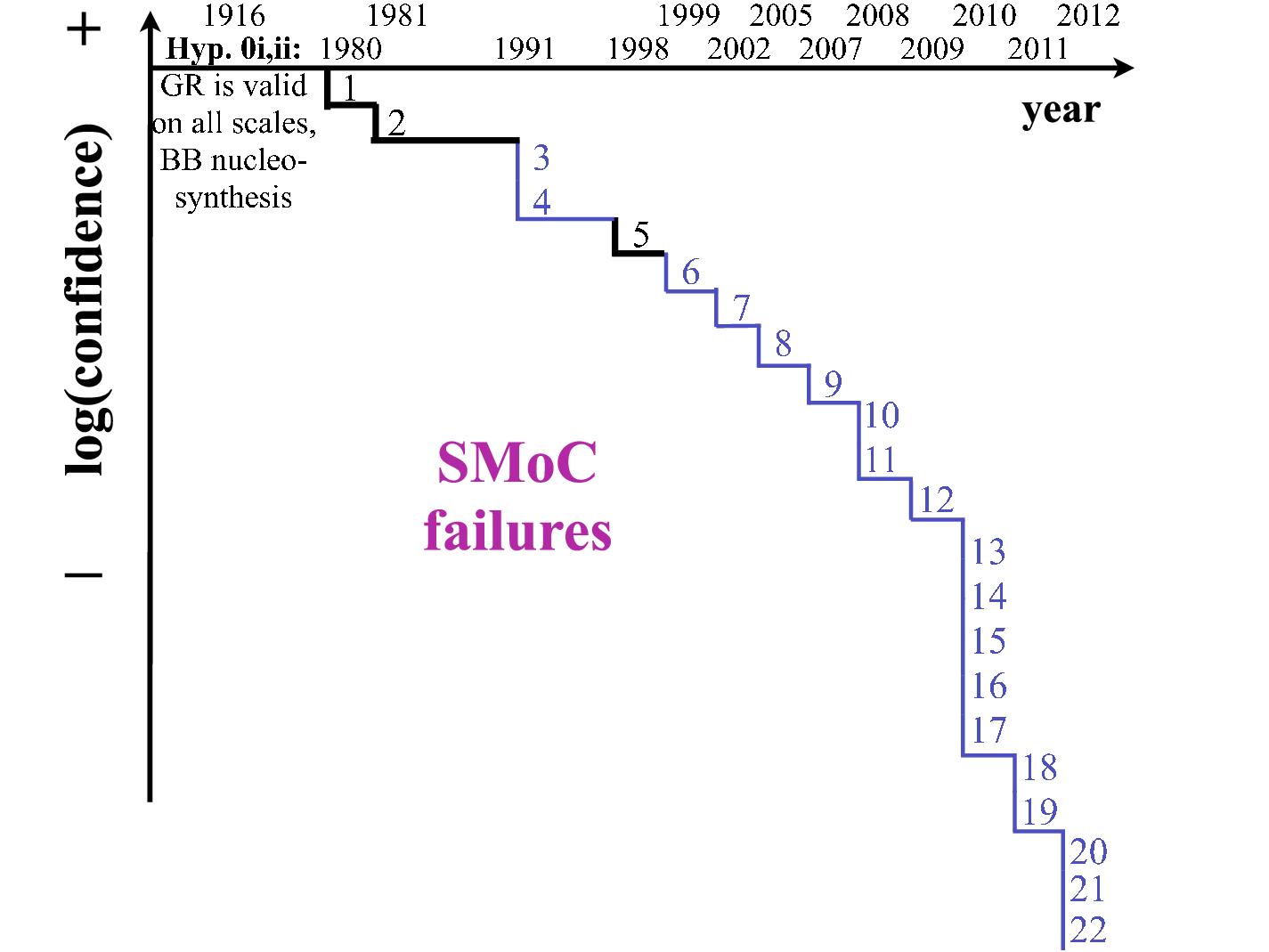}
\caption{The {\sc SMoC-Confidence Graph}: The decline of the SMoC. The
  fundamental assumptions underlying the SMoC are that Hypothesis 0i
  and 0ii are valid. BB nucleosynthesis is taken to be a generic
  property of any realistic cosmological model.  Each {\it additional}
  hypothesis which needs to be invoked to solve a significant
  discrepancy of the fundamental assumptions with the observational
  data leads to a decline in the confidence of the model (thick black
  steps downwards). Each Failure of the model computed within the set
  of hypothesis valid until that time also leads to a decline in the
  confidence of the model (blue steps downwards).  The Failures are
  listed with the relevant references in
  Sec.~\ref{sec:failures_list}. The time axis shows time consecutively
  but not to exact scale. The steps downwards are taken here to be
  equal, although not every failure necessarily has an equal weight. A
  statistically rigorous quantification of the model confidence lies
  beyond the scope of this contribution. It is likely to be
  subjective, because no agreement in the community would be reached
  as to the significance of a particular failure.  The intent of this
  graph is to provide a visual impression of the overall development
  of the SMoC confidence.  It is evident that the currently standard
  cosmological model based on Hypothesis~0i and~0ii has a long history
  of failures without convincing remedies such that the confidence
  that can be placed in the model has become negligible.
}\label{fig:SMoC_decline}
\end{center}
\end{figure*}

\begin{description}

\item (1) 1980: {\it Curvature and homogeneity:} The BB
  would imply the universe to be highly inhomogeneous and curved in
  disagreement with observations. This is solved by introducing {\it
    inflation} \citep{GT80}.

\item (2) 1981: {\it The super-Keplerian galactic rotation curve:}
  Rotation curves of disk galaxies are observed to remain quite flat
  \citep{RF70,Bosma81}. This is solved by introducing {\it cold} or
  {\it warm DM} \citep{BFPR84}.

\item (3) 1991: {\it Angular momentum:} Disk galaxies forming in the
  C/WDM cosmological model dissipate too much angular momentum by
  virtue of the baryons falling into the DM potential wells, ending up
  being too compact with too little angular momentum in comparison
  with observed disk galaxies (\citealt{NB91}, see also \citealt{PS11,
    Martig12, Dutton12, Scannapieco12}).

\item (4) 1991: {\it The cusp/core:} CDM haloes have cusps
  whereas the observationally deduced DM halo profiles have
  substantial core radii similar to the dimension of the luminous
  galaxy (\citealt{DC91}, see also \citealt{Gilmore07a, Gilmore07b,
    deBlok10, Chen10, JG12}). A possible solution has been suggested by the
  simulations of \cite{Governato12} but relies among other assumptions
  on a bursty star-formation rate (SFR) required to repeatedly blow
  out gas and a steep Kennicutt-Schmidt exponent ($n=1.5$) in
  $SFR\propto \rho_{\rm gas}^n$, where $SFR$ is the star formation
  rate and $\rho_{\rm gas}$ is the local gas density. However, in
  reality it is not clear if the dSph and UFD satellites experienced
  bursty SFRs, $n=1$ \citep{PAK08, PAK09}, the IMF would have had a
  lack of massive stars at the low SFRs of the MW satellites, as is
  inferred by \cite{Tsujimoto11}, and the threshold for SF is lower in
  reality than assumed in the simulations (see further
  below). Repeated gas blow-out which is required to evolve the cusps
  to cores is thus not likely to be possible.  WDM models tuned to
  account for the observed large cores in dwarf galaxies have such
  long DM particle streaming lengths that the dwarf galaxies cannot
  form in the first place \citep{Maccioetal12}.

\item (5) 1998: {\it Dark energy:} The fluxes and redshifts of
  observed type~Ia supernovae (SNIa) do not match the cosmological
  model \citep{Riess98, Schmidt98, Perlmutter99} unless the universe
  is assumed to expand at an ever larger rate. To account for the
  implied accelerated expansion {\it dark energy} (DE) is
  introduced. As with inflation, while mathematically allowed, it
  remains unclear if DE constitutes physics {(see e.g. the discussion
    in \citealt{Afshordi12}).  The SNIa flux--redshift data may at
    least partially be explained with an inhomogeneous universe
    \citep{Wiltshire09,SmaleWiltshire11,Marra12} rather than with DE,
    whereby systematics in SNIa light cur\-ve fitting remain an issue
    \citep{SmaleWiltshire11}. \cite{BullClifton12} find that the
    ``appearance of acceleration in observations made over large
    scales does not necessarily imply or require the expansion of
    space to be accelerating, nor does it require local observables to
    indicate acceleration.''  In fact, it might perhaps be surprising
    that a homogeneous SMoC shou\-ld lead to a perfect agreement with
    the observed SNIa data. In other words, the SNIa data that stem
    from the real inhomogeneous universe \citep{Karachentsev12} {\it
      should} show some deviations from a homogeneous SMoC. If none
    are seen then this may imply an over-const\-rained
    model.}\footnote{That the SMoC with DE does not conserve energy is
    well known (e.g. \citealt{Kroupa10}).}

\item (6) 1999: {\it Missing satellites:} Computations with
  more powerful computers showed that many more DM sub-structures form
  than observed galaxies have satellites (\citealt{Klypin99, Moore99};
  the problem is somewhat reduced with WDM: \citealt{Menci12}).

\item (7) 2002: {\it Hierachical structure formation:} As
  more-massive galaxies are build-up hierarchically from smaller
  building blocks in the SMoC, their [$\alpha$/Fe] ratios ought to
  reflect the sub-solar [$\alpha$/Fe] ratios of the building blocks
  (e.g., dE galaxies have low [$\alpha$/Fe] ratios). In conflict with
  this expectation, observed massive E~galaxies show high near-solar
  [$\alpha$/Fe] values \citep{Thomas02}. This may be partially
  alleviated by a prescription for AGN quenching of star formation in
  massive haloes but not so in the interme\-dia\-te-galaxy-mass regime
  (\citealt{Pipino09}, see also \citealt{NLO05, Recchi09}).

\item (8) 2005: {\it The Disk of Satellites (DoS/VPOS):} The
  observed satellite galaxies of the MW are arranged in a vast polar
  structure \citep{Kroupa05, Metz07, Metz08, Metz09, Kroupa10,
    Pawlowski12b}.  Of all objects at Galactocentric distances larger
  than 10$\,$kpc, only 4~per cent are not associated with the VPOS
  (Sec.~\ref{sec:VPOS_concs}).  Extragalactic anisotro\-pic satellite
  systems are common, and Andromeda appears to have a flattened
  satellite system seen edge-on (Sec.~\ref{sec:extragal}).

\item (9) 2007: {\it The TDG mass-deficit:} Unexpectedly,
  observed young TDGs show evidence for DM which however is not
  possible if the SMoC were true \citep{BH92} unless they contain
  undetectable gas (\citealt{Bournaud07}, see also
  \citealt{Gentile07}).

\item (10) 2008: {\it Invariant disk galaxies:} Observed disk galaxies
  are too similar following a simple one-parameter scaling law over
  many orders of magnitude in mass in conflict with the expected
  variation in the SMoC due to the chaotic formation history of each
  DM host halo (\citealt{Disney08}, see also \citealt{Hammer07,
    Kroupa10}).

\item (11) 2008: {\it The common mass-scale:} In the SMoC, DM
  sub-haloes are distributed according to a power-law mass
  function. But observed satellite galaxies have too similar DM masses
  (\citealt{Mateo93,Strigari08}, see also \citealt{Kroupa10} and for
  Andromeda \citealt{Tollerud11}).

\item (12) 2009: {\it Constant surface density:}
  Considering the matter distribution in observed galaxies within one
  DM-halo scale radius, \cite{Gentile09} find ``This means that the
  gravitational acceleration generated by the luminous component in
  galaxies is always the same at this radius. Although the total
  luminous-to-dark matter ratio is not constant, within one halo
  scale-length it is constant''. In the SMoC there is no physical
  principle according to which the DM and baryonic densities ought to
  be invariant within this radius.

\item (13) 2010: {\it The luminous sub-halo mass function:}
  The mass function of observed satellite galaxies disagrees with the
  predicted mass function of luminous sub-haloes \citep{Kroupa10}.

\item (14) 2010: {\it Bulge-less disk galaxies}: That the
  bulge-to-disc flux ratios are smaller than those produced by LCDM
  simulations is pointed out by \cite{GW08}. 58–-74 per cent of all
  observed disk galaxies are claimed to not have a classical bulge
  \citep{Kormendy10}. This is in conflict with the heavy merging
  history expected for bright galaxies if the SMoC were true
  \citep{Hammer07}. For attempts to produce bulgeless disk galaxies
  see text below.

\item (15) 2010: {\it Isolated massive galaxies:} In the
  observed Local Volume of galaxies there are three massive disk
  galaxies that are too far off the matter filament \citep{PN10}.

\item (16) 2010: {\it The void:} The Local Void is observed to
  be too empty in comparison to the SMoC expectation
  \citep{Tikhonov09,PN10}.

\item (17) 2010: {\it The Bullet Cluster:} The observed large
  relative velocity of the two interacting galaxy clusters is not
  accountable for in the SMoC \citep{LK10,TN11}.

\item (18) 2011: {\it The missing bright satellites:} The
  predicted mass function of DM sub-haloes implies that a significant
  number of bright satellite galaxies is missing
  \citep{Bovill11,BoylanKolchin11}.  \cite{VC12} suggest this problem
  does not occur if the DM sub-haloes have Einasto rather than NFW
  density profiles. {\cite{WangFrenk12} suggest this problem does not
    occur if the MW DM halo is less massive than about
    $2\times10^{12}\,M_\odot$.  But this is unlikely as the large
    proper motion of the LMC implies the MW to be more massive than
    about $2\times10^{12}\,M_\odot$ and the LMC to be a recent
    acquisition and on its first passage \citep{BKBH11}. Furthermore,
    if this were the case then the question would need to be posed as
    to how likely the LMC happens to pass the MW within the VPOS.}

\item (19) 2011: {\it The thin old disk:} The MW has a thin
  disk which has stars as old as 10~Gyr. Such old thin disks have
  still not been produced in the SMoC \citep{House11}.

\item (20) 2012: {\it The Train-Wreck Cluster:} The galaxy
  cluster A$\,$520 has been shown to contain what appears to be a DM
  core with too few galaxies as well as evidence for a
  cluster--cluster encounter. The C/WDM paradigm cannot account for
  this separation of DM from the luminous matter, which is the
  opposite behaviour to the Bullet Cluster (Failure 17 above,
  \citealt{Jee12}, see also \citealt{Mahdavi07}).

\item (21) 2012: {\it Missing Dark Matter:} Over spatial scales
  of~100~Mpc extend the density of matter fluctuates by~10~per cent if
  the SMoC were valid. By counting up all matter within the local
  sphere with a radius of~50~Mpc, \cite{Karachentsev12} demonstrates
  the actual density to be too low by a factor of~3--4. Most of the
  missing mass is DM. 

\item (22) 2012 {\it Massive Galaxy Clusters:} The most massive most
    distant galaxy clusters are important constraints on cosmological
    theory because the rapidity with which mass assembles to galaxy
    clusters depends on dark matter and/or on modified gravitation
    (Sec.~\ref{sec:MONDcosm}). \cite{Gonzalez12} discover a giant lensed
    arc near the cluster IDCS J1426.5þ3508 and deduce ``For standard
    LCDM structure formation and observed background field galaxy
    counts this lens system should not exist.''

\end{description}
The {\sc Theory-Confidence Graph}, Fig.~\ref{fig:SMoC_decline},
visualises the development of confidence in the SMoC as a function of
time\footnote{For example, if there are two tests of a model, each
  yielding a confidence of $10^{-4}$ that the model represents two
  different aspects of the data, then the combined confidence is
  $10^{-8}$ if the two tests are independent. Therefore a logarithmic
  confidence scale is indicated in Fig.~\ref{fig:SMoC_decline}.}.
While some of the failures have been taken as major discoveries of new
physics (inflation, DM, DE) because they can be solved by mathematical
formulations in principle allowed by Einstein's field equation, while
other failures are typically discussed away as minor problems, the
consistent appearance of ever new failures suggests rather that the
whole construction based on Hypothesis~0i and~0ii needs to be revised.
Since their discovery, each of the problems has received
attention. Nevertheless, despite important seminal work no convincing
and importantly, {\it no mutually consistent} solutions have
emerged for these problems.

For example, the angular momentum, cusp/core, invariant baryonic
galaxy, the constant surface density and bulge problems
(Failures~3,~4,~10, 12 and~14, respectively) are related and it is
thought that a better understanding of the complex baryonic processes
may solve these. The seminal work by \cite{Brook11} has shown that it
is possible, at least in principle, to grow galactic disks in DM
haloes such that they resemble the real galaxies. But the proposed
solution is for host DM haloes that have a mild history of mergers,
therewith applying to a minority of DM host haloes, while bulgless
disk galaxies are the dominant galactic population \citep{Kormendy10}.
And, the feedback energy required to blow out the baryons such that
they cool and slowly re-accrete is extreme.  The density threshold for
star-forming gas is far higher in the models over a resolution limited
spatial region of extend of about 200~pc in comparison to real
molecular clouds where the density is high in only pc-sized
regions. In the models, the full stellar initial mass function (IMF)
hits the gas once star-formation ensues, while in reality the
dependency of feedback on the star-formation rate (SFR) is smaller at
low SFRs through the dependency of the IMF on the SFR
\citep{WKP11,Kroupa12}.

Also, the presence of the many sub-haloes with satellites in the SMoC
models leads to problems understanding how major disk galaxies such as
our MW can have old thin disks that can extend beyond~20~kpc. After
studying resolution issues in different numerical schemes,
\cite{House11} write in their conclusion about the CDM MW models
studied ``None has a thin disc older than $\sim$6 Gyr, indicating that
it would be difficult to gain a thin disc as old as some estimates for
the Milky Way thin disc within the current cold dark matter
paradigm.''  Similarly, \cite{Kormendy10} state ``It is hard to
understand how bulge-less galaxies could form as the quiescent tail of
a distribution of merger histories.'' And, ``Our Galaxy provides an
additional important conclusion. Its disk stars are as old as~9--10
Gyr ... . Unless our Galaxy is unusual, this suggests: The solution to
the problem of forming giant, pure-disk galaxies is not to use some
physical process like energy feedback to delay star formation until
recently and thereby to give the halo time to grow without forming a
classical bulge''. The existence of massive, pure disk galaxies in the
most massive DM host haloes (e.g. M101), which must have had a significant
growth/merger history, are another problem if not a falsification of
the SMoC. 

That thin self-supporting disk galaxies can form readily if the DM
sub-structures and their merging does not occur has been shown by
\cite{Samland04}. These models still require a slowly growing DM host
halo, but by the absence of the DM sub-structures they are not
consistent with the SMoC, for which the work by Samland had been
critisised.

\subsection{No dark matter but modified gravity}
\label{sec:noDMbutMG}

If the SMoC is falsified and cold or warm DM does not exist, then how
can the dynamics of galaxies be accounted for?  

\begin{figure*}[ht]
\begin{center}
  \includegraphics[scale=0.85, angle=0]{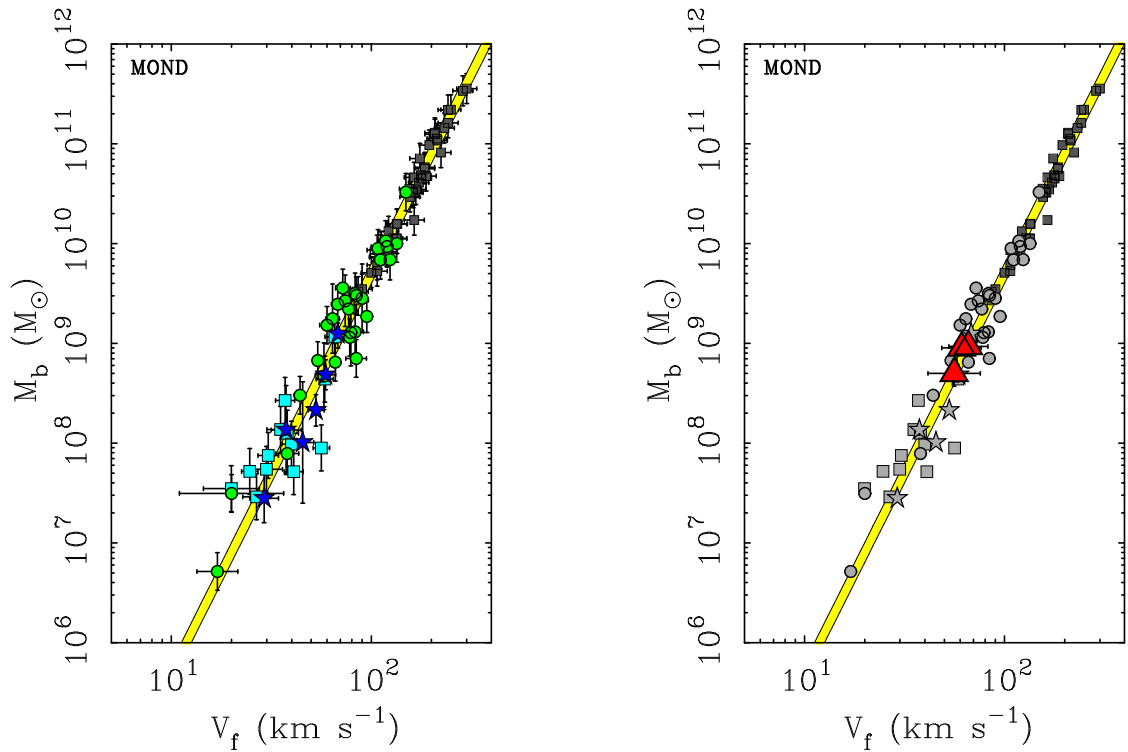}
  \caption{The baryonic Tully-Fisher relation. As
    Fig.~\ref{fig:BTF_noTDGs} and~\ref{fig:BTF_withTDGs} including the
    prediction of the Milgromian (i.e. MONDian) BTF relation (yellow
    region between the two thin solid lines. The left panel is for
    normal galaxies while the right panel includes the three TDGs from
    Fig.~\ref{fig:BTF_withTDGs} as solid (red) triangles. They lie on
    the Milgromian BTF relation disproving the validity of a
    DM-dominated SMoC and being in excellent agreement with a
    Milgromian universe. Note that $V_f$ here is equal to $V_c$ in
    Figs.~\ref{fig:BTF_noTDGs} and~\ref{fig:BTF_withTDGs}.  Note that
    the dwarf DDO~210 ($V_f \approx 15$km/sec) lies on the Milgromian
    relation. It's rotation curve and the Milgromian model are
    displayed in \cite{FM12}.  This figure was kindly made available
    by Stacy McGaugh. }\label{fig:BTF_MOND}
\end{center}
\end{figure*}

Without C/WDM an approach to understanding the physics of galaxies is
to include non-Einsteinian/non-New\-tonian dynamics. All known
observational features of galaxies do demonstrate that the currently
best available dynamics for this purpose is Milgrom's such that below
an accelaration of about~$a_0=3.6\,$pc/ Myr$^2$ gravitation is
effectively larger, as originally proposed by \cite{Milgrom83a,
  Milgrom83b, Milgrom83c, BM84}.  The MDA correlation
(Fig.~\ref{fig:acc_discr}) is one such example, and another
independent falsification of the SMoC because the physics of the
putative DM particles is independent of the local acceleration.  The
MDA correlation is a successful {\it prediction} of Milgromian
dynamics, since the data shown in Fig.~\ref{fig:acc_discr} have been
obtained long after the original formulation of Milgromian dynamics.
The observationally well-defined BTF relation and the observed
super-Keplerian rotation cur\-ves \citep{Sanders09, FM12} are other
such examples (Fig.~\ref{fig:BTF_MOND}).
That dwarf galaxies with circular velocities $V_c=V_f\approx 15$km/sec
and that the TDGs all lie on the BTF relation defined by the more
massive galaxies constitutes a brilliantly successful {\it prediction}
by Milgromian dynamics, given that such data were not available in the
early 1980s. The rotation curve of the gas-rich Local Group dwarf
galaxy DDO~210 is perfectly reproduced with Milgromian dynamics
without adjustment of parameters (fig.~25 in \citealt{FM12}). It is
the lowest point ($V_f\approx 15\,$km/sec) in Fig.~\ref{fig:BTF_MOND}.

In Milgromian dynamics, galaxy evolution and interactions have been
shown, with the first available simulations, to naturallly reproduce 
observational properties of galaxies \citep{Combes10, TC07}, and
TDGs form readily \citep{TC08}. 

Returning to the work by \cite{Samland04}, his slowly-growing DM halo
model leads to an excellent reproduction of the MW galaxy. {\it The
  physical interpretation of the Samland model is that it is
  essentially a model for the emergence of the MW in modified gravity,
  because a slowly growing DM halo without sub-structure is a first
  order approximation of the phantom DM halo associated with a
  Milgromian baryonic galaxy viewed as a Newtonian object
  \citep{Biename09, FM12}.} If anything, the work by \cite{Samland04}
and collaborators had captured a reality that had eluded the
main-stream research community.

How, then does a Milgromian cosmological model fare?

\begin{figure*}[ht]
\begin{center}
\includegraphics[scale=0.6, angle=0]{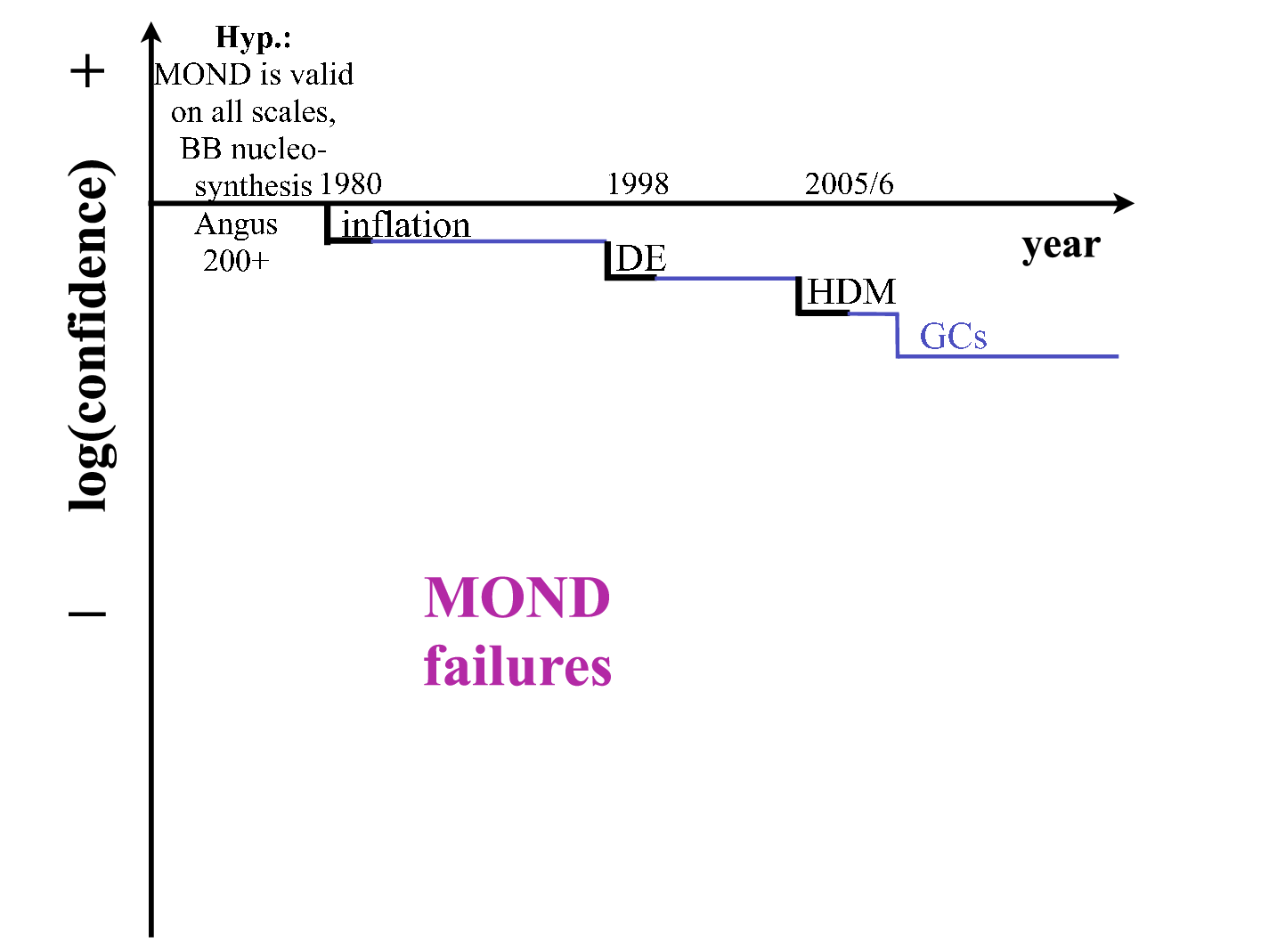}
\caption{Confidence-graph for Milgromian cosmology. As
  Fig.~\ref{fig:SMoC_decline} but assuming gravitation is given by
  Milgromian dynamics. The Milgromian-cosmological model by
  \cite{Angus09,AD11} is adopted. All problems of the SMoC on galactic
  scales vanish in this model, but the necessity of introducing
  inflation, dark energy (DE) and hot dark matter (HDM, see also
  \citealt{Slosar05,Skordis06}) indicates that this cosmological model
  may also not be complete. Also, the Milgromian-prediction of
  super-Keplerian stellar motions in distant fluffy globular clusters
  (GCs, \citealt{Baumgardt05}) has so far not been detected
  (\citealt{FM12} and references therein; see also the Ibata--Sanders
  disagreement: \citealt{Sanders12}).  }\label{fig:MOND_decline}
\end{center}
\end{figure*}

Fig.~\ref{fig:MOND_decline} shows the confidence in a Milgromian
Angus-cosmological model universe in an equivalent plot as
Fig.~\ref{fig:SMoC_decline} for the confidence in the Einstein-based
model.
The situation is definitely better for the Angus-cosmo\-lo\-gi\-cal
model, and if sterile neutrinos with a mass of about 11~eV were to be
discovered then this would massively boost confidence in
Milgromian-based cosmologies. Also, Milgromian cosmological models do
require DE. But, as emphasised in e.g. \cite{FM12}, an interesting
correspondence emerges in this model (but not in the SMoC) for the
acceleration scale $a_0$ (Fig.~\ref{fig:acc_discr}): $a_0^2/c^2
\approx \Lambda$, where $c$ is the speed of light. Why should this be
the case? Remember that $a_0$ is derived from galactic-scale problems
(e.g. using one single rotation curve). This correspondence may hint
at deeper physics of space-time we are yet to discover. Thus,
Milgromian models do not show as massive a drop in confidence as the
SMoC. The situation remains tense since the distant and extended MW
globular cluster data have until now failed to unambiguously show the
expected Milgromian behaviour, although \cite{Scarpa11} consistently
report to have observed super-Keplerian stellar motions in the
outskirts of GCs for which such data are available (see also
\citealt{Hernandez12}).

While the SMoC has been demonstrated to reproduce the large-scale
distribution of matter well, agreement with data is never a proof of a
model, and we do not yet know if a Milgromian cosmology will not also
be able to achieve the same level of agreement. The reason is that
baryonic physics is far more dominant in a Milgromian cosmology but
the available computer power is not available to achieve numerical
resolution high enough to compute the distribution of galaxies, for
example.  At the present we do know that structures form more rapidly
in a Milgromian universe than in SMoC, in consistency with the
observations (Sec.~\ref{sec:MONDcosm}), but it is not known what the
smallest structures that emerge in a Milgromian cosmology may be (dIrr
galaxies?). But {\it any} galaxy which forms in a Milgromian universe
falls on the Tully-Fisher relation, independently of whether it forms
as a TDG or as a primordial dwarf (Fig.~\ref{fig:BTF_MOND}). Indeed,
both will later appear indistinguishable to an observer, given that
the dominant fraction of TDGs would have been born soon after the BB
when the forming gas-rich galaxies were assembling and encountering
each other. A TDG forming at a late cosmological epoch may, however,
be identified by it being metal rich and lying above the
metallicity--luminosity relation of most dwarf galaxies.

How does {\sc The Dual Dwarf Galaxy Theorem} fare in a Milgromian
cosmology? Only a weak form of the theorem would be valid
(Sec.~\ref{sec:dual_theorem}): In a Milgromian universe gas-rich,
rotationally supported TDGs would lie on the BTF relation together
with gas-dominated primordial dwarf disk galaxies and primordial
star-dominated major galaxies, as is observed. And old,
pressure-supported TDGs would be identical dynamically to dE and dSph
galaxies, as is observed. Thus, in a Milgromian universe, TDGs and
primordial galaxies would only differ by TDGs that formed at later
cosmological epochs having relatively young stellar populations and
perhaps being relatively metal-rich, if they formed from pre-enriched
material.

Another alternative, Modified Gravity (MOG) (e.g. \citealt{Moffat06,
  MT09}) has been suggested and eliminates all need for DM and
DE. But, MOG must effectively become Milgromian on galaxy scales.  In
addition it has been shown that flat rotation curves of galaxies can
be accounted for without DM by brane-world models \citep{Gergely11} as
well as by $f(R)$-gravity models \citep{Capozziello09} whereby lensing
constraints are also being studied in these and other theories.

\subsection{Summarising}

Taking all the evidence together, it emerges that all arguments
converge consistently to the result that the SMoC is falsified and
that dark-matter sub-structures do not exist. If they do not exist,
then {\it DM particles that are dynamically relevant on galactic
  scales cannot exist}. This is consistent with such particles not
appearing within the SMoPP, which is at present the most successful
existing theory of physics, and the simple empirical fact that they
have not been found despite a massive world-wide search for their
existence (Sec.~\ref{sec:noDM}).

Noteworthy is that the SMoPP cannot, however, account for mass. Since
mass is the one property of particles that couples to space-time it is
perhaps not surprising that our lack of understanding of this coupling
becomes evident on the astronomical arena as a failure of our popular
formulation of gravitation when confronted with post-Einsteinian
observational data. Perhaps interesting in this context is the
complimentary principle (a material object is both a particle and a
wave): this suggests that matter has also space-time properties and
may hint at a not yet achieved unification of matter, space and time
in the sense that space-time is an emergent property of matter. {\it
  Put another way: that the currently popular SMoC needs to be based
  to more than 96~per cent on unknown physics is nothing more than an
  expression of our present-day ignorance of how mass, space and time
  unify, i.e. of cosmological physics.}  Some proposed quantum-gravity
theories have already been disfavoured through measurements of the
constancy of the speed of light with photon energy \citep{Abdo09}.

\subsection{Future tests}

Independently of the beauty and general acceptance of a model, it must
stand up to observational scrutiny\footnote{The most famous example of
  this is Galileo Galilei's telescopic observations of heavenly bodies
  that instantly shattered the since many generations cherished
  ``truth'' about the origin, structure and functioning of the
  universe. Today it is often amusingly questioned how it was possible
  for educated people to have ignored the evidence shown to them by
  Galilei through his telescope, or even how it was possible for
  people of high rank to deny looking through the telescope in the
  first place.  Then, two major intellectual steps had to be taken
  simultaneously if Galilei's observations were to be grasped by an
  individual, considering the excellent success of the geocentric
  model to account for the observed phenomena and the precise
  predictions it allowed: it had to be accepted that the Sun, and not
  the Earth, was the centre of the then known universe and it had to
  be accepted that orbits were Keplerian rather than perfectly
  circular.  }. This is true also for the conclusions of this
contribution, namely that dynamics is Milgromian {\it and} dwarf
satellite galaxies are mostly of type~B (TDGs and RPDs).

If these conclusions reached here are correct then the following ought
to hold up to future observational tests:

\begin{itemize}

\item ${\rm BTF}_{\rm TDG} = {\rm BTF}_{\rm dIRR}$ must continue to hold.
  That is, it would be important to measure rotation curves of other
  gas-rich TDGs to test if they conform to the BTF relation and thus
  to Milgromian dynamics (Fig.~\ref{fig:BTF_MOND}).

\item TDGs that are older than about one~Gyr should lie on the
  radius--mass relation of dE, dSph and UFD galaxies subject to tidal
  deformation. It would be important to determine the density profiles
  of TDGs in order to place them into the radius--mass or
  radius--luminosity diagramme (Fi\-g. \ref{fig:TDGs}). Expansion
  through gas loss from the observed gas-rich TDGs needs to be
  accounted for in the comparison. {TDGs forming today are likely to
    be subject to stronger tidal fields than in the cosmological past
    (Fig.~\ref{fig:TDGs}).}

\item If satellite galaxies are mostly TDGs then their number should
  scale with the mass of the bulge of the host galaxy. It would be
  important to survey nearby early-type disk galaxies with prominent
  bulges and nearby late-type disk galaxies with similar rotational
  velocities but no bulges to test the
  bulge-mass---number-of-satellite correlation
  (Fig.~\ref{fig:Mbulge_Nsat}).

\item Can the number of observed satellite galaxies be accounted for
  in a realistic cosmological model if they are typically TDGs and
  perhaps RPDGs?  To study this issue it would be important to perform
  high-resolution very gas-rich galaxy--galaxy encounter simulations
  as well as computations of gas-stripping from disk galaxies in
  galaxy clusters to study the formation rate of TDGs and RPDGs,
  respectively, and their survival in Milgromian dynamics.

\item Is it possible to re-create realistic events that created the MW
  VPOS from a tidal arm about 10-11~Gyr ago?

\item Hickson compact groups would not merge efficiently in Milgromian
  dynamics because the dynamical friction on DM halos would not
  exist. If the SMoC were true after all, then such groups would merge
  within about a dynamical time ($\approx1\,$Gyr).  How quickly do
  observed Hickson compact groups merge \citep{Kroupa10}?

\item Does Milgromian dynamics hold on all scales? Observe globular
  clusters and galaxy clusters.

\end{itemize}

\section*{Acknowledgments} 
An anonymous referee be thanked for a rigorous but constructive
report.  I am especially grateful to Sambaran Banerjee, Benoit Famaey,
Stacy McGaugh and Simone Recchi for valuable comments.  I also thank
Klaas de Boer, Joerg Dabringhausen, Gerhardt Hensler, Marcel
Paw\-low\-ski, Jan Pflamm-Altenburg and G\"unther Wuchterl for useful
discussions.  Figures were kindly provided by Joerg Dabringhausen,
Gianfranco Gentile, Fabian L\"ughausen, Stacy McGaugh and Marcel
Pawlowski.  This work was partially suported by the German Research
Foundation (DFG) through grants KR1635/18-1 and HE1487/36-2 within the
priority programme 1177 ``Witnesses of Cosmic History: Formation and
Evolution of Black Holes, Galaxies and Their Environment''.

I would like to express my sincere admiration for those young
researchers who dare follow their curiosity and who publish their
non-conform findings even though this may put their careers at risk.



\begin{thebibliography}{}

\bibitem[Aar\-seth \-et \-al.\-(1979)]{Aarseth79} Aarseth, S.~J., Turner, 
E.~L., \& Gott, J.~R., III 1979, ApJ, 228, 664 

\bibitem[Abdo et al.(2009)]{Abdo09} Abdo, A.~A., Ackermann, 
M., Ajello, M., et al.\ 2009, Nature, 462, 331 

\bibitem[Afshordi(2012)]{Afshordi12} Afshordi, N.\ 2012, 
arXiv:astro-ph/1203.3827 

\bibitem[Ahmed et al.(2010)]{Ahmed10} Ahmed, Z., Akerib, D.~S., et
  al. for the CDMS II Collaboration \ 2010, Science, 327, 1619,
  arXiv:astro-ph/0912.3592

\bibitem[Aliu et al.(2012)]{Aliuetal12} Aliu, E., Archambault, S., 
Arlen, T., et al.\ 2012, PhRvD, 85, 062001 

\bibitem[Amendola \& Tsujikawa(2010)]{Amendola10} Amendola, L., \&
  Tsujikawa, S.\ 2010, Dark Energy: Theory and Observations by Luca
  Amendola and Shinji Tsujikawa.~Cambridge University Press,
  2010.~ISBN: 9780521516006,

\bibitem[Anderson \& Bregman(2010)]{AB10} Anderson, M.~E., \& Bregman,
  J.~N.\ 2010, ApJ, 714, 320

\bibitem[Angloher et al.(2012)]{Angloher11} Angloher, G., Bauer, M.,
  Bavykina, I., et al.\ 2012, EPJC, 72, 1971, arXiv:astro-ph/1109.0702v1

\bibitem[Angus et al.(2006)]{AFZ06} Angus, G.~W., Famaey, B., \& Zhao,
  H.~S.\ 2006, MNRAS, 371, 138

\bibitem[Angus(2008)]{Angus08} Angus, G.~W.\ 2008, MNRAS, 387, 
1481 

\bibitem[Angus \& McGaugh(2008)]{AM08} Angus, G.~W., \& McGaugh,
  S.~S.\ 2008, MNRAS, 383, 417

\bibitem[Angus(2009)]{Angus09} Angus, G.~W.\ 2009, MNRAS, 394, 527

\bibitem[Angus et al.(2010)]{AFD10} Angus, G.~W., Famaey, B., \&
  Diaferio, A.\ 2010, MNRAS, 402, 395

\bibitem[Angus et al.(2011)]{Angus11} Angus, G.~W., Diaferio, 
A., \& Kroupa, P.\ 2011, MNRAS, 416, 1401 

\bibitem[Angus \& Diaferio(2011)]{AD11} Angus, G.~W., \& Diaferio,
  A.\ 2011, MNRAS, 417, 941

\bibitem[Aprile et al.(2011)]{Aprile11} Aprile, E., Arisaka, K., 
Arneodo, F., et al.\ 2011, Physical Review Letters, 107, 131302 

\bibitem[Aubert et al.(2004)]{Aubert04} Aubert, D., Pichon, C., 
\& Colombi, S.\ 2004, MNRAS, 352, 376 

\bibitem[Barnes \& Hernquist(1992)]{BH92} Barnes, J.~E., \& Hernquist,
  L.\ 1992, Nature, 360, 715

\bibitem[Baudis \& for the XENON Collaboration(2012)]{Baudis12}
  Baudis, L., \& for the XENON Collaboration 2012,
  arXiv:astro-ph/1203.1589

\bibitem[Baumgardt et al.(2005)]{Baumgardt05} Baumgardt, H., Grebel,
  E.~K., \& Kroupa, P.\ 2005, MNRAS, 359, L1

\bibitem[Bekenstein(2004)]{Bekenstein04} Bekenstein, J.~D.\ 2004, 
Physical Review D, 70, 083509 

\bibitem[Bekenstein \& Milgrom(1984)]{BM84} Bekenstein,
  J., \& Milgrom, M.\ 1984, ApJ, 286, 7

\bibitem[Bienaym{\'e} et al.(2009)]{Biename09} Bienaym{\'e}, O.,
  Famaey, B., Wu, X., Zhao, H.~S., \& Aubert, D.\ 2009, A\&A, 500, 801

\bibitem[Binney \& Tremaine(1987)]{BT87} Binney, J., \&
  Tremaine, S.\ 1987, Princeton, NJ, Princeton University Press, 1987,
  747 p.

\bibitem[Blanchet \& Le Tiec(2009)]{Blanchet09} Blanchet, L., \& Le
  Tiec, A.\ 2009, PhRvD, 80, 023524

\bibitem[Bludman(1998)]{Bludman98} Bludman, S.~A.\ 1998, ApJ, 508, 535

\bibitem[Blumenthal et al.(1984)]{BFPR84} Blumenthal, G.~R., Faber,
  S.~M., Primack, J.~R., \& Rees, M.~J.\ 1984, Nature, 311, 517

\bibitem[Bovill \& Ricotti(2011)]{Bovill11} Bovill, M.~S., \& Ricotti,
  M.\ 2011, ApJ, 741, 18

\bibitem[Bosma(1981)]{Bosma81} Bosma, A.\ 1981, AJ, 86, 1791 

\bibitem[Bournaud(2010)]{Bournaud10} Bournaud, F.\ 2010, Advances 
in Astronomy, 2010, Article ID 735284, arXiv:astro-ph/0907.3831

\bibitem[Bournaud \& Combes(2003)]{Bournaud03} Bournaud, F., \&
  Combes, F.\ 2003, A\&A, 401, 817

\bibitem[Bournaud et al.(2005)]{Bournaud05} Bournaud, F.,
  Jog, C.~J., \& Combes, F.\ 2005, A\&A, 437, 69

\bibitem[Bournaud et al.(2007)]{Bournaud07} Bournaud, F., Duc, 
P.-A., Brinks, E., et al.\ 2007, Science, 316, 1166 

\bibitem[Bournaud et al.(2008)]{Bournaud08} Bournaud, F., Duc, 
P.-A., \& Emsellem, E.\ 2008, MNRAS, 389, L8 

\bibitem[Bournaud et al.(2011)]{Bournaud11} Bournaud, F., Chapon, 
D., Teyssier, R., et al.\ 2011, ApJ, 730, 4 

\bibitem[Bousso(2008)]{Bousso08} Bousso, R.\ 2008, General 
Relativity and Gravitation, 40, 607 

\bibitem[Bousso(2012)]{Bousso12} Bousso, R.\ 2012, To appear in the
  proceedings of "Subnuclear Physics: Past, Present and Future",
  Pontificial Academy of Sciences, Vatican (October 2011),
  arXiv:astro-ph/1203.0307v1

\bibitem[Boylan-Kolchin et al.(2011a)]{BoylanKolchin11} Boylan-Kolchin, 
M., Bullock, J.~S., \& Kaplinghat, M.\ 2011a, MNRAS, 415, L40 

\bibitem[Boylan-Kolchin et al.(2011b)]{BKBH11} Boylan-Kolchin, 
M., Besla, G., \& Hernquist, L.\ 2011b, MNRAS, 414, 1560 

\bibitem[Brada \& Milgrom(2000)]{BM00} Brada, R., \& Milgrom, M.\
  2000, ApJ, 541, 556

\bibitem[Brandenberger(2012)]{Brandenberger12} Brandenberger, R.~H.\
  2012, 6th Aegean Summer School "Quantum Gravity and Quantum
  Cosmology", arXiv:astro-ph/1203.6698 (see also
  arXiv:astro-ph/1204.6108)

\bibitem[Brasseur et al.(2011)]{Brasseur11} Brasseur, C.~M., 
Martin, N.~F., Rix, H.-W., et al.\ 2011, ApJ, 729, 23 

\bibitem[Brook et al.(2011)]{Brook11} Brook, C.~B., Governato, 
F., Ro{\v s}kar, R., et al.\ 2011, MNRAS, 415, 1051 

\bibitem[Bruneton \& Esposito-Far{\`e}se(2007)]{Bruneton07} Bruneton,
  J.-P., \& Esposito-Far{\`e}se, G.\ 2007, PhRvD, 76, 124012

\bibitem[Bruneton et al.(2009)]{Bruneton09} Bruneton, J.-P.,
  Liberati, S., Sindoni, L., \& Famaey, B.\ 2009, Journal of Cosmology
  and Astroparticle Physics, 3, 21

\bibitem[Br{\"u}ns et al.(2011)]{Bruens11} Br{\"u}ns,
  R.~C., Kroupa, P., Fellhauer, M., Metz, M., \& Assmann, P.\ 2011,
  A\&A, 529, A138

\bibitem[Bull \& Clifton(2012)]{BullClifton12} Bull, P., \& Clifton,
  T.\ 2012, PhRvD, 85, 103512

\bibitem[Bullock(2002)]{Bullock02} Bullock, J.~S.\ 2002, The Shapes of
  Galaxies and their Dark Haloes, Proceedings of the Yale Cosmology
  Workshop, ed: Priyamvada Natarajan.  World Scientific Publishing
  Co. Pte. Ltd., 2002. ISBN 9789812778017, p. 109

\bibitem[Busha et al.(2010)]{Busha09} Busha, M.~T., Alvarez, 
M.~A., Wechsler, R.~H., Abel, T., \& Strigari, L.~E.\ 2010, ApJ, 710, 408 

\bibitem[Capozziello et al.(2009)]{Capozziello09} Capozziello,
  S., Piedipalumbo, E., Rubano, C., \& Scudellaro, P.\ 2009, A\&A,
  505, 21

\bibitem[Carlin et al.(2012)]{Carlin12} Carlin, J.~L., Majewski,
  S.~R., Casetti-Dinescu, D.~I., et al.\ 2012, ApJ, 744, 25

\bibitem[Casas et al.(2012)]{Casas12} Casas, R.~A., Arias, V., Pena
  Ramirez, K., \& Kroupa, K.\ 2012, MNRAS, in press,
  arXiv:astro-ph/1205.5029


\bibitem[Chen \& McGaugh(2010)]{Chen10} Chen, D.-M., \& McGaugh, S.\
  2010, Research in Astronomy and Astrophysics, 10, 1215

\bibitem[Chilingarian et al.(2011)]{Chilingarian11} Chilingarian, 
I.~V., Mieske, S., Hilker, M., \& Infante, L.\ 2011, MNRAS, 412, 1627 

\bibitem[Coc et al.(2012)]{Coc12} Coc, A., Goriely, S., Xu, Y.,
  Saimpert, M., \& Vangioni, E.\ 2012, ApJ, 744, 158

\bibitem[Coleman et al.(2007)]{Coleman07} Coleman, M.~G., de 
Jong, J.~T.~A., Martin, N.~F., et al.\ 2007, ApJL, 668, L43 

\bibitem[Combes(2002)]{Combes02} Combes, F.\ 2002, NewAR, 46, 755 

\bibitem[Com\-bes \& Tiret(2010)]{Combes10} Combes, F., \&
  Tiret, O.\ 2010, American Institute of Physics Conference Series,
  1241, 154

\bibitem[Cooper et al.(2010)]{Cooper10} Cooper, A.~P., Cole, S., 
Frenk, C.~S., et al.\ 2010, MNRAS, 406, 744 

\bibitem[Dabringhausen et al.(2008)]{Dabringhausen08} Dabringhausen, 
J., Hilker, M., \& Kroupa, P.\ 2008, MNRAS, 386, 864 

\bibitem[Dabringhausen \& Kroupa (2012)]{Dabringhausen12}
  Dabringhausen, J., \& Kroupa, P. \ 2012, in prep.

\bibitem[de Lucia(2012)]{deLucia12} de Lucia, G.\ 2012, Reviews in
  Modern Astronomy Vol.24, arXiv:astro-ph/1203.5208

\bibitem[Deason et al.(2011)]{Deason11} Deason, A.~J., McCarthy, 
I.~G., Font, A.~S., et al.\ 2011, MNRAS, 415, 2607 

\bibitem[de Blok(2010)]{deBlok10} de Blok, W.~J.~G.\ 2010, 
Advances in Astronomy, 2010,  article id 789293

\bibitem[{{Dekel} \& {Silk}(1986)}]{DekelSilk86}
{Dekel}, A., \& {Silk}, J. 1986, ApJ, 303, 39

\bibitem[{{Dekel} \& {Woo}(2003)}]{DekelWoo03}
{Dekel}, A., \& {Woo}, J. 2003, MNRAS, 344, 1131

\bibitem[Demers et al.(1994)]{Demers94} Demers, S., Irwin, 
M.~J., \& Kunkel, W.~E.\ 1994, AJ, 108, 1648 

\bibitem[Desmond(2012)]{Desmond12} Desmond, H.\ 2012, MNRAS,
  submitted, arXiv:astro-ph/1204.1497

\bibitem[Diemand et al.(2004)]{Diemand04} Diemand, J., Moore, B., \&
  Stadel, J.\ 2004, MNRAS, 352, 535

\bibitem[Diemand et al.(2008)]{Diemand08} Diemand, J., Kuhlen, 
M., Madau, P., et al.\ 2008, Nature, 454, 735 


\bibitem[Disney et al.(2008)]{Disney08} Disney, M.~J., Romano, 
J.~D., Garcia-Appadoo, D.~A., et al.\ 2008, Nature, 455, 1082 

\bibitem[{{D'Onghia} \& {Lake}(2008)}]{DOnghia08}
{D'Onghia}, E., \& {Lake}, G. 2008, ApJL, 686, L61

\bibitem[Drinkwater \& Gregg(1998)]{Drinkwater98} Drinkwater,
  M.~J., \& Gregg, M.~D.\ 1998, MNRAS, 296, L15

\bibitem[Drinkwater et al.(2000)]{Drinkwater00} Drinkwater, M.~J., 
Jones, J.~B., Gregg, M.~D., \& Phillipps, S.\ 2000, PASA, 17, 227 

\bibitem[Drink\-water et al.(2004)]{Drinkwater04} Drinkwater, M.~J., 
Gregg, M.~D., Couch, W.~J., et al.\ 2004, PASA, 21, 375 

\bibitem[Dubinski \& Carlberg(1991)]{DC91} Dubinski,
  J., \& Carlberg, R.~G.\ 1991, ApJ, 378, 496

\bibitem[Dubinski(1994)]{Dubinski94} Dubinski, J.\ 1994, ApJ, 
431, 617 

\bibitem[Duc \& Mirabel(1998)]{Duc98} Duc, P.-A., \&
  Mirabel, I.~F.\ 1998, A\&A, 333, 813

\bibitem[Duc et al.(2007)]{Duc07} Duc, P.-A., Braine, J., Lisenfeld,
  U., Brinks, E., \& Boquien, M.\ 2007, A\&A, 475, 187

\bibitem[Duc et al.(2011)]{Duc11} Duc, P.-A., Cuillandre, J.-C.,
  Serra, P., et al.\ 2011, MNRAS, 417, 863

\bibitem[Dutton \& van den Bosch(2012)]{Dutton12} Dutton, A.~A., \&
  van den Bosch, F.~C.\ 2012, MNRAS, 421, 608

\bibitem[Einstein(1916)]{Einstein16} Einstein, A.\ 1916, Annalen 
der Physik, 354, 769 

\bibitem[Epelbaum et al.(2011)]{Epelbaum11} Epelbaum, E., Krebs, H.,
  Lee, D., \& Mei{\ss}ner, U.-G.\ 2011, Physical Review Letters, 106,
  192501

\bibitem[Esposito-Farese(2009)]{Esposito09} Esposito-Farese, G.\ 
2009, arXiv:astro-ph/0905.2575 

\bibitem[Faber \& Gallagher(1979)]{FG79} Faber, S.~M., \& Gallagher,
  J.~S.\ 1979, Annual Reviews in A\&A, 17, 135

\bibitem[Famaey \& McGaugh(2012)]{FM12} Famaey, B., \& McGaugh, S.\
  2012, Living Reviews in Relativity, arXiv:astro-ph/1112.3960

\bibitem[Fanelli(2010)]{Fanelli10} Fanelli, D., 2010, PLoS ONE 5(4): e10271.
Doi:10.1371/journal.pone.0010271

\bibitem[Ferguson \& Binggeli(1994)]{FB94} Ferguson,
  H.~C., \& Binggeli, B.\ 1994, Astronomy and Astrophysics Review, 6, 67

\bibitem[Ferrero et al.(2011)]{Ferrero11} Ferrero, I., Abadi, M.~G.,
  Navarro, J.~F., Sales, L.~V., \& Gurovich, S.\ 2011,
  arXiv:astro-ph/1111.6609v1

\bibitem[Font et al.(2011)]{Font11} Font, A.~S., Benson, 
A.~J., Bower, R.~G., et al.\ 2011, MNRAS, 417, 1260 

\bibitem[Forbes et al.(2008)]{Forbes08} Forbes, D.~A., Lasky, P.,
  Graham, A.~W., \& Spitler, L.\ 2008, MNRAS, 389, 1924

\bibitem[Forbes et al.(2011)]{Forbes11} Forbes, D.~A., Spitler, 
L.~R., Graham, A.~W., et al.\ 2011, MNRAS, 413, 2665 

\bibitem[Forbes \& Kroupa(2011)]{FK11} Forbes, D.~A., \& Kroupa, P.\
  2011, PASA, 28, 77

\bibitem[Forero-Romero et al.(2011)]{FR11} Forero-Romero, J.~E.,
  Hoffman, Y., Yepes, G., et al.\ 2011, MNRAS, 417, 1434

\bibitem[Freire \& Wex(2010)]{Freire10} Freire, P.~C.~C., \& Wex, N.\
  2010, MNRAS, 409, 199

\bibitem[Galianni et al.(2010)]{Galianni10} Galianni, P., Patat, F.,
  Higdon, J.~L., Mieske, S., \& Kroupa, P.\ 2010, A\&A, 521, A20

\bibitem[Gao et al.(2011)]{Gao11} Gao, L., Frenk, C.~S., 
Jenkins, A., Springel, V., \& White, S.~D.~M.\ 2012, MNRAS, 419, 1721

\bibitem[Gentile et al.(2007)]{Gentile07} Gentile, G., Famaey, B.,
  Combes, F., et al.\ 2007, A\&A, 472, L25

\bibitem[Gentile et al.(2009)]{Gentile09} Gentile, G., Famaey, 
B., Zhao, H., \& Salucci, P.\ 2009, Nature, 461, 627 

\bibitem[Gergely et al.(2011)]{Gergely11} Gergely, L.~{\'A}., 
Harko, T., Dwornik, M., Kupi, G., \& Keresztes, Z.\ 2011, MNRAS, 415, 3275 

\bibitem[Ghigna et al.(1998)]{Ghigna98} Ghigna, S., Moore, B., 
Governato, F., et al.\ 1998, MNRAS, 300, 146 

\bibitem[Gilmore et al.(2007a)]{Gilmore07a} Gilmore, G., Wilkinson, 
M., Kleyna, J., et al.\ 2007a, Nuclear Physics B Proceedings Supplements, 
173, 15 

\bibitem[Gilmore et al.(2007b)]{Gilmore07b} Gilmore, G., Wilkinson, 
M.~I., Wyse, R.~F.~G., et al.\ 2007b, ApJ, 663, 948 

\bibitem[Gonzalez et al.(2012)]{Gonzalez12} Gonzalez, A.~H., Stanford,
  S.~A., Brodwin, M., et al.\ 2012, ApJ, in press,
  arXiv:astro-ph/1205.3788

\bibitem[Governato et al.(2012)]{Governato12} Governato, F., Zolotov,
  A., Pontzen, A., et al.\ 2012, MNRAS, 422, 1231

\bibitem[Graham \& Worley(2008)]{GW08} Graham, A.~W., \& Worley,
  C.~C.\ 2008, MNRAS, 388, 1708

\bibitem[Guth \& Tye(1980)]{GT80} Guth, A.~H., \& Tye, S.-H.~H.\ 1980,
  Physical Review Letters, 44, 631

\bibitem[Hammer et al.(2005)]{Hammer05} Hammer, F., Flores, H., Elbaz,
  D., et al.\ 2005, A\&A, 430, 115

\bibitem[Hammer et al.(2007)]{Hammer07} Hammer, F., Puech, M., 
Chemin, L., Flores, H., \& Lehnert, M.~D.\ 2007, ApJ, 662, 322 

\bibitem[Hammer et al.(2009)]{Hammer09} Hammer, F., Flores, H., Yang,
  Y.~B., et al.\ 2009, A\&A, 496, 381

\bibitem[Hammer et al.(2010)]{Hammer10} Hammer, F., Yang, Y.~B., 
Wang, J.~L., et al.\ 2010, ApJ, 725, 542 

\bibitem[Harris(1996)]{Harris96} Harris, W.~E.\ 1996, AJ, 112, 1487

\bibitem[Haud(1988)]{Haud88} Haud, U.\ 1988, A\&A, 198, 125 

\bibitem[Hayashi et al.(2003)]{Hayashi03} Hayashi, E., Navarro, 
J.~F., Taylor, J.~E., Stadel, J., \& Quinn, T.\ 2003, ApJ, 584, 541 

\bibitem[Hayes et al.(2011)]{Hayes12} Hayes, B., Brunner, R., \& Ross,
  A.\ 2012, arXiv:astro-ph/1112.5723

\bibitem[Hernandez et al.(2010)]{Hernandez10} Hernandez, X.,
  Mendoza, S., Suarez, T., \& Bernal, T.\ 2010, A\&A, 514, A101

\bibitem[Hernandez \& Jim{\'e}nez(2012)]{Hernandez12} Hernandez, X.,
  \& Jim{\'e}nez, M.~A.\ 2012, ApJ, 750, 9

\bibitem[Hilker et al.(1999)]{Hilker99} Hilker, M., Infante, L.,
  Vieira, G., Kissler-Patig, M., \& Richtler, T.\ 1999, A\&AS, 134, 75

\bibitem[House et al.(2011)]{House11} House, E.~L., Brook, C.~B.,
  Gibson, B.~K., et al.\ 2011, MNRAS, 415, 2652

\bibitem[Hubble(1929)]{Hubble29} Hubble, E.~P.\ 1929, ApJ, 69, 103

\bibitem[Hunter et al.(2000)]{Hunter00} Hunter, D.~A., 
Hunsberger, S.~D., \& Roye, E.~W.\ 2000, ApJ, 542, 137 

\bibitem[Ibata et al.(2001)]{Ibata01} Ibata, R., Lewis, G.~F., 
Irwin, M., Totten, E., \& Quinn, T.\ 2001, ApJ, 551, 294 

\bibitem[Jardel \& Gebhardt(2012)]{JG12} Jardel, J.~R., \& Gebhardt,
  K.\ 2012, ApJ, 746, 89

\bibitem[Jee et al.(2012)]{Jee12} Jee, M.~J., Mahdavi, A., 
Hoekstra, H., et al.\ 2012, ApJ, 747, 96 

\bibitem[Karachentsev(1996)]{Karachentsev96} Karachentsev, I.\ 1996,
  A\&A, 305, 33

\bibitem[{{Karachentsev} {et~al.}(2005){Karachentsev},
    {Karachentseva}, \& {Sharina}}]{Karachentsevetal05}
  {Karachentsev}, I.~D., {Karachentseva}, V.~E., \& {Sharina},
  M.~E. 2005, in IAU Colloq. 198: Near-fields cosmology with dwarf
  elliptical galaxies, ed.  H.~{Jerjen} \& B.~{Binggeli}, 295--302

\bibitem[Karachentsev(2012)]{Karachentsev12}  Karachentsev, I. D.\ 2012
     Astrophys.\ Bull.\ 67, 123, arXiv:astro-ph/1204.3377
 
\bibitem[Kaviraj et al.(2012)]{Kaviraj12} Kaviraj, S., Darg, D., 
Lintott, C., Schawinski, K., \& Silk, J.\ 2012, MNRAS, 419, 70 

\bibitem[Kazantzidis et al.(2004)]{Kazantzidis04} Kazantzidis, S., 
Mayer, L., Mastropietro, C., et al.\ 2004, ApJ, 608, 663 

\bibitem[{{Kirby} {et~al.}(2009){Kirby}, {Guhathakurta}, {Bullock}, {Frebel},
  {Geha}, {Gilbert}, {Kalirai}, {Kaplinghat}, {Kuhlen}, {Majewski},
  {Robertson}, {Simon}, \& {Zemp}}]{Kirby09}
{Kirby}, E.~N., {Guhathakurta}, P., {Bullock}, J.~S., {et~al.} 2009, Astro2010: 
The Astronomy and Astrophysics Decadal Survey, Science White Papers, no. 156

\bibitem[Klessen \& Kroupa(1998)]{KK98} Klessen, R.~S., \& Kroupa, P.\
  1998, ApJ, 498, 143

\bibitem[Kleyna et al.(1998)]{Kleyna98} Kleyna, J.~T., Geller, 
M.~J., Kenyon, S.~J., Kurtz, M.~J., 
\& Thorstensen, J.~R.\ 1998, AJ, 115, 2359 

\bibitem[Kleyna et al.(2003)]{Kleyna03} Kleyna, J.~T., Wilkinson,
  M.~I., Gilmore, G., \& Evans, N.~W.\ 2003, ApJL, 588, L21

\bibitem[Klypin et al.(1999)]{Klypin99} Klypin, A., Kravtsov, A.~V.,
  Valenzuela, O., \& Prada, F.\ 1999, ApJ, 522, 82

\bibitem[Knebe et al.(2008)]{Knebe08} Knebe, A., Arnold, B., 
Power, C., \& Gibson, B.~K.\ 2008, MNRAS, 386, 1029 

\bibitem[Koch \& Grebel(2006)]{KG06} Koch, A., \& Grebel, E.~K.\ 2006,
  AJ, 131, 1405

\bibitem[{{Koposov} {et~al.}(2009){Koposov}, {Yoo}, {Rix}, {Weinberg},
  {Macci{\`o}}, \& {Escud{\'e}}}]{Koposovetal09}
{Koposov}, S.~E., {Yoo}, J., {Rix}, H.-W., {et~al.} 2009, ApJ, 696, 2179

\bibitem[Koposov et al.(2010)]{KRH10} Koposov, S.~E., Rix, 
H.-W., \& Hogg, D.~W.\ 2010, ApJ, 712, 260 

\bibitem[Kormendy et al.(2010)]{Kormendy10} Kormendy, J., Drory, 
N., Bender, R., \& Cornell, M.~E.\ 2010, ApJ, 723, 54 

\bibitem[Kowalski et al.(2008)]{Kowalski08} Kowalski, M., Rubin, 
D., Aldering, G., et al.\ 2008, ApJ, 686, 749 

\bibitem[Klimentowski et al.(2007)]{Klimentowski07} Klimentowski, J., 
{\L}okas, E.~L., Kazantzidis, S., et al.\ 2007, MNRAS, 378, 353 

\bibitem[Kravtsov et al.(2004)]{Kravtsov04} Kravtsov, A.~V., 
Gnedin, O.~Y., \& Klypin, A.~A.\ 2004, ApJ, 609, 482 

\bibitem[Kroupa(1997)]{Kroupa97} Kroupa, P.\ 1997, NewA, 2, 139 

\bibitem[Kroupa(1998)]{Kroupa98} Kroupa, P.\ 1998, MNRAS, 300, 200

\bibitem[Kroupa(2008)]{Kroupa08} Kroupa, P.\ 2008, The Cambridge 
N-Body Lectures, LNP, 760, 181 

\bibitem[Kroupa(2011)]{Kroupa11} Kroupa, P.\ 2011, Stellar Clusters \&
  Associations: A RIA Workshop on Gaia, 17, arXiv:1111.5613

\bibitem[Kroupa et al.(2005)]{Kroupa05} Kroupa, P., Theis, C., \&
  Boily, C.~M.\ 2005, A\&A, 431, 517

\bibitem[Krou\-pa et al.(2010)]{Kroupa10} Kroupa, P., Famaey, B., de
  Boer, K.~S., et al.\ 2010, A\&A, 523, A32

\bibitem[Kroupa et al.(2012)]{Kroupa12} Kroupa, P., Weidner, C.,
  Pflamm-Altenburg, J., et al.\ 2012, arXiv:astro-ph/1112.3340

\bibitem[K{\"u}pper et al.(2010)]{Kuepper10} K{\"u}pper, A.~H.~W.,
  Kroupa, P., Baumgardt, H., \& Heggie, D.~C.\ 2010, MNRAS, 401, 105

\bibitem[Kunkel(1979)]{Kunkel79} Kunkel, W.~E.\ 1979, ApJ, 228, 
718 

\bibitem[Lada \& Lada(2003)]{LL03} Lada, C.~J., \& Lada, E.~A.\ 2003,
  ARA\&A, 41, 57

\bibitem[Leauthaud et al.(2012)]{Leauthaud12} Leauthaud, A., Tinker,
  J., Bundy, K., et al.\ 2012, ApJ, 744, 159

\bibitem[Lee \& Komatsu(2010)]{LK10} Lee, J., \& Komatsu, E.\ 2010,
  ApJ, 718, 60

\bibitem[Lokas(2011)]{Lokas11} Lokas, E.~L.\ 2011, proceedings of XXXV
  International Conference of Theoretical Physics "Matter to the
  Deepest: Recent Developments in Physics of Fundamental
  Interactions", held in Ustron, Poland in September 2011; Acta
  Physica Polonica B, 42, 2185, arXiv:astro-ph/1112.0438

\bibitem[{{Li} \& {Helmi}(2008)}]{Li08} {Li}, Y.-S., \& {Helmi},
  A. 2008, MNRAS, 385, 1365

\bibitem[Li \& Zhao(2009)]{LiZhao09} Li, B., \& Zhao, H.\ 2009, PhRvD,
  80, 064007

\bibitem[Liu \& Li(2012)]{LiuLi12} Liu, H., \& Li, T.-P.\ 2012,
  arXiv:astro-ph/1203.5720

\bibitem[Libeskind et al.(2009)]{Libeskind09} Libeskind, N.~I., Frenk,
  C.~S., Cole, S., Jenkins, A., \& Helly, J.~C.\ 2009, MNRAS, 399, 550

\bibitem[Libeskind et al.(2011)]{Libeskind11} Libeskind, N.~I.,
Knebe, A., Hoffman, Y., et al.\ 2011, MNRAS, 411, 1525 

\bibitem[Liesenborgs et al.(2008)]{Liesenborgs08} Liesenborgs, J., de
  Rijcke, S., Dejonghe, H., \& Bekaert, P.\ 2008, MNRAS, 389, 415

\bibitem[Lisker(2009)]{Lisker09} Lisker, T.\ 2009, Astronomische 
Nachrichten, 330, 1043 

\bibitem[Llinares et al.(2008)]{Llinares08} Llinares, C., Knebe, 
A., \& Zhao, H.\ 2008, MNRAS, 391, 1778 

\bibitem[Lovell et al.(2011)]{Lovell11} Lovell, M.~R., Eke, 
V.~R., Frenk, C.~S., \& Jenkins, A.\ 2011, MNRAS, 413, 3013 

\bibitem[Lynden-Bell(1976)]{LyndenBell76} Lynden-Bell, D.\ 1976, 
MNRAS, 174, 695 

\bibitem[Macci{\`o} \& Fontanot(2010)]{MF09} Macci{\`o}, A.~V., \&
  Fontanot, F.\ 2010, MNRAS, 404, L16

\bibitem[{{Macci{\`o}} {et~al.}(2009){Macci{\`o}}, {Kang}, \& {Moore}}]{MKM09}
{Macci{\`o}}, A.~V., {Kang}, X., \& {Moore}, B. 2009, ApJL, 692, L109 

\bibitem[Macci{\`o} et al.(2010)]{Maccioetal10} Macci{\`o}, A.~V., 
Kang, X., Fontanot, F., Somerville, R.~S., Koposov, S., 
\& Monaco, P.\ 2010, MNRAS, 402, 1995 

\bibitem[Maccio' et al.(2012)]{Maccioetal12} Maccio', A.~V., Paduroiu,
  S., Anderhalden, D., Schneider, A., \& Moore, B.\ 2012, MNRAS,
  in press, arXiv:astro-ph/1202.1282

\bibitem[Mackey \& van den Bergh(2005)]{Mackey05} Mackey, A.~D., \&
  van den Bergh, S.\ 2005, MNRAS, 360, 631

\bibitem[Mahdavi et al.(2007)]{Mahdavi07} Mahdavi, A., Hoekstra, 
H., Babul, A., Balam, D.~D., \& Capak, P.~L.\ 2007, ApJ, 668, 806 

\bibitem[Marino et al.(2010)]{Marino10} Marino, A., Bianchi, L.,
  Rampazzo, R., Buson, L.~M., \& Bettoni, D.\ 2010, A\&A, 511, A29

\bibitem[Marra et al.(2012)]{Marra12} Marra, V.,
  P{\"a}{\"a}kk{\"o}nen, M., \& Valkenburg, W.\ 2012,
  arXiv:astro-ph/1203.2180

\bibitem[Mayer et al.(2001)]{Mayer01} Mayer, L., Governato, F., 
Colpi, M., et al.\ 2001, ApJ, 559, 754 

\bibitem[Majewski et al.(2003)]{Majewski03} Majewski, S.~R.,
  Skrutskie, M.~F., Weinberg, M.~D., \& Ostheimer, J.~C.\ 2003, ApJ,
  599, 1082

\bibitem[Marks \& Kroupa(2010)]{MK10} Marks, M., \& Kroupa, P.\ 2010,
  MNRAS, 406, 2000

\bibitem[Martig et al.(2012)]{Martig12} Martig, M., Bournaud, F.,
  Croton, D.~J., Dekel, A., \& Teyssier, R.\ 2012,
  arXiv:astro-ph/1201.1079

\bibitem[Mart{\'{\i}}nez-Delgado et al.(2004)]{Martinez04} 
Mart{\'{\i}}nez-Delgado, D., G{\'o}mez-Flechoso, M.~{\'A}., Aparicio, A., 
\& Carrera, R.\ 2004, ApJ, 601, 242 

\bibitem[Mart{\'{\i}}nez-Delgado et al.(2010)]{Martinez10} 
Mart{\'{\i}}nez-Delgado, D., Gabany, R.~J., Crawford, K., et al.\ 2010, 
AJ, 140, 962 

\bibitem[Martinez-Delgado et al.(2012)]{Martinez11} Martinez-Delgado,
  D., Romanowsky, A.~J., Gabany, R.~J., et al.\ 2012, ApJ, 748, 24

\bibitem[Mateo(1998)]{Mateo98} Mateo, M.~L.\ 1998, Ann.Rev. of
  A. \&A., 36, 435

\bibitem[Mateo et al.(1993)]{Mateo93} Mateo, M., Olszewski, 
E.~W., Pryor, C., Welch, D.~L., \& Fischer, P.\ 1993, AJ, 105, 510 

\bibitem[Maciejewski et al.(2011)]{Mac11} Maciejewski, M., 
Vogelsberger, M., White, S.~D.~M., \& Springel, V.\ 2011, MNRAS, 415, 2475 

\bibitem[McGaugh(2004)]{McGaugh04} McGaugh, S.~S.\ 2004, ApJ, 609, 652

\bibitem[McGaugh(2005)]{McGaugh05} McGaugh, S.~S.\ 2005, ApJ, 632, 859

\bibitem[McGaugh(2011)]{McGaugh11} McGaugh, S.~S.\ 2011, Physical 
Review Letters, 106, 121303 

\bibitem[McGaugh(2012)]{McGaugh12} McGaugh, S.~S.\ 2012, AJ, 
143, 40 

\bibitem[McGaugh \& de Blok(1998)]{McGaugh98} McGaugh, S.~S., \& de
  Blok, W.~J.~G.\ 1998, ApJ, 499, 41

\bibitem[McGaugh \& Wolf(2010)]{McGaugh10} McGaugh, S.~S.,
  \& Wolf, J.\ 2010, ApJ, 722, 248

\bibitem[McGaugh et al.(2010)]{McGaugh2010b} McGaugh, S.~S.,
  Schombert, J.~M., de Blok, W.~J.~G., \& Zagursky, M.~J.\ 2010, ApJL,
  708, L14

\bibitem[Menci et al.(2012)]{Menci12} Menci, N., Fiore, F., \&
  Lamastra, A.\ 2012, MNRAS, 421, 2384

\bibitem[Merrifield(2002)]{Merrifield02} Merrifield, M.~R.\ 2002, The
  Shapes of Galaxies and their Dark Haloes, Proceedings of the Yale
  Cosmology Workshop, ed: Priyamvada Natarajan.  World Scientific
  Publishing Co. Pte. Ltd., 2002. ISBN 9789812778017, p. 170

\bibitem[Metz \& Kroupa(2007)]{MK07} Metz, M., \& Kroupa, P.\ 2007,
  MNRAS, 376, 387

\bibitem[Metz et al.(2007)]{Metz07} Metz, M., Kroupa, P., 
\& Jerjen, H.\ 2007, MNRAS, 374, 1125 

\bibitem[Metz et al.(2008)]{Metz08} Metz, M., Kroupa, P., 
\& Libeskind, N.~I.\ 2008, ApJ, 680, 287 

\bibitem[Metz et al.(2009a)]{Metz09} Metz, M., Kroupa, P., 
\& Jerjen, H.\ 2009a, MNRAS, 394, 2223 

\bibitem[Metz et al.(2009b)]{Metz09b} Metz, M., Kroupa, P., 
Theis, C., Hensler, G., \& Jerjen, H.\ 2009b, ApJ, 697, 269 

\bibitem[Mieske et al.(2002)]{Mieske02} Mieske, S., Hilker,
  M., \& Infante, L.\ 2002, A\&, 383, 823

\bibitem[Mieske et al.(2012)]{Mieske12} Mieske, S., Hilker,
  M., \& Misgeld, I.\ 2012, A\&A, 537, A3

\bibitem[Milgrom(1983a)]{Milgrom83a} Milgrom, M.\ 1983a, ApJ, 270, 
365 

\bibitem[Milgrom(1983b)]{Milgrom83b} Milgrom, M.\ 1983b, ApJ, 270, 
371 

\bibitem[Milgrom(1983c)]{Milgrom83c} Milgrom, M.\ 1983c, ApJ, 270, 
384 

\bibitem[Milgrom(1999)]{Milgrom99} Milgrom, M.\ 1999, Physics Letters
  A, 253, 273

\bibitem[Milgrom(2007)]{Milgrom07} Milgrom, M.\ 2007, ApJL, 667, 
L45 

\bibitem[Milgrom(2009)]{Milgrom09} Milgrom, M.\ 2009, PhRvD, 80, 
123536 

\bibitem[Milgrom(2010)]{Milgrom10} Milgrom, M.\ 2010, MNRAS, 
403, 886 

\bibitem[Mirabel et al.(1992)]{Mirabel92} Mirabel, I.~F.,
  Dottori, H., \& Lutz, D.\ 1992, A\^A, 256, L19

\bibitem[Miralles-Caballero et al.(2012)]{MCA12} Miralles-Caballero,
  D., Colina, L., \& Arribas, S.\ 2012, A\&A, 538, A61

\bibitem[Misgeld \& Hilker(2011)]{Misgeld11} Misgeld, I., \& Hilker,
  M.\ 2011, MNRAS, 414, 3699

\bibitem[Modesto \& Randono(2010)]{ModestoRandano10} Modesto, L., \&
  Randono, A.\ 2010, arXiv:hep-th/1003.1998

\bibitem[Moffat(2006)]{Moffat06} Moffat, J.~W.\ 2006, Journal of
  Cosmology and Astroparticle Physics, 3, 4

\bibitem[Moffat \& Toth(2009a)]{MT09a} Moffat, J.~W., \& Toth, V.~T.\
  2009a, MNRAS, 397, 1885

\bibitem[Moffat \& Toth(2009b)]{MT09} Moffat, J.~W., \&
  Toth, V.~T.\ 2009b, Classical and Quantum Gravity, 26, 085002

\bibitem[Moore et al.(1999)]{Moore99} Moore, B., Ghigna, S., 
Governato, F., et al.\ 1999, ApJL, 524, L19 

\bibitem[Murray(2009)]{Murray09} Murray, N.\ 2009, ApJ, 691, 946

\bibitem[Nagashima et al.(2005)]{NLO05} Nagashima, M., Lacey, C.~G.,
  Okamoto, T., et al.\ 2005, MNRAS, 363, L31

\bibitem[Navarro \& Benz(1991)]{NB91} Navarro, J.~F., \& Benz, W.\
  1991, ApJ, 380, 320

\bibitem[Nichols \& Bland-Hawthorn(2011)]{Nichols11} Nichols, M., \&
  Bland-Hawthorn, J.\ 2011, ApJ, 732, 17

\bibitem[Nusser(2002)]{Nusser02} Nusser, A.\ 2002, MNRAS, 331, 909

\bibitem[Nusser et al.(2005)]{Nusser05} Nusser, A., Gubser, S.~S., \&
  Peebles, P.~J.\ 2005, PhRvD, 71, 083505

\bibitem[Okabe \& Umetsu(2008)]{OU08} Okabe, N., \& Umetsu, K.\ 2008,
  PASJ, 60, 345

\bibitem[Okamoto \& Frenk(2009)]{OF09} Okamoto, T., \&
  Frenk, C.~S.\ 2009, MNRAS, 399, L174

\bibitem[Okazaki \& Taniguchi(2000)]{OT00} Okazaki, T., \& Tani\-gu\-chi,
  Y.\ 2000, ApJ, 543, 149

\bibitem[Olling \& Merrifield(2000)]{OM00} Olling, R.~P., \&
  Merrifield, M.~R.\ 2000, MNRAS, 311, 361

\bibitem[Olling \& Merrifield(2001)]{OM01} Olling, R.~P., \&
  Merrifield, M.~R.\ 2001, MNRAS, 326, 164

\bibitem[Opik(1922)]{Opik22} Opik, E.\ 1922, ApJ, 55, 406 

\bibitem[Pawlow\-ski et al.(2011)]{Pawlowski11} Pawlowski, M.~S.,
  Kroupa, P., \& de Boer, K.~S.\ 2011, A\&A, 532, A118

\bibitem[Pawlow\-ski et al.(2012a)]{Pawlowski12} Pawlowski, M.,
  Kroupa, P., Angus, G., de Boer, K.S., Famaey, B., Hensler, G. \
  2012a, MNRAS, in press, arXiv:astro-ph/1204.6039 

\bibitem[Pawlow\-ski et al.(2012b)]{Pawlowski12b} Pawlowski, M.,
  Pflamm-Altenburg, J., Kroupa, P. \ 2012b, MNRAS, in press,
  arXiv:astro-ph/1204.5176

\bibitem[Peebles \& Nusser(2010)]{PN10} Peebles, P.~J.~E., \& Nusser,
  A.\ 2010, Nature, 465, 565

\bibitem[Perlmutter et al.(1999)]{Perlmutter99} Perlmutter, S., 
Aldering, G., Goldhaber, G., et al.\ 1999, ApJ, 517, 565 

\bibitem[Pflamm-Altenburg \& Kroupa(2008)]{PAK08} Pflamm-Altenburg,
  J., \& Kroupa, P.\ 2008, Nature, 455, 641

\bibitem[Pflamm-Altenburg \& Kroupa(2009)]{PAK09}
  Pflamm-Altenburg, J., \& Kroupa, P.\ 2009, ApJ, 706, 516

\bibitem[Piontek \& Steinmetz(2011)]{PS11} Piontek, F., \& Steinmetz,
  M.\ 2011, MNRAS, 410, 2625

\bibitem[Pipino et al.(2009)]{Pipino09} Pipino, A.,
  Devriendt, J.~E.~G., Thomas, D., Silk, J., \& Kaviraj, S.\ 2009,
  A\&A, 505, 1075

\bibitem[Popper(1934)]{Popper34} Popper, K.\ 1934, Logik der
Forschung, Mohr Siebeck (publ.), ISBN 3-16-148410-X

\bibitem[Popper(1957)]{Popper57} Popper, K.\ 1957, The Poverty of
  Historicism, Routledge Classics (publ.), ISBN 0-415-27845-7 9 780415 2784541

\bibitem[{{Primack}(2009)}]{Primack09} {Primack}, J.~R. 2009, American
  Institute of Physics Conference Series, 3, 1166

\bibitem[Recchi et al.(2007)]{Recchi07} Recchi, S., Theis, C., Kroupa,
  P., \& Hensler, G.\ 2007, A\&A, 470, L5

\bibitem[Recchi et al.(2009)]{Recchi09} Recchi, S., Calura, F., \&
  Kroupa, P.\ 2009, A\&A, 499, 711

\bibitem[Reshetnikov \& Combes(1997)]{Reshetnikov97} Reshetnikov, V.,
  \& Combes, F.\ 1997, A\&A, 324, 80

\bibitem[Reverte et al.(2007)]{Reverte07} Reverte, D., V{\'{\i}}lchez,
  J.~M., Hern{\'a}ndez-Fern{\'a}ndez, J.~D., \& Iglesias-P{\'a}ramo,
  J.\ 2007, AJ, 133, 705

\bibitem[Riess et al.(1998)]{Riess98} Riess, A.~G., Filippenko, 
A.~V., Challis, P., et al.\ 1998, AJ, 116, 1009 

\bibitem[Rubin \& Ford(1970)]{RF70} Rubin, V.~C., \& Ford, W.~K., Jr.\
  1970, ApJ, 159, 379

\bibitem[Ruzicka et al.(2007)]{RPT07} Ruzicka, 
A., Palous, J., \& Theis, C.\ 2007, A\&A, 461, 155

\bibitem[Sale et al.(2010)]{Sale10} Sale, S.~E., Drew, J.~E., Knigge,
  C., et al.\ 2010, MNRAS, 402, 713

\bibitem[Samland(2004)]{Samland04} Samland, M.\ 2004, PASA, 21, 
175 

\bibitem[Sand et al.(2011)]{Sand11} Sand, D.~J., Strader, J., 
Willman, B., et al.\ 2011, arXiv:astro-ph/1111.6608 

\bibitem[Sanders(2005)]{Sanders05} Sanders, R.~H.\ 2005, MNRAS, 363,
  459

\bibitem[Sanders(2009)]{Sanders09} Sanders, R.~H.\ 2009, Advances in
  Astronomy, 2009, article id. 752439

\bibitem[Sanders(2010a)]{Sanders10a} Sanders, R.~H.\ 2010a, The Dark 
Matter Problem: A Historical Perspective by Robert H.~Sanders.~Cambridge 
University Press, 2010.~ISBN: 9780521113014  

\bibitem[Sanders(2010b)]{Sanders10b} Sanders, R.~H.\ 2010b, MNRAS, 
407, 1128 

\bibitem[Sanders(2012)]{Sanders12} Sanders, R.~H.\ 2012, MNRAS, 422, 21

\bibitem[Sato(1981)]{Sato81} Sato, K.\ 1981, MNRAS, 195, 467 

\bibitem[Scannapieco et al.(2012)]{Scannapieco12} Scannapieco, C.,
  Wadepuhl, M., Parry, O.~H., et al.\ 2012, MNRAS, in press,
  arXiv:astro-ph/1112.0315

\bibitem[Scarpa et al.(2011)]{Scarpa11} Scarpa, R., Marconi, G.,
  Carraro, G., Falomo, R., \& Villanova, S.\ 2011, A\&A, 525, A148

\bibitem[Schmidt et al.(1998)]{Schmidt98} Schmidt, B.~P., 
Suntzeff, N.~B., Phillips, M.~M., et al.\ 1998, ApJ, 507, 46 

\bibitem[{{Shaya} {et~al.}(2009){Shaya}, {Olling}, {Ricotti}, {Vogel},
  {Majewski}, {Patterson}, {Allen}, {van der Marel}, {Brown}, {Bullock},
  {Burkert}, {Combes}, {Gnedin}, {Grillmair}, {Kulkarni}, {Guhathakurta},
  {Helmi}, {Johnston}, {Kroupa}, {Lake}, {Moore}, \& {Tully}}]{Shaya2009}
{Shaya}, E., {Olling}, R., {Ricotti}, M., {et~al.} 2009, Astro2010: 
The Astronomy and Astrophysics Decadal Survey, Science White Papers, no. 274

\bibitem[Skordis et al.(2006)]{Skordis06} Skordis, C., Mota, D.~F.,
  Ferreira, P.~G., \& B{\oe}hm, C.\ 2006, Physical Review Letters, 96,
  011301

\bibitem[Skordis(2009)]{Skordis09} Skordis, C.\ 2009, Classical 
and Quantum Gravity, 26, 143001 

\bibitem[Slosar et al.(2005)]{Slosar05} Slosar, A., Melchiorri, A., \&
  Silk, J.~I.\ 2005, PhRvD, 72, 101301

\bibitem[Smale \& Wiltshire(2011)]{SmaleWiltshire11} Smale, P.~R., \& Wiltshire,
  D.~L.\ 2011, MNRAS, 413, 367

\bibitem[Smolin(2006)]{Smolin06} Smolin, L.\ 2006, The trouble 
with physics : the rise of string theory, the fall of a science, and what 
comes next, by Lee Smolin.~Boston: Houghton Mifflin, 2006 xxiii, 392 p.~ 
ISBN 9780618551057  

\bibitem[Starkman et al.(2012)]{Starkman12} Starkman, G.~D., Copi, 
C.~J., Huterer, D., \& Schwarz, D.\ 2012, arXiv:astro-ph/1201.2459 

\bibitem[Stoehr et al.(2002)]{Stoehr02} Stoehr, F., White, S.~D.~M.,
  Tormen, G., \& Springel, V.\ 2002, MNRAS, 335, L84

\bibitem[Strigari et al.(2008)]{Strigari08} Strigari, L.~E., 
Bullock, J.~S., Kaplinghat, M., et al.\ 2008, Nature, 454, 1096 

\bibitem[Strigari \& Wechsler(2012)]{SW11} Strigari, L.~E., \&
  Wechsler, R.~H.\ 2012, ApJ, 749, 75

\bibitem[Swaters et al.(2011)]{Swaters11} Swaters, R.~A., Sancisi, R.,
  van Albada, T.~S., \& van der Hulst, J.~M.\ 2011, ApJ, 729, 118

\bibitem[Thomas et al.(2002)]{Thomas02} Thomas, D., Maraston, 
C., \& Bender, R.\ 2002, Reviews in Modern Astronomy, 15, 219 

\bibitem[Thompson \& Nagamine(2012)]{TN11} Thompson, R., \& Nagamine,
  K.\ 2012, MNRAS, 419, 3560

\bibitem[Tikhonov et al.(2009)]{Tikhonov09} Tikhonov, A.~V.,
  Gottl{\"o}ber, S., Yepes, G., \& Hoffman, Y.\ 2009, MNRAS, 399, 1611

\bibitem[Tiret \& Combes(2007)]{TC07} Tiret, O., \& Combes, F.\ 2007,
  A\&A, 464, 517

\bibitem[Tiret \& Combes(2008)]{TC08} Tiret, O., \& Combes, F.\ 2008,
  Formation and Evolution of Galaxy Disks, ASP Conference Series, 396,
  259

\bibitem[Tegmark et al.(2002)]{Tegmark02} Tegmark, M., Dodelson, 
S., Eisenstein, D.~J., et al.\ 2002, ApJ, 571, 191 

\bibitem[Tollerud et al.(2008)]{Tollerud08} Tollerud, E.~J., 
Bullock, J.~S., Strigari, L.~E., \& Willman, B.\ 2008, ApJ, 688, 277 

\bibitem[Tollerud et al.(2012)]{Tollerud11} Tollerud, E.~J., Beaton,
  R.~L., Geha, M.~C., et al.\ 2012, ApJ, 752, 45

\bibitem[Trachternach et al.(2009)]{Trachternach09} Trachternach, C.,
  de Blok, W.~J.~G., McGaugh, S.~S., van der Hulst, J.~M., \& Dettmar,
  R.-J.\ 2009, A\&A, 505, 577

\bibitem[Tran et al.(2003)]{Tran03} Tran, H.~D., Sirianni, M., 
Ford, H.~C., et al.\ 2003, ApJ, 585, 750 

\bibitem[Tsujimoto(2011)]{Tsujimoto11} Tsujimoto, T.\ 2011, ApJ, 736,
  113

\bibitem[Tsujimoto \& Bekki(2012)]{Tsujimoto12} Tsujimoto, T., \&
  Bekki, K.\ 2012, ApJ, 747, 125

\bibitem[Tully et al.(2006)]{Tully06} Tully, R.~B., Rizzi, L., 
Dolphin, A.~E., et al.\ 2006, AJ, 132, 729 

\bibitem[van den Aarssen et al.(2012)]{Aarssen12} van den Aarssen,
  L.~G., Bringmann, T., \& Pfrommer, C.\ 2012,
  arXiv:astro-ph/1205.5809

\bibitem[Vera-Ciro et al.(2011)]{VC11} Vera-Ciro, C.~A., 
Sales, L.~V., Helmi, A., et al.\ 2011, MNRAS, 416, 1377 

\bibitem[Vera-Ciro et al.(2012)]{VC12} Vera-Ciro, C.~A., Helmi, A.,
  Starkenburg, E., \& Breddels, M.~A.\ 2012, MNRAS, submitted,
  arXiv:astro-ph/1202.6061

\bibitem[Verlinde(2011)]{Verlinde11} Verlinde, E.\ 2011, Journal of
  High Energy Physics, 4, 29

\bibitem[Walcher et al.(2003)]{Walcher03} Walcher, C.~J.,
  Fried, J.~W., Burkert, A., \& Klessen, R.~S.\ 2003, A\&A, 406, 847

\bibitem[Wang \& White(2012)]{WW12} Wang, W., \& White, S.~D.~M.\
  2012, MNRAS, arXiv:astro-ph/1203.0009v1

\bibitem[Wang et al.(2012)]{WangFrenk12} Wang, J., Frenk, C.~S., 
Navarro, J.~F., \& Gao, L.\ 2012, arXiv:astro-ph/1203.4097v1

\bibitem[Wei et al.(2010)]{Wei10} Wei, L.~H., Kannappan, 
S.~J., Vogel, S.~N., \& Baker, A.~J.\ 2010, ApJ, 708, 841 

\bibitem[Weidner et al.(2004)]{WKL04} Weidner, C., Kroupa, 
P., \& Larsen, S.~S.\ 2004, MNRAS, 350, 1503 

\bibitem[Weidner et al.(2011)]{WKP11} Weidner, C., Kroupa, 
P., \& Pflamm-Altenburg, J.\ 2011, MNRAS, 412, 979 

\bibitem[Weilbacher et al.(2002)]{Weilbacher02} Weilbacher, P.~M., 
Fritze-v.~Alvensleben, U., Duc, P.-A., 
\& Fricke, K.~J.\ 2002, ApJL, 579, L79 

\bibitem[Wess \& Zumino(1974)]{WZ74} Wess, J., \& Zumino, B.\ 1974,
  Nuclear Physics B, 70, 39

\bibitem[Wess \& Bagger(1992)]{WB92} Wess, J., Bagger, J., 1992,
  Supersymmetry and Supergravity, Princeton University Press,
  Princeton. ISBN 0-691-02530-4

\bibitem[Wess \& Akulov(1998)]{WA98} Wess, J., \&
  Akulov, V.~P.\ 1998, Supersymmetry and Quantum Field Theory, Vol. 509

\bibitem[Wetzstein et al.(2007)]{Wetzstein07} Wetzstein, M., Naab, 
T., \& Burkert, A.\ 2007, MNRAS, 375, 805 

\bibitem[Whitmore et al.(1990)]{Whitmore90} Whitmore, B.~C., 
Lucas, R.~A., McElroy, D.~B., et al.\ 1990, AJ, 100, 1489 

\bibitem[Willman \& Strader(2012)]{WS12} Willman, B., \& Strader, J. \
  2012, AJ, submitted, arXiv:astro-ph/1203.2608 

\bibitem[Wilt\-shire\-(2009)]{Wiltshire09} Wiltshire, D.~L.\ 2009,
  PhRvD, 80, 123512

\bibitem[Wolf et al.(2010)]{Wolf10} Wolf, J., Martinez, G.~D., 
Bullock, J.~S., et al.\ 2010, MNRAS, 406, 1220 

\bibitem[Yagi et al.(2010)]{Yagi10} Yagi, M., Yoshida, M., Komiyama,
  Y., et al.\ 2010, AJ, 140, 1814

\bibitem[Yang 
\& Hammer(2010)]{YH10} Yang, Y., \& Hammer, F.\ 2010, ApJL, 725, L24 

\bibitem[Yao(2006)]{Yao06} Yao, W.-M., et al.\ 2006, J. Phys. G:
  Nucl. Part. Phys. 33, 1, doi:10.1088/0954-3899/33/1/001

\bibitem[Yoshida et al.(2008)]{Yoshida08} Yoshida, M., Yagi, M.,
  Komiyama, Y., et al.\ 2008, ApJ, 688, 918

\bibitem[Zentner \& Bullock(2003)]{ZB03} Zentner, A.~R., \& Bullock,
  J.~S.\ 2003, ApJ, 598, 49

\bibitem[Zhao(1998)]{Zhao98} Zhao, H.\ 1998, ApJL, 500, L149 

\bibitem[Zhao(2008)]{Zhao08} Zhao, H.~S.\ 2008, Journal of Physics
  Conference Series, 140, 012002

\bibitem[Zhao(2007)]{Zhao07} Zhao, H.\ 2007, ApJL, 671, L1 

\bibitem[Zhao \& Famaey(2010)]{ZF10} Zhao, H., \& Famaey, B.\ 2010,
  PhRvD, 81, 087304

\bibitem[Zhao \& Li(2010)]{ZhaoLi10} Zhao, H., \& Li, B.\ 2010, ApJ,
  712, 130

\bibitem[Zlosnik et al.(2007)]{Zlosnik07} Zlosnik, T.~G., 
Ferreira, P.~G., \& Starkman, G.~D.\ 2007, PhRvD, 75, 044017 

\bibitem[Zwicky(1937)]{Zwicky37} Zwicky, F.\ 1937, ApJ, 86, 217 

\bibitem[Zwicky(1956)]{Zwicky56} Zwicky, F.\ 1956, Ergebnisse 
der exakten Naturwissenschaften, 29, 344 


\end{thebibliography}
\end{document}